\RequirePackage[l2tabu, orthodox]{nag}

\documentclass[12pt,phd,a4paper,oneside]{ucl_thesis}

\usepackage{blindtext}

\usepackage{emptypage}

\usepackage{graphicx}

\usepackage{float}

\usepackage{amsmath}

\usepackage{gensymb}
\usepackage{textcomp}

\usepackage{setspace}
\usepackage{multirow}

\usepackage[backend=biber,style=numeric-comp,url=false,doi=false,isbn=false,backref=false,hyperref,maxbibnames=99]{biblatex}

\usepackage[format=hang,font=small,labelfont=bf]{caption}

\usepackage{etoolbox}

\usepackage{xspace}
\usepackage{url}
\usepackage{breakurl}
\usepackage{enumitem}
\usepackage{booktabs}
\usepackage{multirow}
\usepackage{balance}
\usepackage{xstring}
\usepackage{xcolor}
\newif \ifcomments \commentstrue
\ifcomments
    \newcommand{\sarah}[1]{\textsf{\color{violet}{[Sarah: {#1}]}}}
    \newcommand{\h}[1]{\textsf{\color{purple}{[H: {#1}]}}}
    \newcommand{\todo}[1]{\begin{large}{\textbf{\color{red}{ [TODO: {#1}] }}}\end{large}}
\else
    \newcommand{\sarah}[1]{}
    \newcommand{\h}[1]{}
    \newcommand{\todo}[1]{}
\fi

\usepackage[utf8]{inputenc}
\usepackage{xspace}
\usepackage{url}

\usepackage{enumitem}
\usepackage{booktabs}
\usepackage{multirow}
\usepackage{balance}
\usepackage{csvsimple}
\usepackage{array, makecell}
\usepackage{mhchem}
\dblfloatsep=\floatsep
\dbltextfloatsep=\textfloatsep
\usepackage{amsthm}
\theoremstyle{definition}

\usepackage{tabularx}
\usepackage{makecell}
\usepackage{array}
\usepackage{graphicx}
\usepackage{xstring}
\usepackage{xcolor}
\usepackage{supertabular}
\newcommand{\ethaddr}[1]{\textcolor{blue}{\href{https://etherscan.io/address/#1}{\StrLeft{#1}{4}...}}}%
\newcommand{\trxaddr}[1]{\textcolor{blue}{\href{https://tronscan.org/\#/address/#1}{\StrLeft{#1}{4}...}}}%
\newcommand{\bscaddr}[1]{\textcolor{blue}{\href{https://bscscan.org/\#/address/#1}{\StrLeft{#1}{4}...}}}

\input{LinksAndMetadata}

\begin{document}

\title{Investigating transactions in cryptocurrencies}
\author{Haaroon M. Yousaf}
\department{Department of Computer Science}

\maketitle
\makedeclaration

\begin{abstract} %

This thesis presents techniques to investigate transactions in uncharted cryptocurrencies and services. 
Cryptocurrencies are used to securely send payments online. 
Payments via the first cryptocurrency, Bitcoin, use pseudonymous addresses that have limited privacy and anonymity guarantees. 
Research has shown that this pseudonymity can be broken, allowing users to be tracked using clustering and tagging heuristics. 
Such tracking allows crimes to be investigated. 
If a user has coins stolen, investigators can track addresses to identify the destination of the coins. 
This, combined with an explosion in the popularity of blockchain, has led to a vast increase in new coins and services.
These offer new features ranging from coins focused on increased anonymity to scams shrouded as smart contracts. 
In this study, we investigated the extent to which transaction privacy has improved and whether users can still be tracked in these new ecosystems.  
We began by analysing the privacy-focused coin Zcash, a Bitcoin-forked cryptocurrency, that is considered to have strong anonymity properties due to its background in cryptographic research. We revealed that the user anonymity set can be considerably reduced using heuristics based on usage patterns. 
Next, we analysed cross-chain transactions collected from the exchange ShapeShift, revealing that users can be tracked as they move across different ledgers. 
Finally, we present a measurement study on the smart-contract pyramid scheme Forsage, a scam that cycled \$267 million USD (of Ethereum) within its first year, showing that at least 88\% of the participants in the scheme suffered a loss. 
The significance of this study is the revelation that users can be tracked in newer cryptocurrencies and services by using our 
new heuristics, which informs those conducting investigations and developing these technologies.

\end{abstract}

\begin{impactstatement} %

The work presented in this thesis is intended to inform those conducting cryptocurrency investigations and may facilitate the design and development of projects within distributed ledger ecosystems, particularly with respect to privacy and security. 

The work shown in Chapter~\ref{chapterlabelZcash} was responsibly disclosed to the founders of Zcash, who responded with updated privacy recommendations and best practices for users~\cite{zcashcoprivacy, zcachcobestpractice}. It was covered by multiple media outlets~\cite{Cox2018, Floyd2018,BitcoinExchangeGuideNewsTeam2018, AdrianColyer2018} and is mentioned in the later revisions of the Zcash protocol specification~\cite{Hopwood2017}. 

The work in Chapter~\ref{chapterlabelTracing} was responsibly disclosed to ShapeShift and was covered by the MIT Technology Review\cite{Orcutt2019}. The work introduces new tracing heuristics and highlights criminal activity with case studies. These heuristics can be applied not just to ShapeShift, but to platforms which offer cross-currency trading. Weeks before we published the research, The Wall Street Journal published an investigation into ShapeShift which paralleled our work~\cite{Scheck2018}, and which was responded to by ShapeShift's CEO\cite{Erik2018}. When the work was originally published, the ShapeShift exchange did not have a Know-Your-Customer (KYC)/Anti-money laundering policy, however this has since been introduced.

The analysis in Chapter~\ref{chapterlabelForsage} will inform those investigating cryptocurrency scams. The work presents a multi-angled analysis of a large pyramid scheme operating on the Ethereum cryptocurrency. Since publication, we have been contacted by a law enforcement investigator requesting advice on how to proceed with active investigations on similar scams.

All the work published in this thesis has been uploaded to open conference proceedings and open access repositories. As of November 27, 2021 the work in this thesis had over 147 citations. The relevant source code for each project has been shared publicly to Github~\cite{githubsourcecode}. Talks at each of the conference proceedings are freely available to view online. This will facilitate impact for those researching into improving privacy and combatting scams in distributed ledgers.

\end{impactstatement}

\begin{acknowledgements}
This thesis would not have been possible without support from my primary supervisor, Professor Sarah Meiklejohn.
Since our first run-in at Euston Station, there was never a discussion where I was not left with increased motivation and knowledge. I am truly indebted for all the support. A special thanks to my closest collaborator, George Kappos.

I have been privileged to work with many across the globe and would also like to thank all of my co-authors and contributors (in alphabetical order): Sarah Allen, Sarah Azouvi, Ben Steer, Sergi Delgado-Segura, Bernhard Haslhofer, Alex Hicks, Ari Juels, Sanket Kanjalkar, Tyler Kell, Mary Maller, Andrew Miller, Ania Piotrowska, Pierre Reibel, Sofia Rollet and Rainer Stuetz.

This research was funded with a PhD scholarship which I would like to thank the European Commission and The Initiative For Cryptocurrencies \& Contracts (IC3) for providing. 

Finally, I would like to especially thank my parents, Farzana and Muhammad Yousaf, for keeping my stomach full, my head dry and letting me skip years of chores in the name of science. Without their support, I would not have come this far. It is true when they say that no one is self-made. 

\end{acknowledgements}

\setcounter{tocdepth}{2} 
\phantomsection %
\tableofcontents
\listoffigures
\listoftables

\chapter{Introduction}
\label{chapterlabel1}

Bitcoin alleviates the problems of centralisation and censorship within a financial system. 
Anyone has the freedom to create a cryptocurrency wallet and have coins sent to them, simply by providing their public key. 
No user identification, passport or verification is needed to create such a wallet. 
This bypasses traditional financial Know Your Customer (KYC) principles, which inform a set of rules used by financial services in order to identify users prior to conducting business, and thereby anticipate and prevent crime. %
With this freedom, coins can be freely sent to any address without discrimination and free from censorship. These coins cannot be returned or reversed (unless the new owner explicitly does so). By default the system offers pseudonymity, as mentioned in the Bitcoin white paper, transaction privacy is supposedly preserved~\cite{Nakamoto2008b}. %

Notably, users in oppressed circumstances can purchase and freely send coins without the fear of banks or governments impeding the transaction, and have done so with Bitcoin~\cite{anna_baydakova_danny,robinson_2021}. 
There is no risk from hyper-inflation as there is only a fixed number of mintable coins. 
Donations to charitable organisations can be made without the sender revealing who they are~\cite{bergman2021revealing,jayasinghe2017philanthropy}. However, such a system has flaws, and some may argue that those flaws are within its pseudonymity, which allows users to obscure their identity. 

For some, this acts as a layer of privacy, preventing their financial transactions from being monitored. 
For others, this acts as a shroud which allows them to conceal their crimes, such as stealing coins or selling illegitimate goods. 
Such crimes may need to be investigated. 
For example, a victim wants to track coins stolen from them to discover if they were sent to an exchange, in the hopes of identifying the criminal and collecting their funds. 
However, if KYC has not been adopted by that particular exchange, it is much more difficult to discover the identity of the criminal. 
Alternatively, cryptocurrency exchanges may want to ensure the coins they accept from users 
were neither stolen nor gained nefariously. 
These instances, amongst many others, illustrate the need to be able to track coins. 

Scientific research has shown that Bitcoin is not private, and entities can be identified and tracked through the use of clustering and wallet identification techniques~\cite{Meiklejohn2013a,Herrera-Joancomart2015,ober2013structure,6113303}. 
This, combined with the openness of the Bitcoin source code, has inspired others to create so called privacy coins. 
Privacy coins are alternative cryptocurrencies to Bitcoin which improve the underlying user privacy with new features. For example: Zcash has introduced the notion of a shielded pool which uses zero-knowledge proofs to obscure the properties of transactions~\cite{zcash}; Dash uses Coinjoins (as PrivateSpend) which allows users to perform transactions together, to make it difficult to identify which sender paid which recipient~\cite{dash,coinjoin}; and Monero uses ring signatures to create mix-ins allowing users to include keys of other users within their own transaction to increase their anonymity set~\cite{monero,noether2015ring}. 

Parallel to privacy coins, the ecosystem has seen an introduction in companies allowing users 
to freely trade between different coins. 
July 2014 saw the announcement of cryptocurrency exchange ShapeShift, a service that allowed users to trade coins across different cryptocurrency ledgers without the need for KYC. The service officially launched in 2015, and for the first three years operated without any identity checks, until forced to do so by regulators~\cite{ss-id}. 
In 2017, the Wannacry ransomware hackers reportedly used ShapeShift to move their illicitly acquired cryptocurrencies~\cite{wannacry}. 

With respect to privacy coins, one should also ask whether the claims by the developers of these hold true. 
Given the avenue for cross-currency trading and potential for crime, one must ask whether cross-currency trading has any affect on user privacy.

The advancements of cryptocurrency technology has resulted in an increase in new coins and services in the ecosystem. 
As of June 15, 2021, CoinMarketCap lists over 10,000 different cryptocurrencies~\cite{coinmarketcap}. 
Given these opportunities, criminals are quick to exploit new technologies to scam those who are not as technologically literate. 
For example, Bitconnect, a ponzi scheme, was introduced in 2016 and sold a blockchain-based coin that claimed to offer a high rate of return~\cite{bitconnectfinder,bitconnectlending}.
The scheme was ultimately closed after regulators ruled that it was a scam, causing the value of the coin to tumble by 92\%~\cite{bitconnect26b}. 
Similarly Wotoken, a ponzi scheme claiming large profits due to advanced trading bots, raised over \$1.1 billion USD before being shutdown by law enforcement~\cite{helms_2020}. 
As scams continue to appear, it becomes important to understand the magnitude and dynamics of these schemes.

In this thesis, we empirically analyse transactions in new cryptocurrencies and services. We investigate transaction privacy on the blockchain by empirically measuring the privacy coin Zcash, developing strategies to trace coins on the cross-trading service ShapeShift, and reveal the scale of a modern day smart-contract pyramid scheme, Forsage. We question whether coins offer the privacy they promise and, if not, what techniques can be introduced to defeat them.

\section{Scope and Contributions}

Our study analysed public transactions that occurred on-chain within the public ledger. 
Off-chain privacy, for example in software such as the lightning network~\cite{poon2016bitcoin} (an off-chain payment protocol), has been implemented but was beyond the scope of this research. 
At the start of this study there was no evidence of any published  %
techniques that analysed Zcash~\cite{217535}, cross-currency trading~\cite{236358} or thoroughly investigated pyramid schemes in cryptocurrencies~\cite{kell2021forsage}. 

First, this work offers the reader a foundation in (Chapter~\ref{chapterlabel2}). 
We start with presenting a background on cryptocurrencies, in particular Bitcoin and its relevant components. 
Then we explain the anonymity of Bitcoin and state-of-the-art techniques used to defeat anonymity/privacy via address clustering, along with countermeasures proposed by the community. 
We end with some of the core concepts used in privacy coins. 

In Chapter~\ref{chapterlabelLitReview} we present a literature review of the ecosystem. We start with an overview of the research that preceded cryptocurrencies. We then discuss the literature used to defeat and improve anonymity in both Bitcoin and privacy coins and end with a review of the various aspects of crime in the blockchain ecosystem. 
 
Our first contribution to analysing transaction privacy is presented in Chapter~\ref{chapterlabelZcash}. 
Here we analyse the privacy coin Zcash, finding that address clustering and tagging is a very viable technique, 
privacy guarantees are severely limited due to a small anonymity set, 
and users perform unwise transactions with traceable patterns that damage the privacy of themselves and others.
We end with a case study of a prominent hacker collective that used the coin to sell security vulnerabilities.  

With the rise of alternative cryptocurrencies, exchanges began to provide trading services, allowing users to directly swap coins between different cryptocurrencies. In Chapter~\ref{chapterlabelTracing} we present one of the first academic analyses of cross-currency trading. We show techniques for tracing users moving across chains, heuristics clustering cross-chain user addresses and multiple case studies showcasing criminal use.

Having been around for over a century, pyramid schemes have scammed users out of billions of dollars~\cite{hidajat2020predator}. 
In Chapter~\ref{chapterlabelForsage} we contribute to other analyses of this crime in blockchain and present an in-depth empirical study of Forsage, a smart-contract pyramid scheme on Ethereum. 
We explain how the obfuscated smart contract , using a purpose-built transaction simulator, quantify the gains and losses, and study the promotional videos showing how the promoters leverage the new technology of smart contracts to lure users. 

The question that then arises is about who are the 'others' who would implement such work. We define 'others' as the following; Fellow scientists who discover and implement mediation's which would improve user privacy, thus thwarting our attacks. Scientists who produce follow-up work with even more attacks, as well as investigators who use our heuristics to measure risk in their services, or trace transactions to combat crime. 

\section{Included Work}
\label{sectionscheduleofwork}

Parts of this thesis have been published in the following papers. All papers are joint work unless otherwise stated. 

\begin{itemize}
 \item  George Kappos, Haaroon Yousaf, Mary Maller, and Sarah Meiklejohn. An Empirical Analysis of Anonymity in Zcash.  In 27th USENIX Security Symposium (USENIX Security 18),  pages  463–477,  Baltimore,  MD,  8  2018, USENIX Association, \url{https://www.usenix.org/conference/usenixsecurity18/presentation/kappos}. Source code: \url{https://github.com/manganese/zcash-empirical-analysis/}. Included in \textbf{Chapter~\ref{chapterlabelZcash}.}
 
 \item  Haaroon Yousaf,  George Kappos, and Sarah Meiklejohn. Tracing Transactions Across Cryptocurrency Ledgers. In 28th USENIX Security Symposium (USENIX Security 19), pages 837–850, Santa Clara, CA, 8 2019. USENIX Association. Paper:  \url{https://www.usenix.org/conference/usenixsecurity19/presentation/yousaf}. Source code: \url{https://github.com/manganese/tracingTransactionsAcrossCryptocurrencyLedgers}. Included in \textbf{Chapter~\ref{chapterlabelTracing}.}
 
\end{itemize}
 
Other parts of the thesis are under submission to conferences, and have been published in pre-print. 

\begin{itemize}
 \item Tyler Kell, Haaroon Yousaf, Sarah Allen, Sarah Meiklejohn and Ari Juels. Forsage: Anatomy of a Smart-Contract Pyramid Scheme. In: \textit{arXiv preprint arXiv:2105.04380}, 2021. Paper: 
  \url{https://arxiv.org/abs/2105.04380}. Source code: \url{https://github.com/initc3/forsage}. Included in \textbf{Chapter~\ref{chapterlabelForsage}.}
\end{itemize}

\section{Additional Work}

The following papers were published, as part of my research, and are not included in this thesis. 

\begin{itemize}
	\item  George  Kappos,  Haaroon  Yousaf,  Ania  M.  Piotrowska,  Sanket  Kanjalkar, Sergi Delgado-Segura, Andrew Miller, and Sarah Meiklejohn.  An Empirical Analysis of Privacy in the Lightning Network. In International Conference on Financial Cryptography and Data Security. Springer, 2021. Paper:  \url{https://fc21.ifca.ai/papers/130.pdf}. %
    \item Pierre Reibel, Haaroon Yousaf, and Sarah Meiklejohn. Short Paper:  An Exploration  of  Code  Diversity  in  the  Cryptocurrency  Landscape.   In International Conference on Financial Cryptography and Data Security, pages 73–83. Springer, 2019. Paper:  \url{http://fc19.ifca.ai/preproceedings/134-preproceedings.pdf}.
    \item George Kappos, Haaroon Yousaf, Rainer Stuetz, Sofia Rollet, Bernhard Haslhofer and Sarah Meiklejohn. How to Peel a Million: Validating and Expanding Bitcoin Clusters. arXiv pre-print 2021, \url{TBC}.
\end{itemize}

\section{Work Done in Collaboration}

A large part of the work in this thesis was completed in collaboration with researchers across the globe, all of whom are listed in Section~\ref{sectionscheduleofwork}.

In \textbf{Chapter~\ref{chapterlabelZcash}}, Sarah Meiklejohn discovered that the Zcash shielded pool was leaking non-trivial information that could damage the privacy of users.  
I managed and processed blockchain data used for the project, general statistics, %
tagging analysis and leading the case study on the hacker collective with Mary Maller. 
Joint work includes the clustering heuristic with Sarah Meiklejohn and tag collection with Mary Maller. George Kappos analysed the interactions with the pool and shielded pool. 

In \textbf{Chapter~\ref{chapterlabelTracing}}, I wrote the scraping tool to collect all data (including all blockchain nodes and exchanges), statistics, cross-currency tracing via pass-through and patterns of ShapeShift Usage (excluding Trading bots). With regards to joint work, all authors contributed to identifying blockchain transactions and I worked on the clustering analysis with Sarah Meiklejohn.  George Kappos worked on tracking cross-currency u-turns, round trip and the trading bots.  

In ~\textbf{Chapter~\ref{chapterlabelForsage}}, I processed the data from the Ethereum nodes and performed the contract measurements study. All authors contributed to the  proposed solutions. Tyler Kell worked on contract deconstruction. Sarah Allen and Tyler Kell worked on the community-dynamics study. 

With regards to work not included in this thesis. In the paper, \textit{Why is a Ravencoin Like a TokenDesk? An Exploration of Code Diversity in the Cryptocurrency Landscape}~\cite{reibel2019short}, I came up with the study and proposed the breakdown of work. Sarah Meiklejohn led and wrote the paper and worked with Pierre Rebel who performed the scraping and analysis. 

In the paper, \textit{An Empirical Analysis of Privacy in the Lightning Network}~\cite{Kappos2021}, I ran and managed our Bitcoin and lightning nodes, created and performed the property heuristic and extended the previous version of the balance discovery attack based upon the original from Sergi Delgado-Segura. 

In the paper, \textit{How to Peel a Million: Validating and Expanding Bitcoin Clusters} (under re-submission), I worked on identifying the transaction and address features, and jointly worked on the algorithms with George Kappos.

\chapter{Background} \label{chapterlabel2}

\section{Cryptocurrencies}

This chapter details the basic components needed to gain an understanding of cryptocurrencies. 

\subsection{Bitcoin}

The first decentralised electronic cryptocurrency, Bitcoin, was
created by Satoshi Nakamoto in 2008~\cite{Nakamoto2008b}.  
A white paper explaining the architecture and design was originally sent as a link to the metzdowd cryptography email list in October 2008, and was shortly followed by the release of Bitcoin open source software in January 2009~\cite{satoshifirstmail}. 
That paper lays the foundation for a decentralised financial system through a digital asset class called cryptocurrencies. 

By definition, cryptocurrencies are a form of digital currency coupled with cryptography and a blockchain. 
A blockchain is a distributed ledger which contains a record of all transactions. 
In Bitcoin, the complete record is public, viewable and verifiable by all participants and thus reduces the need 
for a central authority. 
All participants can choose to have a copy of the ledger, allowing it to be decentralised. 
Public key cryptography is used in Bitcoin to maintain the integrity and authenticity of each transaction. %

Digital coins are used as a medium of exchange and allowed to be freely traded between users without fear of censorship.    
All coins are stored in wallets each of which has one or many public key(s), or wallet address(es), and has one or many private key(s). 
Users can generate as many wallets and keys as needed, without the requirement for any verification of identity.

Coin minting and supply is fixed within the protocol, and awarded to users who maintain security by mining.  
Transaction finality is protected by this consensus of miners.  

This was the first system of its kind that allows users to transfer digital cash without the fear of double spending and need for any intermediaries, effectively a non-custodial, trustless and decentralised global payments network. 
On 8th November 2021, the value of Bitcoin soared to the highest currently recorded: \$67,566.83 USD per Bitcoin. 
This innovation has led to the development of new research fields and new cryptocurrencies (e.g., Ethereum, ZCash and Monero). 
According to CoinMarketCap, as of November 27th 2021 there are over 14,000 cryptocurrencies with a total market capitalisation of 2.4 trillion USD~\cite{coinmarketcap}.

\subsection{Blockchain components}

In this section we explore the core foundational components used in blockchain technologies. We begin with Keys and Wallets which are responsible for holding, sending and receiving coins. Then we discuss the format of transactions which are used to send and receive coins. Finally, we discuss the structure of blocks and role of miners. Further information about the cryptography can be found in various textbooks~\cite{narayanan2016bitcoin,boneh2017graduate}.

\subsubsection*{Keys and Wallets}
Bitcoin's foundation is built with its use of public key cryptography. 
This allows the user to create two related but different keys: a private key and a (derived) public key. 

The private key is the secret component that controls and signs transactions, akin to a bank PIN code which is used to authorise money that is spent.
However, unlike a PIN code, the private key also acts as the vault to store coins, and if leaked allows all coins to be compromised. 
If a user looses their private key, this would also mean a loss of access to the coins belonging to that key. 

Public keys are derived from private keys.  
These are the user's Bitcoin address and digital fingerprint, with workings akin to a bank account number. \textit{19Yq6pRM3mRUZMbWZoBpRWhNehiQHqznGR} is an example of a public key used in Bitcoin.
Users send coins to public keys, similar to the way in which cash is transferred to bank account numbers.
When sending coins, the user must sign the transaction with a unique signature created from the associated private key. 
The transaction, signature and public address combination is then verified by users of the network to ensure the user owns the coins being spent. 
Both keys, public and private, are generated using the Elliptic Curve Digital Signature Algorithm (ECDSA)~\cite{johnson2001elliptic}. 
Key management is performed via digital wallets. Just like a physical wallet which can hold one or more cards/cash, digital wallets hold private and public keys. 
The wallet is controlled by the software the user chooses to run, for example in Bitcoin this is done via the \textit{bitcoind} program. 

A user can spend funds from multiple private keys in the same transaction, and thus signs the transaction which each of the corresponding keys. 
The signature is stored within the transaction inside the input field, which allows the network to verify that the coins have been rightly spent. 
Keys are pseudonymous and are not tied to any physical identity. However, if the identity of a public key is revealed, then blockchain analytics and heuristics can be used to identify all transactions that involved the said key, thus tying and identifying the financial habits of the user.  

\subsubsection*{Transactions}

A transaction is the structure used to transfer coins. 
It has inputs, outputs and features.
Inputs are the addresses from which coins are received, and outputs are addresses to which coins are sent.  The number of inputs and outputs is determined by the user, but there must be at least one input and one output in a transaction. Features are attributes on transactions which trigger certain conditions. For example, a Bitcoin transaction can set a feature called locktime which means the transaction is only valid once the specific time has passed.

\begin{figure}[h]
	\centering
	\includegraphics[width=0.8\linewidth]{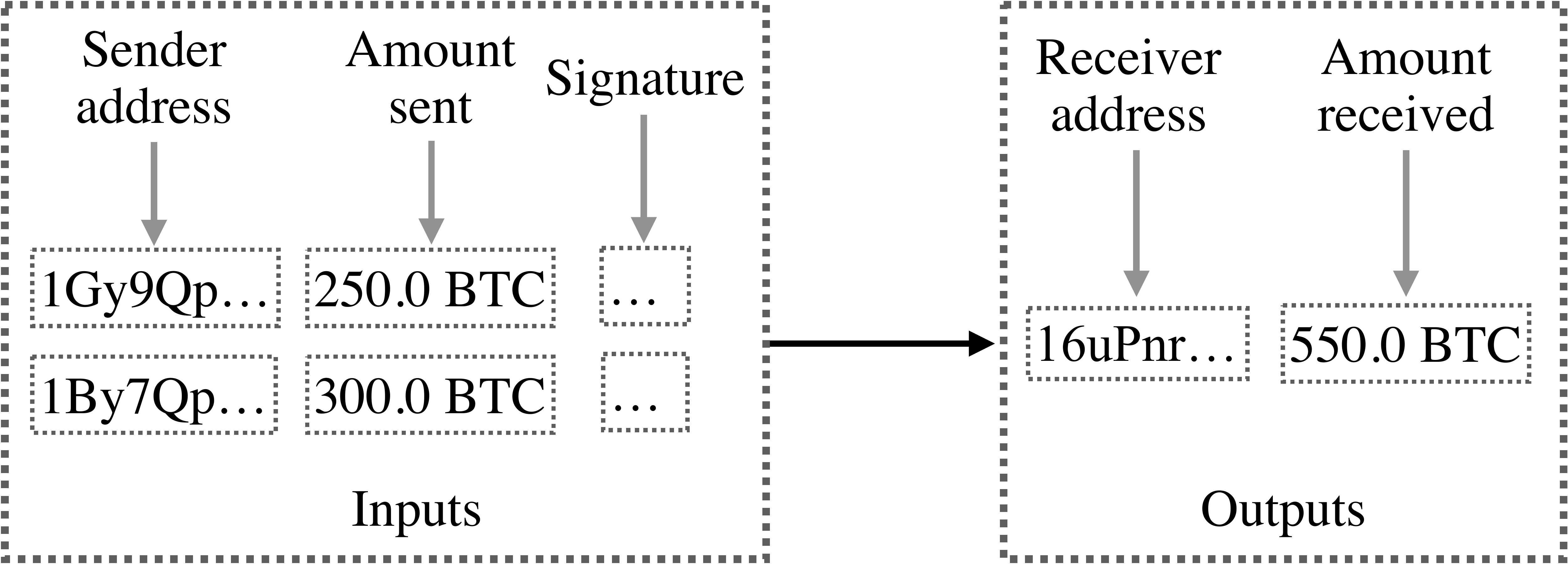}
	\caption[A simple diagram illustrating a transaction]{A simple diagram illustrating a transaction. The left side shows two inputs sending 250 and 300 BTC respectively. Each is attached with a signature that proves the coins were spent by the senders and can be used by the network to verify the transaction. The right side shows a single output receiving these coins.}
	\label{fig:backgroundSimpleTx}
\end{figure}

Figure~\ref{fig:backgroundSimpleTx} illustrates a simple example of a two input and one output transaction. 
On the input side, two addresses \textit{1Gy9Qp...} and \textit{1By7Qp...} are sending 250~BTC and 300~BTC respectively to address \textit{16uPnr...}, the output. In this instance a single entity may own both of the input addresses or it may be two users performing a transaction together to send coins to the same recipient. 
Only public addresses are shown on the blockchain and only public addresses are used to send and receive coins. 
The sending user(s) sign the transaction with the corresponding private key. 
This produces signature(s) to be included within the transaction to allow for verification. 
Once the transaction is confirmed and verified by the network, it is stored on the blockchain within blocks. 

In order to maintain security and incentivise miners to validate and publish transactions, users can optionally add a transaction fee. 
This fee represents a portion of the bitcoins transferred, calculated by deducting the total bitcoins in the output  the total bitcoins in the input. 
The fee is collected by the miner who publishes the transaction onto the chain. 
As space within a block is limited, miners can choose the transactions they would like to include in their block. 
Thus, there is an economic incentive to choose transactions with high fees.  
This creates a market for transaction fees, and thus users must be aware of the current fees being accepted to ensure their transaction is published.

\subsubsection*{Blocks and Miners}

\begin{figure}[h]
	\centering
	\includegraphics[width=0.8\linewidth]{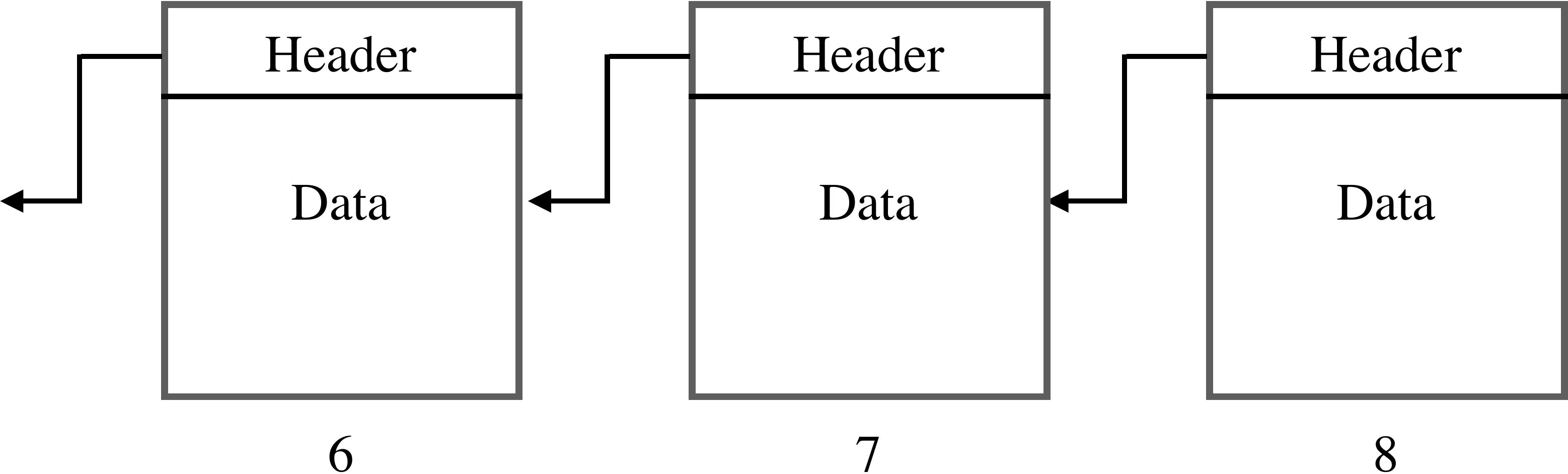}
	\caption{A simple diagram illustrating a block chain. }
	\label{fig:backgroundBlock}
\end{figure}

A block is a data structure that contains both a header and block data.  
The data in a block is made of newly confirmed transactions (can hold zero or more transactions). 
The header contains block metadata and a cryptographic hash of the  previous block.
As shown in Figure~\ref{fig:backgroundBlock}, multiple blocks are chained together with 
their respective hashes forming a block chain. 
This is akin to a linked list, using hash pointers instead of traditional pointers. 

As Bitcoin is an append-only ledger, the cryptographic hash pointers of previous blocks are a vital and necessary core
component of maintaining security. 
Any modification to the data in any part of the chain changes the cryptographic hash of the modified block, subsequently causing all following hashes to change. 
Thus, data is only appended to the end in the form of new blocks. 
Each block is stamped with a block number indicating its position in the chain. 
This, combined with the chain of hashes, makes blocks immutable to change. 

All blocks are created (in blockchain terminology \emph{mined}) by miners who follow the Nakamoto consensus~\cite{Nakamoto2008b}. 
One part is a proof-of-work (PoW) algorithm which uses a set of rules that govern the network and the another is the heavy emphasis on following the longest chain. 

In the case of Bitcoin, miners solve a difficult mathematical puzzle in return for the reward of newly minted coins and transaction fees. 
This ensures that nodes generate proof to show they have spent some computational power/energy. 
The difficulty of the puzzle prevent miners from spamming the network with fake blocks to quickly obtain new coins/fees as puzzle solutions require some computational power to generate but are very easy to verify. 
This allows block generation and transaction confirmation to be secure and decentralised within a trustless network, as trust is obtained from verifying a miners puzzle and the process incentivizes nodes to behave honestly in order to obtain the rewards. 

In Bitcoin proof-of-work, the puzzles are difficult to solve, easy to verify and have varying levels of difficulty depending on the overall power of the network, with the answer to the puzzle being a hash within the target difficulty. 
To generate this hash, miners use the hash of the previous block, data from transactions they choose to validate, a time stamp and a nonce. The nonce is the value that is iterated in order to find the hash. 
Once found, the miner then enters their public key and pushes the block to the network. 
Full nodes in the network verify the block contents and the answer to the puzzle. 
If no issues are found, this new block is added to their chain, the miner earns the reward containing both newly minted coins and transaction fees from pending transactions they included (which are now marked as confirmed) and the entire process repeats. 
Rewards and fees act as incentives for miners to follow the rules and participate in the network. 

On average this routine takes 10 minutes which provides proof that the majority of the power in the network participated. 
In the case where two miners find a block at the same time, causing a chain split, the miners are all
programmed to follow the longest chain which removes the need for a third party to direct the system. 
The system is protected against double spend attacks, as the attackers would require more than 50\% of the
network hash power to take control of the chain. 

The longest chain rule, states that nodes in the network must follow and build upon the blockchain with the most blocks as this is the most legitimate as it has had the most computational power spent upon it. 
New users in the network simply follow the longest chain and begin to mine on top of it, allowing users to join and leave the network without having any negative impacts. 
This also prevents the network from needing to delegate authority, as by default all nodes follow the authority of the longest chain.

\subsection{Accounting models}\label{sec:back-accounting}

In this section we explain the two accounting models used in the blockchain technologies we explored, these being the Unspent Transaction Output model (used in systems such as Bitcoin, Dash and Zcash) and Account-based model (used in smart-contract based cryptocurrencies such as Ethereum). 

\subsubsection*{Unspent Transaction Output} 

In the Unspent Transaction Output (UTXO) accounting model transactions have one or more inputs and one or more outputs. 
The outputs determine the receiving amount (the coins) and condition. Each output is a UTXO. 

The receiving condition is a script, that when true, allows the coins to be spent. 
In Bitcoin, one example of this script consists of the address of the recipient who can spend the coins, provided they 
can sign the follow-on transaction with the associated private key. 
At a high level, this output is seen simply as the address. 
For example, a user who mined a block would receive newly minted coins sent to a script only redeemable with their address. 
To spend this coin, the user must turn this into an input within a new transaction.  

The input contains a signature which proves the corresponding output to be true. 
For example, a user wants to send their newly minted coins to a cryptocurrency exchange.
With their private key they create a signature that solves the script that was sent to them,
thus allowing them to spend their coins, and also marking the output as spent. 
With this they repeat the process by creating a new output with the recipient amount and condition.
The spent input combined with the new output creates a transaction. 

Inputs cannot be split; when spent, the entire input is used. 
This ensures that each is spent only once. 
If users want to partially spend their UTXO, they would create an additional change output to themselves.
This process of change is usually handled by the wallet software.  
This process is akin to spending physical cash. 
For example, to spend a £20 note one must give the note in its whole physical form to the recipient, who then keeps the note and in exchange returns any change. One does not physically tear a note.  

\begin{figure}[h]
	\centering
	\includegraphics[width=1\linewidth]{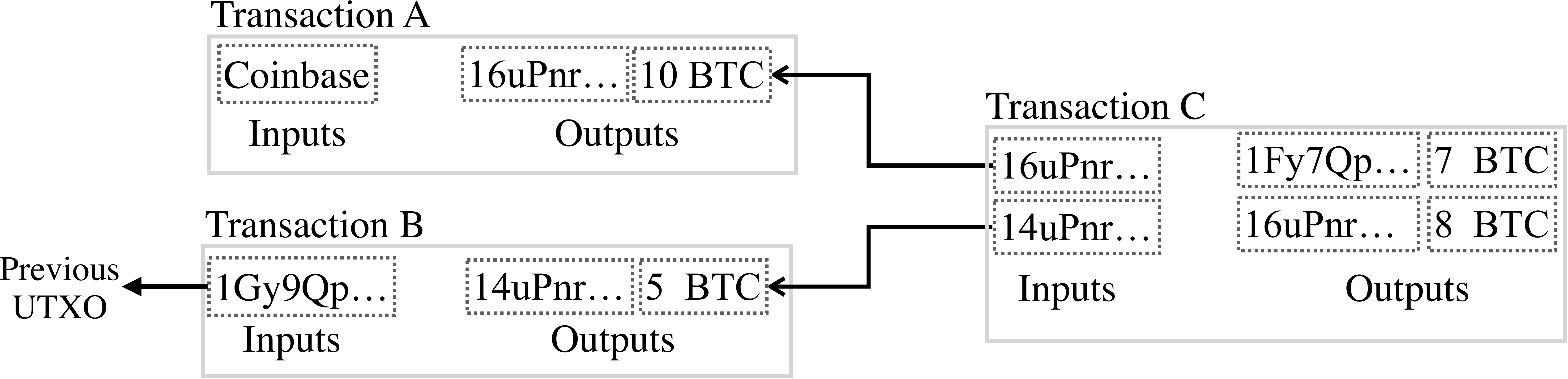}
	\caption{A diagram illustrating UTXO transactions.}
	\label{fig:backgroundUtxo}
\end{figure}

Figure~\ref{fig:backgroundUtxo} shows a high level diagram of three UTXO-based transactions. Here, user Alice has received two transactions: Transaction A with newly minted coins sent to address \textit{16uPnr...} and Transaction B with 5 BTC sent to her second address \textit{14uPnr...}. She spends both of these in Transaction C by spending  both as inputs, sending 7 BTC to Bob \textit{1Fy7Qp...} and the remainder to herself as change. 

The UTXO model ensures that no user is able to double spend, as every UTXO can only be 
spent once and must be completely used. Users can spend UTXOs independently of one another. 
The network can determine who owns what coins by simply taking a list of all UTXOs. 
From an analytics perspective, UTXOs make it easier for users to trace coins to their source. 

\subsubsection*{Account-Based}

Account-based models are an alternative and used in cryptocurrencies such as Ethereum, a smart-contract-based blockchain. 
An address acts as an account (with a private and public key) and coins are deposited directly to the address. 
Accounts have a balance and keep a record of all transactions. 
In Ethereum, accounts can be controlled by a user or by code within a smart-contract. 

When a user wants to send funds, the transaction reduces their (sender) balance and increases the recipient's balance. 
Each account has a public nonce which acts as a protection against malicious users attempting to replay transactions. This nonce is incremented and attached to each transaction sent from the sending account. 
Compared to UTXOs, coins in accounts can be split and do not need to be fully spent. 
This means account-based blockchains do not create change as in Bitcoin.

\section{Anonymity}\label{chapter:back}

As previously mentioned, Bitcoin is designed to operate with pseudonymity on an open ledger that maintains the public history of all transactions. 
This transparency allows transactions to be monitored. 
By analysing transaction patterns, researchers have created heuristics able to reduce the effectiveness of pseudonymity by linking transactions and addresses to real world entities. 

\subsection{Multi-input heuristic}\label{chapter:back:multiinput}

A multi-input (also known as co-spend or clustering) heuristic is the foundational process used to cluster Bitcoin addresses, linking together addresses that may be owned by the same entity. This, when combined with address tagging, reveals the real world identity behind the transactions. 
The effectiveness of this technique has been demonstrated by many researchers~\cite{EvaluatingUserPrivacy,Meiklejohn2013a,6113303,QuantitativeAnalBitcoin} and has formed the basis for commercial cryptocurrency analytics and surveillance companies (e.g., Chainalysis and Elliptic). 

A user must have access to the private key in order to spend coins. 
Inputs in a transaction are spent as the owner(s) signs the transaction(s).  It can therefore be deduced that addresses spent together in the same inputs are probably owned or at least controlled by the same entity. 
By cascading this process across all transactions, one is left with clusters of addresses that have been used together. 

\begin{figure}[h]
	\centering
	\includegraphics[width=0.8\linewidth]{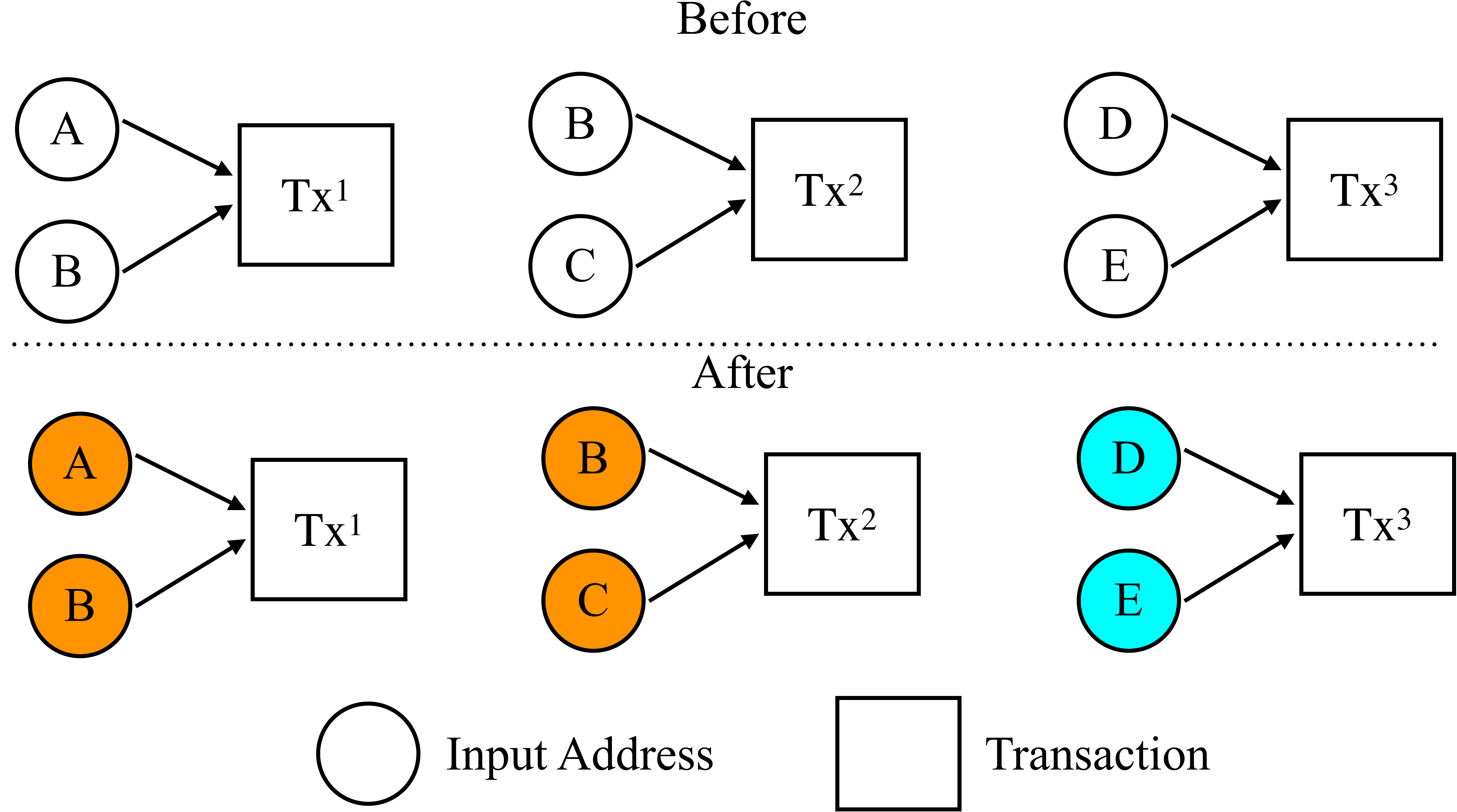}
	\caption{Colouring nodes based upon the multi-input heuristic.}
	\label{fig:backgroundMultiInput}
\end{figure}

Figure~\ref{fig:backgroundMultiInput} shows a visual example of this process, with three transactions that each have two input addresses.
Before, we can see unclustered transactions, indicated by a white circle. 
To conduct the heuristic we execute the following steps. 
We look at the inputs of the first transaction and label these addresses, A and B, with the colour orange. 
In the next transaction, we look at the inputs and check if they have been coloured (allocated a cluster) before, if so we give them the existing colour, if not we assign a new colour. 
In this instance the second transaction has input B, which was labelled previously. Thus these inputs are given the colour orange. 
We repeat this process until all transactions have been labelled. 
Finally, we can see that in the final transaction input addresses D and E have not been labelled before and are given a new colour, cyan. 
This completes the heuristic and the 
"after" section shows the final result containing two clusters, cluster orange with addresses A, B and C and cluster cyan with addresses D and E. 

In practise this algorithm can be modelled in a number of ways. The method mentioned above is akin to the disjoint set (union find) algorithm~\cite{QuantitativeAnalBitcoin}. Alternatively this can be modelled by extracting clusters from connected components~\cite{rowland_weisstein} within a graph, where nodes are addresses and are connected by edges to other nodes if they were involved as inputs in a transaction. 

\subsection{Countermeasures}\label{sec:back:counter}
A variety of counter-measures have arisen since the development of the clustering heuristic, namely coin mixing and privacy coins. 

Mixing is the process of combining coins with users to attain increased privacy and anonymity. CoinJoin, a form of mixing, was proposed by Maxwell in the Bitcoin forums in 2013~\cite{coinjoin}. They present a method for users to mix their coins together in a single transaction. This obfuscates coin movement by severing the link between inputs and outputs.  Clusters are then combined, triggering incorrect results when using clustering techniques.  

\begin{figure}[h]
	\centering
	\includegraphics[width=0.8\linewidth]{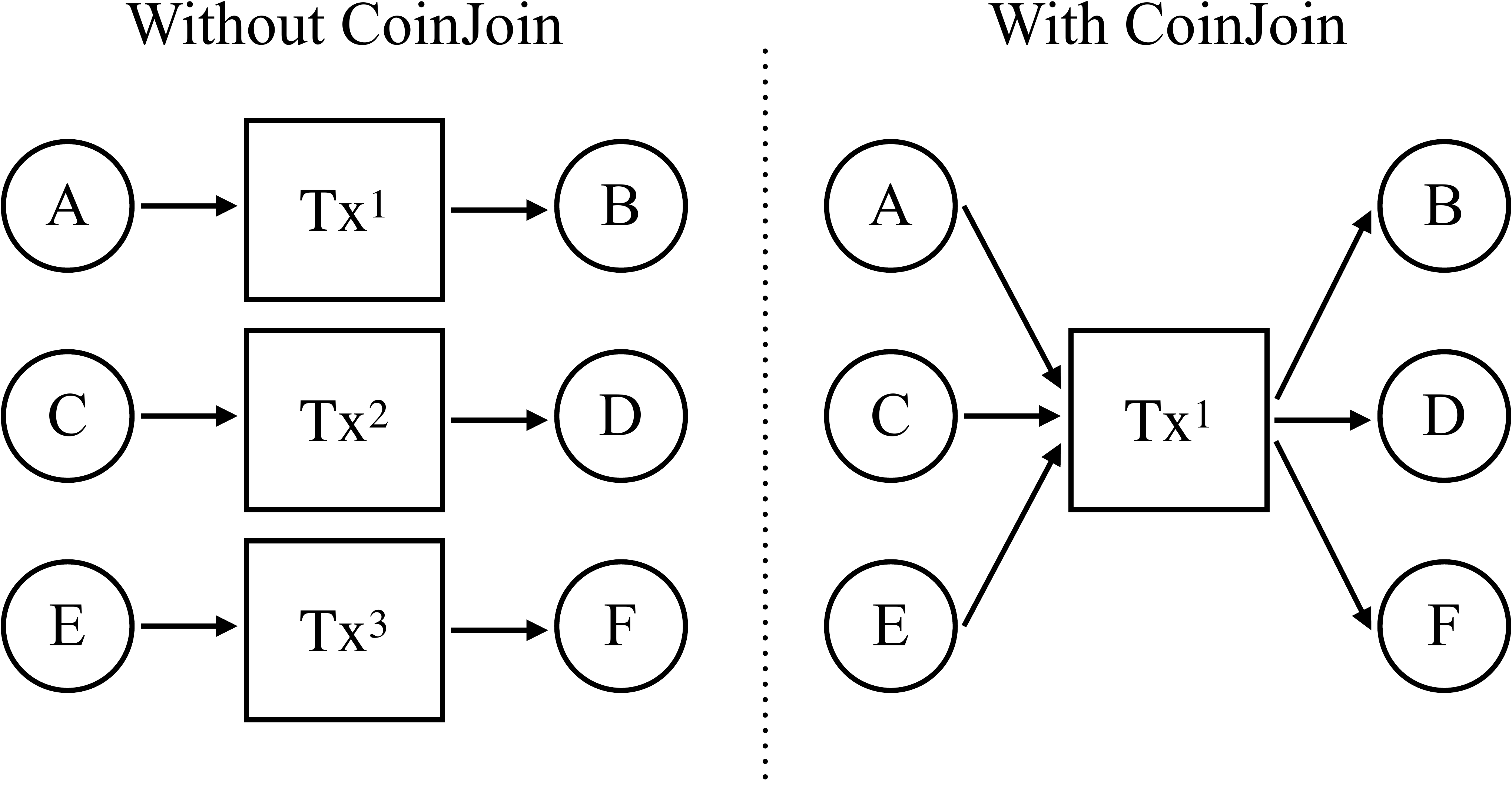}
	\caption{Transactions with and without CoinJoin.}
	\label{fig:backgroundCoinJoin}
\end{figure}

For example, let us say that user Alice wants to pay Bob, Charlie wants to pay Dennis and Eve wants to pay Francis. This can either be recorded as three separate transactions on the chain, shown in `without CoinJoin` in Figure~\ref{fig:backgroundCoinJoin}. This approach makes it trivial to identify payees and recipients. Alternatively, users can perform a CoinJoin together and join their coins in a single transaction. This would appear as Alice, Charlie and Eve are sending coins to Bob, Dennis and Francis and hides the intricate details of exactly who paid whom. 

Alternatively, let us say that Alice, Bob and Charlie unfortunately had their wallet addresses leaked and tied to their real-world identity, or had stolen Bitcoins and wanted to hide their trail. They could perform multiple CoinJoins one after the other with other users, in an attempt to pool and mix all their tainted coins together with non-tainted coins. This can be performed using a tumbler or mixing service, which given a transaction fee, would automate the entire process and allow the user to select a degree of privacy, e.g. mix and return funds after 10 CoinJoins.  

In practice, performing CoinJoins is cumbersome. Users need to find other willing users and perform the advanced task of signing and merging their transactions. 
Developers have created tools that automate this by automatically finding users and performing the CoinJoin, such as  Wasabi Wallet~\cite{wasabi_wallet} and Samourai Wallet~\cite{Samourai}. Tumblers, such as Bitcoin.Fog, allow users to mix their coins, but services like this are often shutdown by law enforcement agencies due to criminal usage and money laundering~\cite{lyons_2021}.  %

\section{Privacy Coins}

Bitcoins pseudonymity was short lived as two years after release the attacks on its protocol came to fruition~\cite{Meiklejohn2013a,QuantitativeAnalBitcoin} %
Since there has been an increase in privacy-preserving coins (privacy coins) which attempt to solve the privacy flaws of Bitcoin, such as Dash~\cite{dash,coinjoin}), Zcash~\cite{zcash,SP:BCGGMT14} and Monero~\cite{monero,cryptonote}, each of which provide different guarantees of anonymity. 

\subsection{ZCash}\label{sec:back-zcash}

\newcommand{\ttot}{t-to-t\xspace}
\newcommand{\ttoz}{t-to-z\xspace}
\newcommand{\ztot}{z-to-t\xspace}
\newcommand{\ztoz}{z-to-z\xspace}
\newcommand{\vjoinsplit}{\textsf{vJoinSplit}\xspace}
\newcommand{\vjoinsplits}{\textsf{vJoinSplits}\xspace}
\newcommand{\vin}{\textsf{zIn}\xspace}
\newcommand{\vout}{\textsf{zOut}\xspace}
\newcommand{\realvin}{\textsf{vIn}\xspace}
\newcommand{\realvout}{\textsf{vOut}\xspace}

Zcash (ZEC) is an alternative cryptocurrency developed as a (code) fork of 
Bitcoin that
aims to break the link between senders and recipients in a transaction.  In
Bitcoin, recipients receive funds into addresses (referred to as the \realvout
in a transaction), and when they spend them they do so from these addresses
(referred to as the \realvin in a transaction).  The act of spending bitcoins 
thus creates a
link between the sender and recipient, and these links can be followed as
bitcoins continue to change hands.  It is thus possible to track any given 
bitcoin from its creation to its current owner.

\begin{figure}[t]
	\centering
	\includegraphics[width=0.8\linewidth]{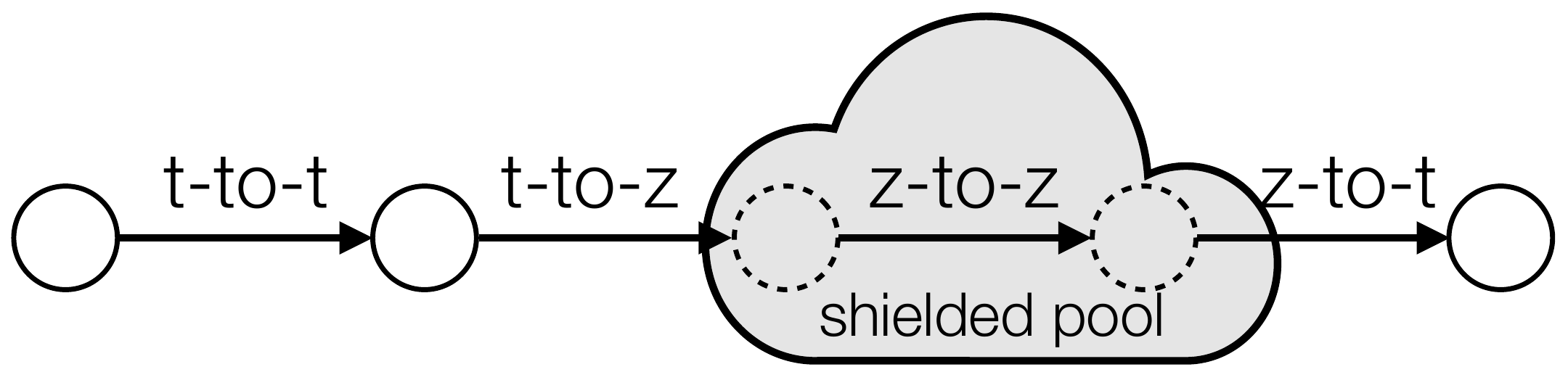}
	\caption[A diagram illustrating the different types of Zcash 
	transactions]{A simple diagram illustrating the different types of Zcash 
		transactions.  All transaction types are depicted and described with
		respect to a single input and output, but can be generalised to handle
		multiple inputs and outputs.  In a \ttot transaction, visible quantities of 
		ZEC move between visible t-addresses (\vin,\vout$\neq\emptyset$).  In a 
		\ttoz transaction, a visible amount of ZEC moves from a visible t-address into 
		the shielded pool, at which point it belongs to a hidden z-address
		(\vout$=\emptyset$).  In a \ztoz transaction, a hidden
		quantity of ZEC moves between hidden z-addresses (\vin,\vout$=\emptyset$).  
		Finally, in a \ztot transaction, a hidden quantity of ZEC moves from a hidden 
		z-address out of the shielded pool, at which point a visible quantity of it 
		belongs to a visible t-address (\vin$=\emptyset$).}
	\label{fig:diagram}
\end{figure}

Any transaction which interacts with the so-called shielded pool in Zcash does 
so through the inclusion of a \emph{\vjoinsplit}, which specifies where the
coins are coming from and where they are going.  To 
receive funds, users can provide either a transparent address (t-address) or 
a shielded address (z-address).  Coins that are held in z-addresses are said
to be in the shielded pool.

To specify where the funds are going, a \vjoinsplit contains (1) a list 
of output t-addresses with funds assigned to them (called
\emph{\vout}), (2) two shielded outputs, and (3) an encrypted memo field.  
The \vout can be empty, in which case the transaction is either
\emph{shielded} (\ttoz) or \emph{private} (\ztoz), depending on the inputs.
If the \vout list contains a quantity of ZEC not assigned to any address, then
we still consider it to be empty (as this is simply the allocation of the
miner's fee).  Each shielded output contains an unknown quantity of ZEC as
well as a hidden double-spending token.  The shielded output can be a dummy
output (i.e., it contains zero ZEC) to hide the fact that there is no shielded
output.  The encrypted memo field can be used to send private messages to the
recipients of the shielded outputs.

To specify where the funds are coming from, a \vjoinsplit also contains (1) a 
list of input t-addresses (called \emph{\vin}), (2) two double-spending tokens, 
and (3) a zero-knowledge proof.  The \vin can be empty, in
which case the transaction is either \emph{deshielded} (\ztot) if \vout is not
empty, or private (\ztoz) if it is.  Each double-spending token is either a 
unique token belonging to some previous
shielded output, or a dummy value used to hide the fact that there is no
shielded input.  The double-spending token does not reveal to which shielded
output it belongs.  The zero-knowledge proof guarantees two things.  First, it
proves that
the double-spending token genuinely belongs to some previous shielded output.
Second, it proves that the sum of (1) the values in the addresses in \vin plus 
(2) the values 
represented by the double-spending tokens is equal to the sum of (1) the values 
assigned to the addresses in \vout plus (2) the values in the 
shielded outputs plus (3) the miner's fee.  A summary of the different types 
of transactions is in Figure~\ref{fig:diagram}.

\subsection{Dash}\label{sec:back-dash}

As in Zcash, the ``standard'' transaction in Dash is similar to a Bitcoin
transaction in terms of the information it reveals.  Its main anonymity
feature \emph{PrivateSend} transactions are a type of 
CoinJoin~\cite{coinjoin}.  

A CoinJoin is specifically designed to
invalidate the multi-input clustering heuristic described in
Section~\ref{sec:back:counter}, as it allows multiple users to come together and
send coins to different sets of recipients in a single transaction.  
If each sender in a CoinJoin sends the same number of coins to their recipient, then it is difficult to determine which input address corresponds to which output address, thus severing the link between an individual sender and recipient.

In a traditional CoinJoin, users must find each other in some offline manner
(e.g., an IRC channel) and form the transaction together over several rounds 
of communication.  
This process is often centralised, as users become aware of one another and may use a service to find each other.
Dash aims to simplify this for users by
automatically finding other users for them and chaining multiple mixes
together.  In order to ensure that users cannot accidentally de-anonymize
themselves by sending uniquely identifiable values, these PrivateSend
transactions are restricted to specific denominations: 0.01, 0.1, 1, and
10~DASH.

\chapter{Literature Review}\label{chapterlabelLitReview}

\section{Cryptocurrencies}

\subsection{Early digital cash}
One of the earliest research publications on digital cash is \textit{eCash}, created in 1983 by David Chaum~\cite{chaum1983blind}.  %
Banks issue eCash to users, who are able to store it on their local machines. 
eCash uses a concept called blind signatures, introduced in 1983, which blinds contents of message before being signed, hiding the contents from the signer. 
In the context of eCash, this allows users to hide their identities from banks, and anonymises the links between spend and withdrawal transactions. 
However, key issues with the system are that it requires banks to both participate and act as the central authority, and
banks have the power to mint digital cash as well as the ability to refuse deposits. 

Following this, in 1998 Wei Dais proposed ~\textit{B-money} an anonymous and distributed cash system~\cite{Dai1998}. 
B-money is similar to Bitcoin, whereby, new transactions are broadcast to all users, who also keep a complete record of the ledger.  
Users are represented by public keys, and mint coins via completing computational puzzles. 
This solved the previous issues raised by eCash as the new features alleviated the need for an issuing bank. 
However, B-money raised a number of other issues. 
One was with the protocol for creating new coins, which required account keepers to decide on the cost of computations. 
This at the time was not feasible, due to the rapid advancement of technology. 
Wei did propose an alternative way to mint money with a four step process, however this seemed complicated.
Secondly, B-money was entirely conceptual, having only been written on paper with no code or tests to support its claims. %

In 2005, Szabo introduced ~\textit{bit gold} ~\cite{Szabo2008, Peck2012}, which used a proof-of-work scheme to mint new coins. 
In this scheme, users would compete to solve computational puzzles and in doing so would earn bit gold. 
Each puzzle solution became a part of the next one, and thus would form a chain. 
However, similarly to B-money, bit gold remained as a concept and was not turned into working code.   

\subsection{Bitcoin and the alt-coins}
In November 2008, Satoshi Nakamoto revealed the Bitcoin white paper to the metzdowd's cryptography mailing list~\cite{satoshifirstmail,Nakamoto2008b}. This system solved previous problems by creating a distributed shared public ledger that prevented double spending, did not include any third parties, allowed pseudonymous user accounts and used a modified version of Hashcash's proof-of-work scheme to generate new coins~\cite{hashcashimplementation}. As of today, Bitcoin is one of the most valuable cryptocurrencies with a market capitalisation of many billions of dollars (USD)~\cite{coinmarketcap,Bonneau2015}. 
Researchers and companies have examined and evolved the underlying software, forking the source code into many new coins or developing new software based on the presented ideas. These coins are named alternative coins or alt-coins, and examples include   Litecoin~\cite{Lee2011a},  Dash~\cite{dash},  Zcash~\cite{zcash} and  Ethereum~\cite{buterin_buterin_2014}. 
 
Litecoin, an alt-coin forked from Bitcoin, has some changes such as different proof-of-work algorithm, decreased block generation times and an increased maximum number of coins. %
Dash offers faster transaction confirmations and is packaged with a coin mixing service. %
Zcash builds upon Bitcoin by adding privacy preserving transactions that make use of zero-knowledge proofs~\cite{miers2013zerocoin,SP:BCGGMT14,Hopwood2017}.
Ethereum is a blockchain proposed by Buterin in 2013, is not a fork of Bitcoin but it carries many of the same principles. %
The main differences include: an introduction of a Turing-complete (limited by execution costs) programming language named Solidity, allowing user created smart contracts to be published on-chain, introduction of custom user created tokens on-chain and an account-based accounting model. As of July 2021, Ethereum is the second most valuable cryptocurrency, just after Bitcoin. %

\section{Privacy}

The openness of the ledger allows the transaction flows of the network to be monitored. 
A large volume of work is dedicated to analysing the privacy of Bitcoin, 
including improving its anonymity and analysing the privacy of alt-coins. 
Here, we outline the most relevant works. 

\subsection{Anonymity of Bitcoin}\label{sec:litprivanonbtc}

Reid et al. presented one of the first anonymity analyses on Bitcoin in 2011~\cite{6113303}. 
Their work analyses nearly two-years-worth of data across multiple angles. 
Firstly, they model both transaction and wallets as two separate network graphs, revealing
that the network has been increasing over time and that the user network is cyclic with users
sending coins to previously-owned addresses rather than treating addresses as one-time use. 
By focusing on privacy, they also discovered that public-keys can be linked with others via 
passively analysing the monetary flow between addresses, 
Using this, they performed one of the first academic analysis of a theft of bitcoins, 
tracing some of the stolen funds to an online wallet provider. 
However, some of the limitations of this study are that the analysis was conducted 
very early in the Bitcoin timeline and therefore was performed on a small graph, 
with potential users who were early adopters, and thus, this may not have represented a mature graph. 

In 2012, Ron and Shamir~\cite{Ron2013} analysed statistics across the entire blockchain, from genesis until May 2012. Compared to previous analysis, the Bitcoin graph had since grown by a factor of three. 
Using the union-find algorithm, Ron and Shamir cluster 3M addresses as 1.8M entities, identifying that 
some of the most active clusters belong to exchange or wallet providers. 
Furthermore, Ron and Shamir identified that many bitcoins remained unused in sink addresses, and most transactions moved only small quantities of the total bitcoin in circulation, and users created transaction chains in an attempt to hide and weaken the links between their addresses. 
In their paper, the researchers could have made further efforts to identify the larger entities, as only three out of a million or so entities are revealed. Their reasoning and usage of the union-find algorithm is largely unclear, as is the method used to identify clusters.  

In the same conference as Ron and Shamir (Financial Cryptography and Data Security, 2013), Androulakil et al. presented an evaluation of user privacy~\cite{FC:AKRSC13}. 
They offered a short evaluation of privacy using two heuristics: a multi-input heuristic and a change heuristic, as well as a longer evaluation on an account-based behaviour simulator. 
Using the first heuristic they are able classify 1.6M addresses into 1M entities, and this is further reduced 
to 693k entities when using the second heuristic.
With the simulator, they identify that Bitcoin does not do enough to protect the privacy of its users, as despite following the recommended guidelines, behaviour-based clustering can profile 40\% of participants.
The evaluation of the two heuristics is very short, and the paper largely focuses on the data from the simulation rather then the ground truth data from their defined heuristics. 

Meiklejohn et al.~\cite{Meiklejohn2013a} performed an in-depth measurement study on Bitcoin and its anonymity. They significantly expanded upon previous clustering efforts and examined the network from creation to April 2013, revealing network statistics, usage and account heuristics, attacks on pseudo-anonymity and the role of entities in the wider ecosystem. They interacted with a wide variety of services, manually obtaining tags for 1,070 addresses. Using both the multi-input, and introducing, a peel-chain heuristic, they analysed the entire network and expanded their tags across clusters covering over 1.8M addresses. Their work demonstrates the effectiveness of such clustering techniques. 
In addition, they presented multiple real world case studies tracking criminal activities, finding that funds from a Bitcoin ponzi scheme were distributed to a variety of services, and tracking multiple thefts to  exchanges.  
However, Meiklejohn et al. also presented progressive enhancements, as both heuristics had been previously published in some form. The effectiveness of the reidentification attack can be reduced if users use mixing services. Their clusters were not compared against ground truth, however, during this time period in the Bitcoin ecosystem, ground truth data was likely only obtainable by directly contacting exchanges and services that may not have wanted to share their sensitive data. 

Spagnuolo et al. presented BitIodine~\cite{FC:SpaMagZan14}, a modular analytic framework which parses, clusters, classifies and visualises Bitcoin data. 
In their work they describe the workings of the system which uses a variety of open-source tools, including the 
C\+\+ programming language, Neo4J (a graph database) and Gephi (an open source visualiser). 
They implement both the the multi-input and change heuristic, and test these on real-world use cases. 
Notably, they argue that they find a connection between an address belonging to the 
owner of a dark market and a separate address which at one point contained over 111k BTC. In addition, they analysed addresses that sent and received coins from the CryptoLocker ransomware. Using their tool they estimate the scammers obtained over 1.1M USD worth of BTC and identify what is claimed to be a potential "test" transaction that occurred days before the first ransom was paid. 
There are however some limitations with both the work and tool. With regard to the work, the case studies are very short, and stating an address has a "meaningful connection" to another is not indicative of holding any real significance.  The tool itself stores all transactions and clusters in memory, which makes the process memory-intensive and costly, given the increasing size of the Bitcoin ledger. It is limited to exporting static graphs and is not bundled with a user interface. Thus the tool requires users to have a high level of technical knowledge to operate. 

Kalodner et al. have presented an open-source blockchain analytics tool BlockSci~\cite{blocksci}. 
Their paper describes in detail the design choices, and architectural challenges, and then presents multiple real-world case studies. 
When compared to standard graph analytic platforms, the BlockSci program is significantly faster and easier to use than BitIodine, as it is bundled with documentation and a python interface.  
The tool itself features an extensive analytics engine, allowing clustering and tracing across UTXO cryptocurrencies and their corresponding forks. 
Within their use-cases, the authors demonstrate that anonymity is affected by usage patterns in multi-signature wallets, identifying that 5\% of Bitcoin addresses have their privacy affected as users cash out their coins on corresponding forks, such as Bitcoin Cash. 
By clustering these transactions across Bitcoin forks, they reveal information about users that can be linked back to their wallets. 
The paper does not compare the tool against previous blockchain analytics programs such as BitIodone, but instead against graph analytics tools. Since the release of the paper, development of the tool has since discontinued. The tool also requires a machine with very large memory in order to cluster the blockchain.

The research above demonstrates that Bitcoin's anonymity can be breached,  since there have been new proposals to resolve these issues and improve privacy without major changes to the protocol. 
One process is CoinJoin~\cite{coinjoin}, where users create a transaction 
together, merging their inputs and outputs in order to reduce the linkability within their transaction. 
This process is one of the foundational features of the cryptocurrency Dash~\cite{dash} which automates the entire procedure. 
This CoinJoin mechanism is explained in more detail in Chapter~\ref{sec:back:counter}.

Analysis around the anonymity of CoinJoin~\cite{FCW:MeiOrl15, malte-2gen, maurer2017anonymous, michaelcoinjoinflaw, coinjoinsoduko} presents many issues. Atlas affirms that inputs and outputs of a CoinJoin can be linked through brute-forcing all possible summations and more so, if denominations can be uniquely identified~\cite{coinjoinsoduko}. However, the process can quickly become computationally expensive as CoinJoin is a variant of the Knapsack problem~\cite{ross1989stochastic,maurer2017anonymous}.
Dash~\cite{dash} improves the usability of Maxwell's CoinJoin~\cite{coinjoin} through automating the process of finding participants to mix with using fixed denominations, further explained in Chapter~\ref{sec:back-dash}. 
However this adds a time delay to transactions, as users must wait to find others with whom to mix.

\subsection{Anonymity of privacy coins}

In Bitcoin transactions information is entirely public and, as previously shown, allows any entities to freely track the movement of coins.
Protocols have since been developed to integrate anonymity. 
Many of these have been implemented into alternative coins marketed with a privacy focus. 

Zerocash is a protocol, forked from Bitcoin, that adds privacy-preserving transactions~\cite{SP:MGGR13,SP:BCGGMT14}. 
These use zero-knowledge proofs to "shield" coins, adding an anonymity layer that hides the amount, sender and receiver. 
Thus, when coins are spent, no information about the transaction is revealed other than a potential fee. 
Users have the option to use this feature, for anonymous coins they utilise zerocoins, also known as shielded coins/transactions, and for non-anonymous coins they can spend basecoins, also known as transparent coins/transactions. 
Coins can be converted to either type at the users preference. 
This concept is further explained in Chapter~\ref{sec:back-zcash}.
Zerocash was subsequently commercialised into the cryptocurrency ZCash. 
The initial downsides of ZCash were that the original versions required a significant amount of processing power (3GB of RAM and several minutes) to construct the zero-knowledge proof required for a single shielded transaction. This was not suitable for sending quick transaction, nor were users able to perform such computations on their mobile devices. They originally needed to run a full ZCash node, as most wallet software only supported transparent transactions. In doing so, this led to the second issue, whereby running a node and performing shielded transactions required some familiarity with the command-line, as even the software did not come with a user interface.
Given that this type of transaction was optional, not enabled by default or the above issues, the anonymity set of the shielded pool was thus confined to users who were technically advanced. 
Recent advancements have since solved these issue such as reducing the memory usage by 98\%, reducing transaction times by 80\%~\cite{bowe_2019}, and introducing mobile wallets which support all types of transactions such as Zecwallet Lite~\cite{zecwallet_website}.

Prior to the above advancements, researchers had revealed the short-comings of the shielded pool. In 2018, we published the first peer-reviewed study discussing the limitations and privacy flaws that were then present in Zcash. This is explained in detail in Chapter~\ref{chapterlabelZcash}. Whilst we conducted this research, others published studies that had parallels with our work. 

In late 2017, Quesnelle~\cite{zcash-anon} also published a short study analysing the privacy of Zcash. They revealed that transaction meta-data can be used to link coins between deshielding and shielded transactions. 
They identified that the vast majority of coins were not used within the shielded pool, given that only 19.6\% of transactions used some form of a privacy-preserving feature and that 57.7\% of these were used to deshield coins. 
Although Quesnelle is credited with creating round-trip transactions, this is also a concept that we discovered during our research. In this process transparent coins are sent to a shielded address, and then a similar or identical number of shielded coins is returned to a public address within a close time frame. 
Around 31.5\% of shielded coin transactions were conducted within round-trips. 
Using these, Quesnelle linked coins belonging to miners after their respective mining pool had deshielded newly minted coins. 
Quesnelle believes the reasons for these short-comings are due to lack of support within third-party wallets and high computational costs. 
The study has limitations because the work exclusively focuses on brief statistics and round-trips. The background does not clearly explain the workings of Zcash, and the transparent transactions which account for the majority of transactions in Zcash have not been fully discussed.  

Birkyukov et al.~\cite{biryukov2019privacy} published heuristics analysing the behaviour of miners in both transparent and shielded pools. They extended previous work~\cite{217535} by presenting heuristics targeting the two payout strategies deployed by mining pools. The first strategy identifies users of mining pools by following transactions that pay users with public addresses, whereas the second strategy identifies mining pools that directly pay users from a shielded address. Combined, their heuristics are able to link 88.4\% of all mining rewards to shielded addresses. When combined with the Founder heuristic from previous work, presented in Section~\ref{sec:founders}, they are able to link over 84\% of all de-shielded transactions. In terms of downsides, the work does not focus any analysis on transparent transactions and their heuristics primarily target de-anonymising mining pool transactions. 

In 2014, developers created Monero~\cite{monero}, a privacy focused cryptocurrency based on Nicolas van Saberhagen's whitepaper on CryptoNote~\cite{cryptonote,bytecoin}. 
Monero uses ring signatures, which are a digital signature that can be created by a user from a group of users that each have keys. When creating a signature, it is computationally impracticable to identify which member of the group created the signature. 
When a user wants to spend their coins in Monero, they generate a "min-in", which is a ring signature using their output as a key with other public keys taken from previous outputs in the blockchain. 
These all act as decoys within the new transaction, and when created the inputs appear to be equally likely to have been originally spent, which masks the origin of the transaction. 
All aspects of the transaction are obfuscated, hiding the senders, receivers and amounts. 

Multiple research projects in 2017 identified issues which can disrupt the anonymity of the system~\cite{ESORICS:KFTS17,miller2017empirical,tramer2020remote}. 
Kumar et al. presented three heuristics that were used to trace 87\% of inputs~\cite{ESORICS:KFTS17}. 
By using temporal analytics, they identify that it is not as difficult to predict the correct output in a ring signature. Given that over time it is more likely a UTXO had already been spent and thus the most recent output is very likely to be the real one being spent. 
Secondly, by using a technique called leveraging output merging, they analyse user behaviour in transactions where two outputs belong to the same entity. Thirdly, they present that users themselves can disrupt the anonymity of others by choosing to do a mix with zero mix-ins, damaging the anonymity set of others who may have referenced their coins. 
Moser et al. presented similar work, by detailing issues surrounding the coin selection algorithm~\cite{miller2017empirical}. Similarly, they found it more likely to select more recently generated outputs.

These works all reveal that, despite best efforts, so called "privacy coins" are vulnerable to attack. Over time, scientific advancements will identify weak points in systems that were previously thought to be secure. Therefore, it is pertinent to develop new techniques in order to strengthen current practices and create better mitigations.

\section{Blockchain crime}
The decentralised, pseudonymous and uncensorable transaction system has attracted a wide range of users, some of whom exploit the system for nefarious reasons.
Bitcoin's ability to transfer coins across borders (regardless of location), combined with misleading claims that Bitcoin offers anonymity, have attracted those involved in crime. 
Such crimes range from operating a dark market, theft, and ransomware, to scams and money laundering. 
This section details the literature that analyses transactions linked to misdeeds. 

\subsection{Dark markets}
Underground markets are those which sell goods or services that may or may not be forbidden by law. 
Historically, there have been many examples with some predating the second world war~\cite{history_editors_2017,Trueman2015}. 
With the advancement of technology, these markets have naturally made an online appearance, with the earliest cyber markets selling information about goods and services (e.g., credit card numbers, viruses, botnets) via internet relay chats (IRC)~\cite{Franklin2007, Thomas2006}. 
Following take downs by law enforcement, this activity moved to online forums, marketplaces and registration-only websites with research analysing their appearances in China~\cite{zhuge2009studying}. The underground economy is now considered to be an integral piece of cybercrime, due to servicing criminal enterprises with illicit goods and services~\cite{Herley2010}. %

Silk Road, a notorious and commonly known underground market attributed to Bitcoin launched in 2011~\cite{Christin2013a}. It operates similarly to Amazon and eBay, where anonymous sellers are able to sell goods and services to anonymous consumers~\cite{HOUT2013385}. The market ran as a Tor hidden service using Bitcoin as the medium of exchange~\cite{biryukov2014content}. 
The market sold a large variety of goods, such as unlicensed firearms and drugs (both legitimate and illegitimate), and services such as assassination, botnets, malware and targeted hacking. The initial version, 1.0, operated for over two years and the second version, 2.0, for a year before both being  shut down by the FBI~\cite{ANDERSONa, Cook}. 

In 2013, Christin~\cite{Christin2013a} published an in-depth measurement study of Silk Road, analysing the types of products sold, the evolution of sellers including their countries of origin, and economic indicators, including the use of Bitcoin and sales volume. With regards to Bitcoin, they estimated that over a 29-day span the marketplace transacted 1.3M BTC (1.22M USD), which they believe corresponded to between 4.5\% and 9\% of all exchange trades. The Bitcoin analytics in the paper are very short and the authors make no attempts to trace the illicit funds, however this is acknowledged as not being the primary focus. 
In 2014, Spagnuolo et al.~\cite{FC:SpaMagZan14} used their Bitcoin tracing, classification, and verification system to analyse crime and perform an analysis of potential connections between the Silk Road cold wallet and its founder, and measured on-chain crime from the CryptoLocker ransomware. 
We discuss this paper earlier in Chapter~\ref{sec:litprivanonbtc}.%

Many new anonymous market places have been created since the closure of Silk Road, with recent crime studies reporting an increasing trend in dark market revenues and competition. In 2015, Soska et al.~\cite{Soska2015} performed a long term measurement analysis of 35 marketplaces across two years, detailing the growth of the underground ecosystems. 
In their study they scraped and parsed the data from these market places across multiple  snapshots to analyse how the system evolved over time. 
Through analysing sales volumes in Silk Road, they projected the marketplace produced \$100M USD a year, which was in accordance with the amounts projected by the US government. 
Overall, they estimate that the ecosystem transacts over \$500,000 USD per day.  
Market place closures and shutdowns resulted in users moving their business elsewhere, and that, interestingly, the ecosystem was resilient to scams and law enforcement take-downs. 
In terms of limitations, it is unclear whether the researchers have made their datasets available for others to use in their own research. Some of the datasets are censored due to insufficient amounts of data collected. The authors did not appear to directly interact with the services as estimated sales volume is predicted based on user feedback instead of real transactions. 

Markets are, however, actively shut down, either voluntarily (Sheep Market place~\cite{Sankin2013}) or forcefully by law enforcement (Utopia~\cite{Greenberg2014}). Further analysis of underground markets includes the following works~\cite{Holt, holz2009learning, Levchenko2011b, stone2013underground, Biryukov2013}, however these do not focus on cryptocurrency transactions but demonstrate techniques used to collect data. 

\subsection{Thefts, Ransomware and Sextortion} 

Cryptocurrency exchanges and user accounts have been subjected to numerous cases of theft. Opened in July 2010, Mt Gox was one of the worlds largest digital asset exchanges, allowing users to trade via standardised market methods~\cite{trautman2014virtual}. %
At its peak the service was responsible for handling over 70\% of all bitcoin transactions~\cite{frunza2015solving}. 
In 2011, it was reported that someone had been gradually stealing coins from the exchange-owned wallets, siphoning 750,000 bitcoins (\$330 million USD) which subsequently caused the company to file for bankruptcy~\cite{mtgoxbankrupcy}. 

In 2015, the Bitstamp exchange was breached and hackers stole 19,000 bitcoins (\$5 million USD)~\cite{higgins2015details}. Employees from Bitstamp were targeted through phishing campaigns which ran for weeks. 
The attackers distributed malware via Skype and email channels, compromising internal machines. In 2016, Bitfinex, a Hong Kong based exchange, was also breached with hackers stealing 119,756 BTC (\$72 million USD) causing a 20\% drop in the Bitcoin price when announced~\cite{boireau2018securing,chen2021bitcoin}. 

Lazarenko et al.~\cite{lazarenko2018financial} list and classify 48 attacks on blockchain projects which had lead to theft of coins. 
These attacks are grouped into eight categories, some of which include insider attacks, phishing and malware. 
For each project attacked, they list the number of coins lost and detail the cause. 
Their analysis shows that the number of attacks on blockchain projects has increased annually, likely due to the increase in available technologies. 
Exchanges had been subject to the most thefts and attacks on Bitcoin, whereas Ethereum Initial Coin Offerings (when an unregulated entity raises money through distribution of cryptocurrency assets) had the most Ether stolen. 
Overall the report finds that 59\% of blockchain related projects are closed after a cyber attack. 
The analysis in the paper is very brief, and the paper simply lists out a number of attacks, however this is due to the research being classified as a survey rather than an in-depth study. 

Since 2012 a new type of malware has emerged: ransomware. It is malicious software that prevents a user from accessing their computer until they pay a certain amount to the operator.
The payloads are spread through a variety of ways, for example through malicious email attachments~\cite{adriano_2017} or via exploited vulnerabilities~\cite{mohurle2017brief}.  
In May 2017 the ransomware WannaCry attacked Windows Machines worldwide, encrypting user data  and holding it as ransom, requiring users to pay with bitcoin to regain access. 
Some of the infected machines belonged to national hospitals and telecommunication companies~\cite{wannacryavast}. 
The ransomware caused an estimated \$4 billion USD worth of damage~\cite{wannacrydamage}. 
After the outbreak, it was discovered that the attackers had used the cross-currency exchange ShapeShift to convert their tainted Bitcoins into Monero, which was reported to then 'disappear'~\cite{wannacrymonero}. 

Conti et al. published an in-depth study of ransomware on Bitcoin, presenting techniques to identify, collect and analyse transaction data~\cite{conti2018economic}.
They present an identification framework to identify extorted ransom transactions which consists of three components.
First is the identification of addresses from online resources, removal guides and threat reports. These are then clustered using both the multi-input and change heuristics.
Next is the extraction of all transactions which contain these addresses, and finally, the classification of the amounts received as ransom if at least one of their payment conditions is satisfied. 
This methodology is applied to twenty ransomware cases, and for each they detail payment strategies, methods of infection and ransoms extorted. 
Overall ransomware operators earned an excess of 3M USD, and the operators of WannaCry earned 238 payments averaging a total of 86k USD (47 BTC), compared to the operators of 
CryptoWall who obtained 2.2M USD (5,351 BTC).
The work is very thorough and the appendices list all the addresses used in the initial portions of the investigation, which is useful for future research. 
Limitations of the work include that the ransom addresses are collected from public sources and such data quality cannot be guaranteed, however they compare
their results and find that it is similar to previous works. 

Paquet-Clouston et al.~\cite{paquet2019ransomware} present methods identifying illicit Bitcoin transactions by focusing on 35 ransomware families. 
The work obtains, clusters and filters a set of seed addresses obtained from researchers and online sources. 
By following the flow of money, they found that multiple ransomware families interacted with the same actor.  
Across four years they reveal a lower bound of over 22k BTC (\$12M USD) in ransomware payments, with the Locky ransomware receiving over \$7.8M USD and more than 50\% of payments. The researchers concluded that a small number of actors, just three families, dominated this genre of crime and were accountable for 86\% of the marketshare. 

Huang et al.~\cite{huang2018tracking} created a framework to track ransomware end-to-end and applied this to real-world cases across a two-year period. 
They first obtained a list of addresses from public ransomware infection reports, seed addresses and a form of synthetic address (victims they created and tracked by sending operators micro-payments). These were turned into clusters using co-spend heuristics, and transactions to and from the clusters were filtered and analysed. 
Overall, they estimated that operators from 10 ransomware attackers obtained \$16M USD from 20k potential victims. 
Through looking at inflows to clusters, it is found that exchanges account for 40\% of payments.
They also revealed that BTC-e, a Russian exchange, prior to being seized by US law enforcement, was a key exit point used for criminals in the Locky and CryptoDefense ransomwares, as they saw \$3M USD flow through their exchange. 

Some scams feed on the fear, gullibility and technical infancy of the target users. Paquet-Clouston et al.~\cite{Paquet2019} analysed emails sent by criminals to victims of sextortion scams requesting bitcoin as payment. Sextortion is a spam scheme whereby an attacker emails victims, claiming that sensitive and private photos or videos will be leaked to their contacts unless they are paid some bitcoins. The researchers analysed 4.3M emails, bucketing emails into 15 campaigns and extracted all bitcoin addresses. 
They extracted 245 addresses from the emails, which when clustered and filtered came to 485 payment addresses that had received coins. Their analysis revealed that scammers attempt to extort higher amounts of coins based on the language used in the emails, Emails written in the English language asked for a mean of \$745 USD whereas Spanish language scams were asked for \$249, indicating that attackers adjusted their prices based on the victims' perceived language and location. By following cash flows from the spam clusters, they discovered that entities were moving coins to known exchanges. However, this part of tracing is very limited as tags were obtained from an open source website. By analysing the campaigns they identified that multiple clusters appeared across multiple campaigns, suggesting that the majority of revenue was collected by a single real world entity. Untagged  clusters were sent 48 BTC (17\% of revenue), and these were mentioned to possibly be potential cash-out services. In conclusion, they estimated that one entity may be responsible for the majority of crimes, with scammers harnessing a lower bound of over \$1 million USD within 11-months. In terms of limitations, the addresses were obtained from a data set which was caught by a spam filter, and thus may not have reached most recipients, thus the estimations are a lower bound and only cover a small section of this ecosystem. The multi-input heuristic may have clustered in addressess belonging to the same author but perhaps different scams.  The source of the tags is extremely limited, and thus it is unclear exactly how many of the coins were sent to and from exchanges.

\subsection{Investment Programs and Money Laundering}

Past research has quantified and described crime and scams running on blockchain ecosystems that use high return investment programs and ponzi schemes as well as mixing services as a form of money laundering. 

Vasek et al.~\cite{Vasek2015TheresNF} presented analysis of four types of scams (ponzi schemes, mining scams, scam wallets and exchanges) that used Bitcoin, identifying 192 scams and tracking their payments. 
By collecting scams from various data sets online, and after a cleaning process, they extracted all relevant transactions from the Bitcoin blockchain. 
They modelled 42 scams into four categories and analysed each.
One class of scams they analysed was high yield investment programs (HYIPs).
These are schemes that promise investors a high rate of return. Previous investors are paid by new investors, and this process repeats until the scheme closes or collapses. 
By comparing traditional HYIPs to Bitcoin based HYIPs, they found that Bitcoin was not yet widely accepted. However, some act more traditionally using fiat currency and then transitioning into Bitcoin. From these, they identified a user who earned 1.6M USD. 
Another category they analysed was scam wallets. These are online wallets that claim to host Bitcoin for users, but the operators, in-fact, steal the coins that have been deposited. One scam wallet in particular earned 4,100 BTC (\$1M USD). 
Overall, they found that scammers earned \$11 million USD from 13,000 victims. 
The research within this paper is very thorough, covering a wide variety of scams including economic perspectives. The work however only focuses on how many coins were received and they do not track the illicit funds after they have been received. 

In 2018, Bartoletti et al.~\cite{Bartoletti2018} applied data mining techniques and machine learning algorithms to detect Ponzi schemes in Bitcoin. 
Ponzi schemes are a form of HYIP where users of the schemes are paid only by other new users that join the scheme. 
They collected 32 addresses belonging to ponzis and through clustering obtained 1211 addresses. Analyses revealed that these received \$10M USD worth of deposits. 
Using these clusters they derived and extracted features (e.g., statistics, measures of inequality) from both these and random addresses. 
By testing a range of supervised binary classifiers, they found that the random forest algorithm was correctly able to identify 96\% of addresses as ponzi schemes. 
This experiment is very thorough, and the authors released addresses and features used and tested across multiple classifiers and ponzi schemes. The limitations are that this approach was only tested against ponzi schemes, so it would be interesting to see whether it is applicable to other types of scams.

Bitcoin is not the only cryptocurrency prone to being targeted by crime; this behaviour also occurs on Ethereum. Chen et al.~\cite{chen_zheng_cui_ngai_zheng_zhou_2018} used data mining and machine learning to detect Ponzi schemes, scams that fraudulently promise high returns by generating income for previous investors by taking money from later investors. 
They first analyse a known smart contract ponzi scheme by studying the transactions betwen the contract and participants. 
With this they developed sets of features that are key to ponzi schemes. 
The first set of features looks at statistics, including the number of payments into and out, and the proportion of users who sent money to the coin before ever receiving anything in return. 
The second set analyses the use of opcodes, assembly like commands used by contracts extracted from the Ethereum Virtual Machine. 
These features were tested with a machine learning algorithm that classified whether the contract resembled a ponzi scheme. 
The classifier was tested on 54 previously known ponzi schemes and was able to detect 45 schemes. The undetected 9 were manually investigated and found not to be ponzi schemes. 
Out of 280k contracts on Ethereum, they estimate that 434 ponzi scheme smart contracts exist. 
In terms of limitations, the reasoning for the machine learning algorithm is not convincing, it would be clearer to compare this against other algorithms to justify the decision. 
The results detected 386 algorithms that were not checked, and it would have been helpful to take a random sample to manually check whether they could be confirmed. 

Bartoletti et al.~\cite{BARTOLETTI2020259} performed a survey on ponzi schemes in Ethereum and analysed them from multiple perspectives. Similar to prior works, they extracted features from Ethereum ponzi schemes based on their transaction behaviours and smart contract code, using this to create a machine learning algorithm to classify and detect smart contracts. In addition, they also measured multiple statistics, including the gains and losses incurred by users of 23 contracts, identifying that most users never receive any money after investing with only one or two users who earn high amounts of coin. The study is very thorough and ends with raising a number of recommendations that could be used to combat this crime. However, the study does not analyse where the money goes after being sent to the key players in the scheme.

Ponzi schemes, extortion scams and ransomware are not the only crimes occurring on blockchain. A Pump-and-dump (P\&D) is a scheme where users artificially inflate (the ``pump'') the price of an asset by recommending that others purchase it at a specific time in a coordinated manner, in the hopes that its value increases. Once the value reaches a certain point, the holders of the scheme sell that asset to profit from the gain in price (the ``dump'').
This scam was prevalent within the stock market space~\cite{bradshaw2003pump}, and is now performed on cryptocurrencies. 
With cryptocurrencies, organisers use online chat services such as Telegram to coordinate their schemes. Several research articles analysing schemes have been published, identifying how they are run and producing models to detect their activities~\cite{236350, Hamrick2018, Kamps2018, li2020cryptocurrency}. %

Tao et al.~\cite{li2020cryptocurrency} published an in-depth study analysing the early P\&D ecosystem. 
They manually collect data from pump and dump groups online and, after 
filtering and processing, analysed 500 distinct P\&D events from 80 groups involving 239 tokens across 3 cryptocurrency exchanges. 
Their analysis revealed that P\&Ds are regularly scheduled at the advertised times, only lasting for several minutes but being able to return an average of 212\% for investors. 
However, an investor's performance depends on when they receive the message to conduct their trade, with some groups operating with insiders who are aware of the coins beforehand. Jihau et al.~\cite{236350} extend previous works by revealing the entire anatomy of P\&D schemes on cryptocurrencies and by producing a prospective prediction algorithm that able to identify and trade coins that signal movements indicative of P\&Ds. Their algorithm strictly uses market data and is able to generate a 60\% return over two and a half months.

Previous research has addressed the use of money laundering tools in the ecosystem. 
These tools can be used to launder dirty Bitcoins, obscuring the trail of coins and making it difficult to trace stolen money. 
Alternatively, they are sold as anonymisation tools, that protects the users anonymity by merging their coins with others. 

Moser et al.~\cite{moser2013inquiry} explored mixing services designed to anonymise Bitcoin transactions. 
They sent coins to three services and used graph analytics to reverse-engineer the functionality of service, to determine whether they could back-trace their coins.  
Two of the three services were revealed to be providing some level of anonymity, as the researchers were unable to trace back their coins. 
The services from \textit{blockchain.info} and \textit{bitcoin.fog} were found to aggregate small transactions into large bundles before performing any payments. 
However, using \textit{bitlaundry} they were able to find a connection, but only within a small amount of one of their transactions. This service did not bundle transactions, and instead used user inputs to pay back other users, not delivering on the privacy promised. 
As researchers were only able to identify a small amount of transactions from one service, this  highlights the challenges faced by current anti-money laundering techniques when only using graph analytics. 
The study is very limited, with the researchers only covering three services with a small number of transactions over a short period of time. 
It might have been more effective if the researchers had sent more transactions into each of the relevant services over a longer time period. 

Wegberg et al.~\cite{van2018bitcoin} extended the literature by analysing the usability and cash-out effectiveness of mixing services and exchanges. 
In their experiment, they sent coins to five mixing services and three of the five mixers stole the coins. 
By tracing their stolen coins, they identified that two mixers combined their stolen coins in the same transaction after the experiment had concluded, indicating that they might have been collaborating or were the same entity. 
The two working mixers successfully delivered their coins with no taint, meaning that there was no linkage found between the coins sent and the coins received.  
They concluded that  for smaller denominations the mixers offered a cost effective and user friendly service with a minimised risk of being scammed if the user had read reviews prior to using the service.

\chapter{An Empirical Analysis of Anonymity in Zcash}
\label{chapterlabelZcash}
\newcommand*\dash{\ifvmode\quitvmode\else\unskip\kern.16667em\fi---%
\hskip.16667em\relax}

\newenvironment{sarahlist}{
\begin{description}[itemsep=2pt,leftmargin=0.4cm]
}{\end{description}}

\newtheorem{theorem}{Theorem}
\newtheorem{definition}[theorem]{Definition}
\theoremstyle{definition}
\newtheorem{heuristic}[theorem]{Heuristic}

\section{Overview}

Since the introduction of Bitcoin in 2008~\cite{satoshifirstmail}, 
cryptocurrencies have become increasingly popular to the point of reaching 
a near-mania, with thousands of deployed 
cryptocurrencies now collectively attracting trillions of dollars in
investment.  While the broader positive potential of ``blockchain'' (i.e., 
the public decentralized ledger underlying almost all cryptocurrencies) is 
still unclear, despite the growing number of legitimate users there are today
still many people using these cryptocurrencies for
less legitimate purposes.  These range from the purchase of drugs or other
illicit goods on so-called dark markets such as Dream Market, to the payments
from victims in ransomware attacks such as WannaCry, with many other 
crimes in between.  Criminals engaged in
these activities may be drawn to Bitcoin due to the relatively low friction 
of making international payments using only pseudonyms as identifiers, but 
the public nature of its ledger of transactions raises the question of how 
much anonymity is actually being achieved.

Indeed, a long line of research~\cite{reid2013analysis,Ron2013,FC:AKRSC13,sarah-fistfulofbitcoins,FC:SpaMagZan14}
has by now demonstrated that the use of pseudonymous addresses in Bitcoin
does not provide any meaningful level of anonymity.  Beyond academic research,
companies now provide analysis of the Bitcoin blockchain as a 
business~\cite{elliptic-article}.  %
This type
of analysis was used in several arrests associated with the
takedown of Silk Road~\cite{silk-road}, and to identify 
the attempts of the WannaCry hackers to move their ransom earnings from
Bitcoin into Monero~\cite{wannacry}.

Perhaps in response to this growing awareness that most cryptocurrencies do
not have strong anonymity guarantees, a number of alternative
cryptocurrencies or other privacy-enhancing techniques have been deployed with
the goal of improving on these guarantees.  The most notable cryptocurrencies
that fall into this former category are Dash~\cite{dash} (launched in January
2014), Monero~\cite{monero} (April 2014), and Zcash~\cite{zcash} 
(October 2016).  At the time of this
writing all have a market capitalization of over 1 billion
USD~\cite{coinmarketcap}, although this
figure is notoriously volatile, so it should be taken with a grain of salt.

Even within this category of privacy-enhanced cryptocurrencies, and despite
its relative youth, Zcash stands
somewhat on its own.  From an academic perspective, Zcash is backed by highly
regarded research~\cite{SP:MGGR13,SP:BCGGMT14}, and thus comes with seemingly
strong anonymity guarantees.  Indeed, the original papers cryptographically
prove the security of the main privacy feature of Zcash (known as the 
\emph{shielded pool}), in which users can spend shielded coins without 
revealing which coins
they have spent.  %
These strong guarantees have attracted at least some criminal 
attention to Zcash: the underground
marketplace AlphaBay was on the verge of accepting it before their shutdown 
in July 2017~\cite{alphabay}, and the Shadow Brokers hacking group started
accepting Zcash in May 2017 (and in fact for their monthly dumps accepted 
exclusively Zcash in September 2017)~\cite{tsb-zcash}.

Despite these theoretical privacy guarantees, the deployed version of
Zcash does not require all transactions to take place within the shielded pool
itself: it also supports so-called \emph{transparent} transactions, which are
essentially the same as transactions in Bitcoin in that they reveal the
pseudonymous addresses of both the senders and recipients, and the amount 
being sent.  It
does require, however, that all newly generated coins pass through the 
shielded pool before being spent further, thus ensuring that all coins have 
been shielded at least once.  This requirement led the Zcash developers to 
conclude that the anonymity set for users spending shielded coins is in fact
all generated coins, and thus that 
``the mixing strategies that other cryptocurrencies use for anonymity 
provide a rather small [anonymity set] in comparison to Zcash'' and 
that ``Zcash has a distinct advantage in terms of transaction 
privacy''~\cite{zcash-faqs}. 

In this Chapter, we provide the first in-depth empirical analysis of anonymity
in Zcash, in order to examine these claims and more generally provide a
longitudinal study of how Zcash has evolved and who its main
participants are.
We begin in Section~\ref{sec:zcash-usage} by providing a
general examination of the Zcash blockchain, from which we observe that the 
vast majority of Zcash activity is in the transparent part of the 
blockchain, meaning it does not engage with the shielded pool at all.  In
Section~\ref{sec:T-to-T}, we explore this aspect of Zcash by adapting the 
analysis that has already been developed for Bitcoin, and find that exchanges 
typically dominate this part of the blockchain.

We then move in Section~\ref{sec:pool-interaction} to examining interactions
with the shielded pool.  We find that, unsurprisingly, the main actors
doing so are the founders and miners, who are required to 
put all newly generated coins directly into it.  Using newly
developed heuristics for attributing transactions to founders and miners, we
find that 65.6\% of the value withdrawn from the pool can be
linked back to deposits made by either founders or miners.  We also implement
a general heuristic for linking together other types of transactions, and
capture an additional 3.5\% of the value using this.  Our relatively simple
heuristics thus reduce the size of the overall anonymity set by 69.1\%.

In Section~\ref{sec:z-to-z}, we then look at the relatively small percentage
of transactions that have taken place within the shielded pool.  Here, we find
(perhaps unsurprisingly) that relatively little information can be inferred,
although we do identify certain patterns that may warrant further
investigation.  Finally, we perform a small case study of the activities of
the Shadow Brokers within Zcash in Section~\ref{sec:shadow-brokers}, and in
Section~\ref{sec:zec:conclusions} we conclude.

All of our results have been disclosed, at the time of the work's 
submission, to the creators of Zcash, and discussed extensively with them 
since.  This has resulted in changes to both their public 
communication about Zcash's anonymity as well as the transactional behavior 
of the founders.  Additionally, all the code for our analysis is available as
an open-source
repository.\footnote{{\scriptsize
\url{https://github.com/manganese/zcash-empirical-analysis}}}

\section{Background}\label{sec:back-participants}

In this section we describe four types of participants who interact in the 
Zcash network.

Founders took part in the initial creation and release of Zcash, and
will receive 20\% of all newly generated coins (currently 2.5~ZEC out of the
12.5~ZEC block reward).  The founder addresses are specified in the Zcash 
chain parameters~\cite{params}.

Miners take part in the maintenance of the ledger, and in
doing so receive newly generated coins (10 out of the 12.5~ZEC block reward), 
as well as any fees from the transactions
included in the blocks they mine.  Many miners choose not to mine on their
own, but join a mining pool; a list of mining pools can be found in 
Table~\ref{tab:miners}.  One or many miners win each block,
and the first transaction in the block is a \emph{coin generation} (coingen) 
that assigns newly generated coins to their address(es), as well as to the 
address(es) of the founders.

Services are entities that accept ZEC as some form of payment.  These include
exchanges like Bitfinex, which allow users to trade fiat
currencies and other cryptocurrencies for ZEC (and vice versa), and platforms 
like ShapeShift~\cite{shapeshift}, which allow users to trade within
cryptocurrencies and other digital assets without requiring registration.  

Finally, users are participants who hold and transact in ZEC at a more
individual level.  In addition to regular individuals, this category includes
charities and other organizations that may choose to accept donations in
Zcash.  A notable user is the Shadow Brokers, a hacker group who have 
published several leaks containing hacking tools from the NSA and accept 
payment in Zcash.  We explore their usage of Zcash in 
Section~\ref{sec:shadow-brokers}.

\section{General Blockchain Statistics}\label{sec:zcash-usage}

We used the \texttt{zcashd} client to download the Zcash blockchain, and
loaded a database representation of it into Apache Spark.  We then performed 
our analysis using a custom set of Python scripts equipped with \mbox{PySpark}.  
We last parsed the block chain on January 21 2018, at which point 258,472 blocks 
had been mined.  Overall, 3,106,643~ZEC had been generated since the genesis 
block, out of which 2,485,461~ZEC went to the miners and the rest 
(621,182~ZEC) went to the founders. 

\subsection{Transactions}\label{sec:tx-usage}

\begin{figure}[t]
\centering
\includegraphics[width=0.7\linewidth]{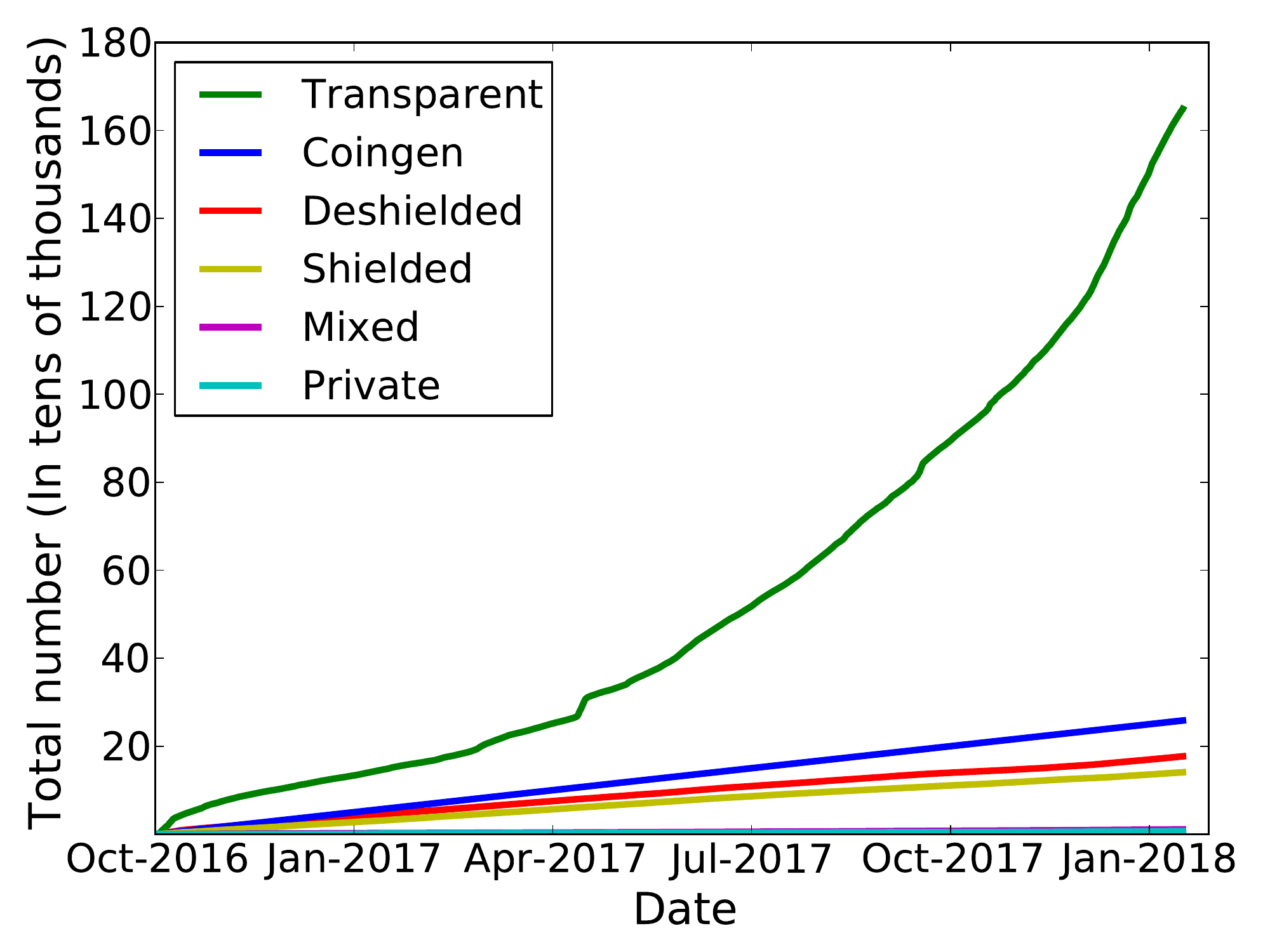}
\caption[Total number of the different types of transactions 
over time.]{The total number of each of the different types of transactions 
over time.}
\label{fig:cumulative_types_of_transactions}
\end{figure}

\begin{figure}[t]
\centering
\includegraphics[width=0.8\linewidth]{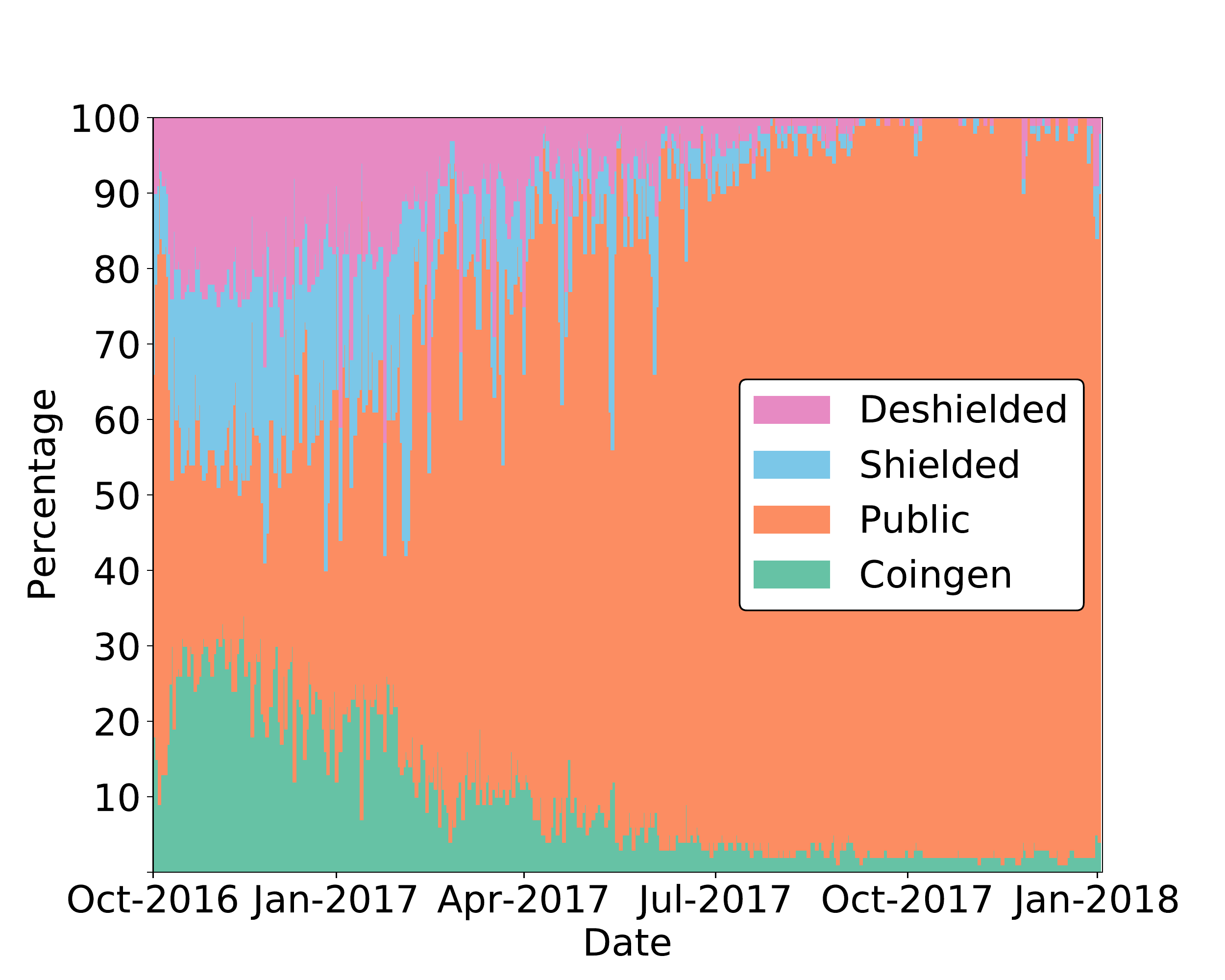}
\caption[Value of Transactions over time.]{The fraction of the value in each block representing each different 
type of transaction over time, averaged daily.  Here, `public' captures both 
transparent transactions and the visible components of mixed transactions.}
\label{fig:fistful_DailyAverage}
\end{figure}

Across all blocks, there were 2,242,847 transactions.  A complete 
breakdown of the transaction types is in Table~\ref{tab:tx-types},
and graphs depicting the growth of each transaction type over time are in
Figures~\ref{fig:cumulative_types_of_transactions}
and~\ref{fig:fistful_DailyAverage}.\footnote{We use the term `mixed' to mean
    transactions that have both a \realvin and a \realvout, and a \vjoinsplit.}    
The vast majority of transactions are public (i.e., either transparent or a 
coin generation).
Of the transactions that do interact with the pool (335,630, or 14.96\%, in
total), only a very small percentage are private transactions; i.e., 
transactions within the pool.
Looking at the types of transactions over time
in Figure~\ref{fig:cumulative_types_of_transactions}, we can see that the
number of coingen, shielded, and deshielded transactions all grow in an
approximately linear fashion.  As we explore 
in Section~\ref{sec:miners}, this correlation is due largely to the
habits of the miners.  Looking at both this figure and
Figure~\ref{fig:fistful_DailyAverage}, we can see that while the number of
transactions interacting with the pool has grown in a relatively linear
fashion, the value they carry has over time become a very small 
percentage of all
blocks, as more mainstream (and thus transparent) usage of Zcash has
increased.

\begin{table}
\centering
\begin{tabular}{lS[table-format=7.0]S[table-format=2.1]}
\toprule
Type & {Number} & {Percentage} \\
\midrule
Transparent & 1648745 & 73.5\\
Coingen & 258472 & 11.5 \\
Deshielded & 177009 & 7.9\\
Shielded & 140796 & 6.3 \\
Mixed & 10891 & 0.5 \\
Private & 6934 & 0.3 \\
\bottomrule
\end{tabular}
\caption[Total number of Zcash transaction types]{The total number of each transaction type.}
\label{tab:tx-types}
\end{table}

\subsection{Addresses}

Across all transactions, there have been 1,740,378 distinct t-addresses used.
Of these, 8,727 have ever acted as inputs in a \ttoz transaction and 330,780 
have ever acted as outputs in a \ztot transaction.  As we explore in
Section~\ref{sec:miners}, much of this asymmetry is due to the behavior of 
mining pools, which use a small number of addresses to collect the block
reward, but a large number of addresses (representing all the individual
miners) to pay out of the pool.  Given the nature of the shielded pool, it is 
not possible to know the total number of z-addresses used.%

\begin{figure}[t]
\centering
\includegraphics[width=0.7\linewidth]{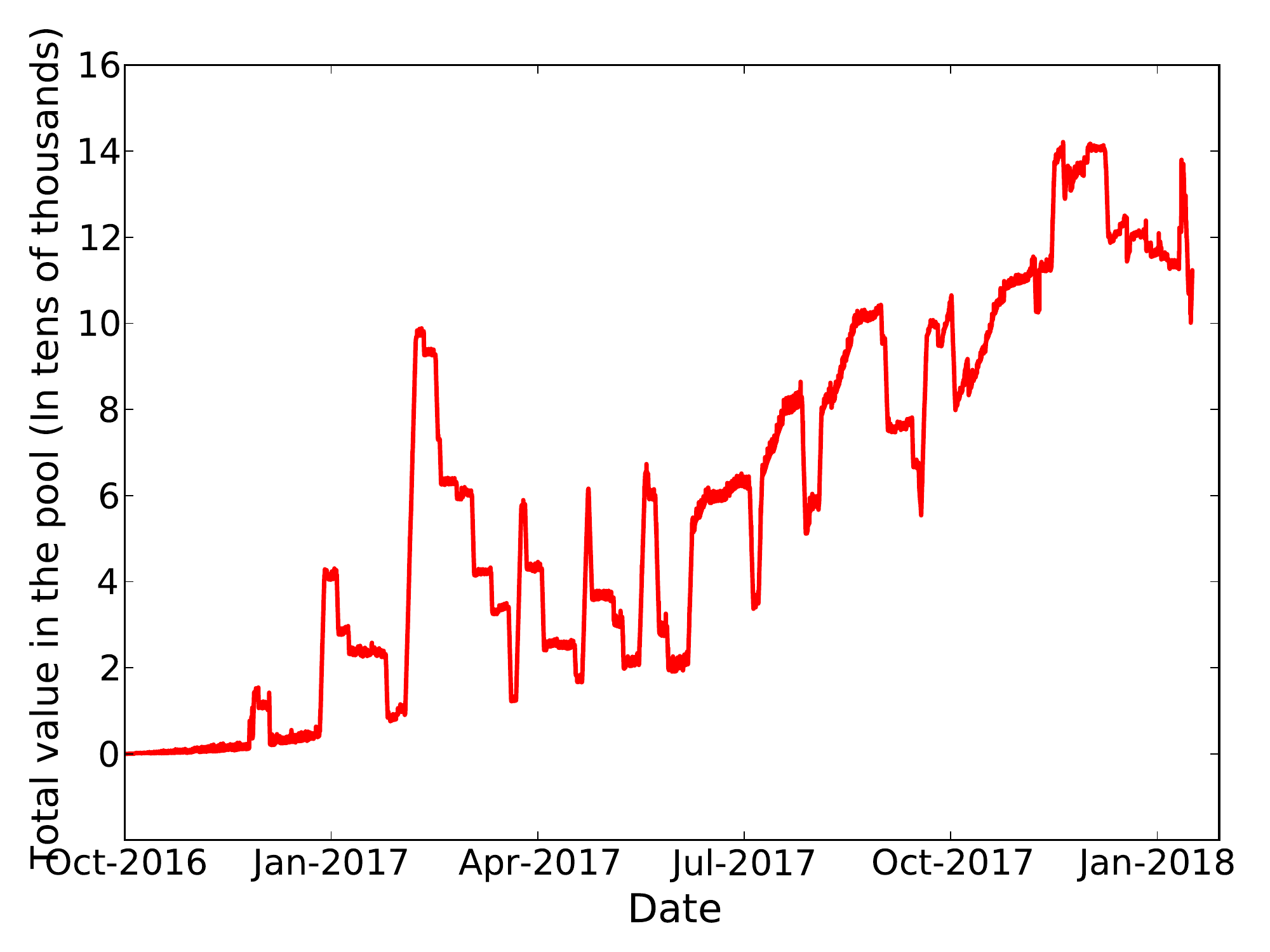}
\caption[Total value in the shielded pool over time]{The total value in the shielded pool over time, in tens of thousands of ZEC.}
\label{fig:total_value_in_pool}
\end{figure}

Figure~\ref{fig:total_value_in_pool} shows the total value in the pool over 
time.  Although the overall value is increasing over time, 
there are certain shielding and de-shielding patterns that create spikes.  As
we explore in Section~\ref{sec:pool-interaction}, these spikes are due largely
to the habits of the miners and founders.  At the time of writing, there are 
112,235 ZEC in the pool, or 3.6\% of the total monetary supply. 

If we rank addresses by their wealth, we first observe that only 25\% of all
t-addresses have a non-zero balance.  Of these, the top 1\% hold 78\% of all
ZEC.  The address with the highest balance had 118,257.75~ZEC, which means
the richest address has a higher balance than the entire shielded pool.

\section{T-Address Clustering} \label{sec:T-to-T}

As discussed in Section~\ref{sec:zcash-usage}, a large proportion of the
activity on Zcash does not use the shielded pool.  This means it is
essentially identical to Bitcoin, and thus can be de-anonymized using the same
techniques discussed for Bitcoin in Section~\ref{chapter:back}.  

\subsection{Clustering addresses}\label{sec:clustering}

To identify the usage of transparent addresses, we begin by recalling the 
``multi-input'' heuristic for clustering Bitcoin
addresses.  In this heuristic, addresses that are used as inputs to the same 
transaction are assigned to the same cluster.  In Bitcoin, this heuristic can 
be applied to all transactions, as they are all transparent.  In Zcash, we 
perform this clustering as long as there are multiple input t-addresses.

\begin{heuristic}\label{heuristic:inputclusters}
If two or more t-addresses are inputs in the same transaction (whether that
transaction is transparent, shielded, or mixed), then they are controlled by 
the same entity. 
\end{heuristic} 

In terms of false positives, we believe that these are
at least as unlikely for Zcash as they are for Bitcoin, as Zcash is a direct
fork of Bitcoin and the standard client has the same behavior.  In fact, we 
are not aware of any input-mixing techniques like CoinJoin~\cite{coinjoin} for 
Zcash, so could argue that the risk of false positives is even lower than it 
is for Bitcoin.  As this heuristic has already been used extensively in
Bitcoin, we thus believe it to be realistic for use in Zcash.

We implemented this heuristic by defining each t-address as a node in a graph,
and adding an (undirected) edge in the graph between addresses that had been
input to the same transaction.  The connected components of the graph then
formed the clusters, which represent distinct entities controlling potentially
many addresses.  The result was a set of 560,319 clusters, of which 97,539
contained more than a single address.

As in Bitcoin, using just this one heuristic is already quite effective but
does not capture the common usage of \emph{change addresses}, in which a
transaction sends coins to the actual recipient but then also sends any coins
left over in the input back to the sender.  Meiklejohn et
al.~\cite{sarah-fistfulofbitcoins} use in their analysis a heuristic based on
this behavior, but warn that it is somewhat fragile.  Indeed, their heuristic
seems largely dependent on the specific behavior of several large Bitcoin
services, so we chose not to implement it in its full form.  Nevertheless, we
did use a related Zcash-specific heuristic in our case study of the Shadow
Brokers in Section~\ref{sec:shadow-brokers}.%

\begin{heuristic}\label{heuristic:others}
If one (or more) address is an input t-address in a \vjoinsplit transaction
and a second address is an output t-address in the same \vjoinsplit
transaction, then if the size of \vout is $1$ (i.e., this is the only
transparent output address), the second address belongs to the same user who 
controls the input addresses. 
\end{heuristic}

To justify this heuristic, we observe that users may not want to deposit
all of the coins in their address when putting coins into the pool, in which 
case they will have to make change.  The only risk of a false positive is if
users are instead sending money to two separate individuals, one using a 
z-address and one using a t-address.  One notable exception to 
this rule is users of the zcash4win wallet.  Here, the address of the wallet 
operator is an output t-address if the user decides to pay the developer fee,
so it would produce exactly this type of transaction for users putting money into
the shielded pool.  This address is identifiable, however, so these types of
transactions can be omitted from our analysis.  Nevertheless, due to concerns
about the safety of this heuristic (i.e., its ability to avoid false
positives), we chose not to incorporate it into our general analysis below.

\subsection{Tagging addresses}\label{sec:tagging}

Having now obtained a set of clusters, we next sought to assign names to them.
To accomplish this, we performed a scaled-down version of
the techniques used by Meiklejohn et
al.~\cite{sarah-fistfulofbitcoins}.  In particular, given that Zcash is still
relatively new, there are not many different types of services that accept
Zcash. We thus restricted ourselves to interacting with exchanges.

We first identified the top ten Zcash exchanges according to 
volume traded~\cite{coinmarketcap}.  We then created an account with each 
exchange and deposited a small quantity of ZEC into it, tagging as we did 
the output t-addresses in the resulting transaction as belonging to the 
exchange.  We 
then withdrew this amount to our own wallet, and again tagged the t-addresses
(this time on the sender side) as belonging to the exchange.  We occasionally
did several deposit transactions if it seemed likely that doing so would tag
more addresses.  Finally, we also interacted
with ShapeShift, which as mentioned in Section~\ref{sec:back}
allows users to move amongst cryptocurrencies without the need to create an
account.  Here we did a single ``shift'' into Zcash and a single shift out.  A
summary of our interactions with all the different exchanges is in
Table~\ref{tab:ourtxs}.

\begin{table}
\centering
\begin{tabular}{lccc}
\toprule
Service & Cluster & \# deposits & \# withdrawals \\ 
\midrule
Binance  &  7      &  1       &   1       \\
Bitfinex &  3      &  4       &   1		  \\
Bithumb  &  14     &  2       &   1       \\
Bittrex  &  1      &  1       &   1       \\
Bit-z    &  30     &  2       &   1       \\
Exmo     &  4      &  2       &   1       \\
HitBTC   &  18     &  1       &   1       \\
Huobi    &  26     &  2       &   1       \\
Kraken   &  12     &  1       &   1       \\
Poloniex &  0      &  1       &   1       \\
\midrule
ShapeShift & 2   &  1  &  1 \\
zcash4win  & 139 &  1  & 2  \\
\bottomrule 
\end{tabular}
\caption[Services and identified clusters after running heuristics]{The services we interacted with, the identifier of the cluster they
were associated with after running Heuristic~\ref{heuristic:inputclusters}, 
and the number of deposits and withdrawals we did with them.  The first ten
are exchanges, ShapeShift is an inter-cryptocurrency exchange, and zcash4win
is a Windows-based Zcash client.}
\label{tab:ourtxs} 
\end{table}

Finally, we collected the publicized addresses of the founders~\cite{params}, 
as well as addresses from known mining pools.  For the latter we started by 
scraping the tags of these addresses from the Zchain explorer~\cite{zchain}.
We then validated them against the blocks advertised on some of the websites of 
the mining pools themselves (which we also scraped) to ensure that they were the 
correct tags; i.e., if the recipient of the coingen transaction in a given 
block was tagged as belonging to a given mining pool, then we checked to see
that the block had been advertised on the website of that mining pool. %
We then augmented these sets of addresses with the addresses tagged as 
belonging to founders and miners according to the heuristics developed in
Section~\ref{sec:pool-interaction}.  We present these heuristics in
significantly more detail there, but they resulted in us tagging 123 founder
addresses and 110,918 miner addresses (belonging to a variety of different
pools).

\subsection{Results}\label{sec:tagging_results}

As mentioned in Section~\ref{sec:clustering}, running
Heuristic~\ref{heuristic:inputclusters} resulted in 560,319 clusters, of 
which 97,539 contained more than a single address.  We assigned each cluster a
unique identifier, ordered by the number of addresses in the cluster, so that 
the biggest cluster had identifier 0.
  
\subsubsection{Exchanges and wallets}

As can be seen in Table~\ref{tab:ourtxs}, many of the exchanges are associated
with some of the biggest clusters, with four out of the top five clusters
belonging to popular exchanges.  In general, we found that the top five
clusters accounted for 11.21\% of all transactions.  Identifying exchanges is
important, as it makes it possible to discover where individual users may have
purchased their ZEC.  Given existing and emerging regulations, they are also 
the one type of participant in the Zcash ecosystem that might know the
real-world identity of users.

In many of the exchange clusters, we also identified large fractions of
addresses that had been tagged as miners.  This implies that individual 
miners use the addresses of their exchange accounts to receive their mining 
reward, which might be expected if their goal is to cash out directly.  We
found some, but far fewer, founder addresses at some of the exchanges as well.

Our clustering also reveals that ShapeShift (Cluster~2) is fairly heavily
used: it had received over 1.1M~ZEC in total and sent roughly the same.  Unlike 
the exchanges, its cluster contained a relatively small number of miner 
addresses (54), which fits with its usage as a way to shift money, rather 
than hold it in a wallet.

\subsubsection{Mining pools and founders}

Although mining pools and founders account for a large proportion of the 
activity in Zcash (as we explore in Section~\ref{sec:pool-interaction}), many 
re-use the same small set of addresses frequently, so do not belong to 
large clusters.  For example, Flypool had three single-address clusters while 
Coinotron, coinmine.pl, Slushpool and Nanopool
each had two single-address clusters. (A list of mining pools can be
found in Table~\ref{tab:miners} in Section~\ref{sec:miners}).
Of the coins that we
saw sent from clusters associated with mining pools, 99.8\% of it went into
the shielded pool, which further validates both our clustering and tagging
techniques.

\subsubsection{Philanthropists}

Via manual inspection, we identified three large organizations that accept 
Zcash donations: the Internet Archive, \url{torservers.net}, and Wikileaks.   
Of these, \url{torservers.net} accepts payment only via a z-address, so we
cannot identify their transactions (Wikileaks accepts payment via a z-address
too, but also via a t-address).  Of the 31 donations to the Internet Archive
that we were able to identify, which totaled 17.3~ZEC, 9 of them were made
anonymously (i.e., as \ztot transactions).  On the other hand, all of the
20 donations to Wikileak's t-address were made as \ttot transactions. 
None of these belong to clusters, as they have never sent a transaction. 

Most of the donations are small quantities of ZEC.  For example, the transparent donations to wikileaks ranged between 0.00065 ZEC and 1.4 ZEC with a median donation of  0.035 ZEC.

\section{Interactions with the Shielded Pool}\label{sec:pool-interaction}

What makes Zcash unique is of course not its t-addresses (since these
essentially replicate the functionality of Bitcoin), but its shielded pool.
To that end, this section explores interactions with the pool at its
endpoints, meaning the deposits into (\ttoz) and withdrawals out of the pool 
(\ztot).  We then explore interactions within the pool (\ztoz transactions) in
Section~\ref{sec:z-to-z}.

\begin{figure}
\centering
\includegraphics[width=0.75\linewidth]{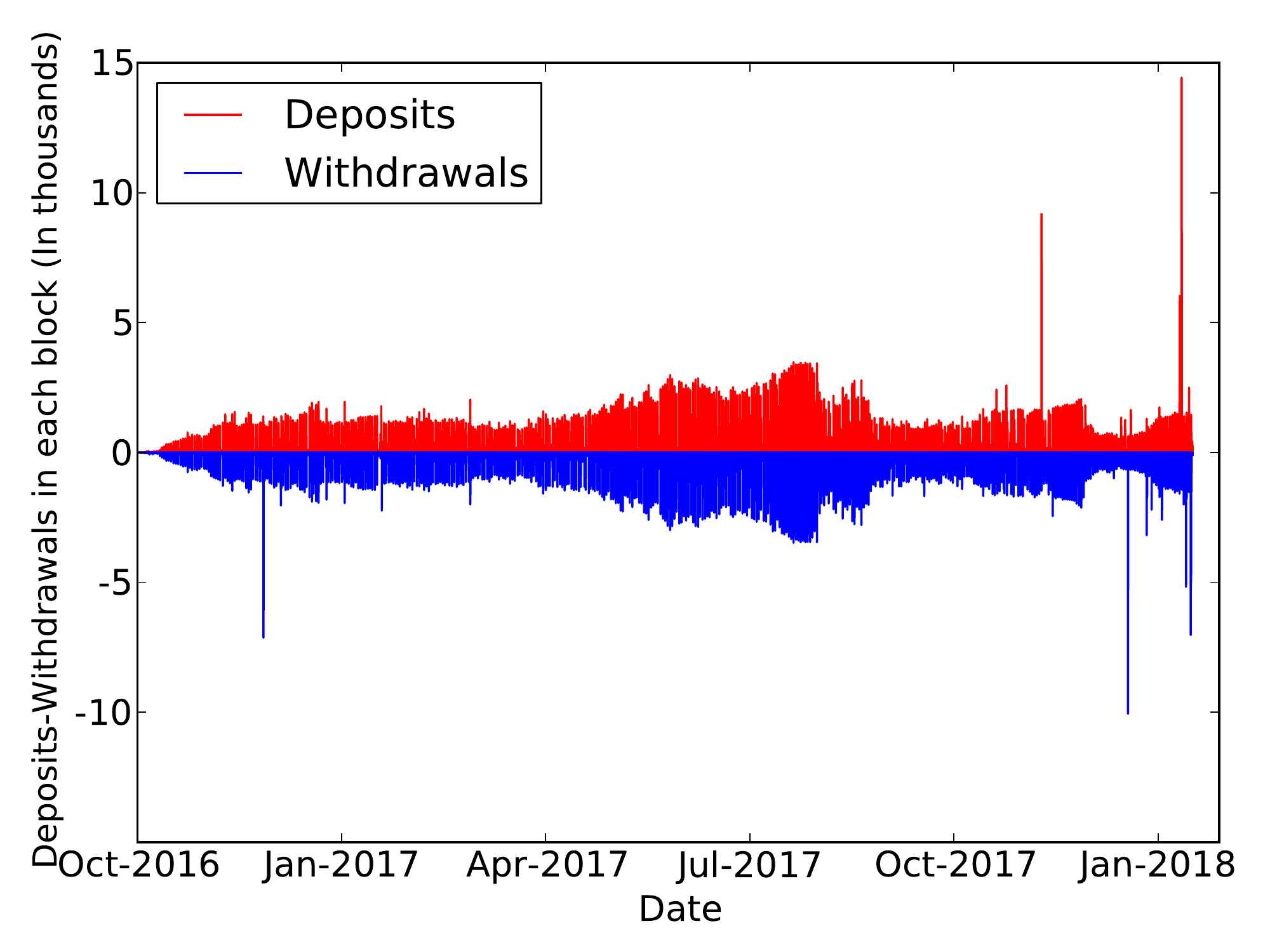}
\caption[Amount of ZEC deposited and withdrawn from the shielded pool]{Over time, the amount of ZEC put into the shielded pool (in red) and 
the amount taken out of the pool (in blue).}
\label{fig:InAndOutEachBlock}
\end{figure}

To begin, we consider just the amounts put into and taken out of the pool.  
Over time, 3,901,124~ZEC have been deposited into the
pool,\footnote{This is greater than the total number of generated coins, as
    all coins must be deposited into the pool at least once, by the miners or
    founders, but may then go into and out of the pool multiple times.}
and 3,788,889 have been withdrawn.  Figure~\ref{fig:InAndOutEachBlock} plots
both deposits and withdrawals over time.

This figure shows a near-perfect reflection of deposits and
withdrawals, demonstrating that most users not only withdraw the exact number
of ZEC they deposit into the pool, but do so very quickly after the initial
deposit.  As we see in Sections~\ref{sec:founders} and~\ref{sec:miners}, this
phenomenon is accounted for almost fully by the founders and miners.  Looking
further at the figure, we can see that the symmetry is broken occasionally,
and most notably in four ``spikes'': two large withdrawals, and two large
deposits.  Some manual investigation revealed the following:

\begin{sarahlist}
\item[``The early birds''] The first withdrawal spike took place at block
height 30,900, which was created in December 2016.  The cause of the spike was
a single transaction in which 7,135~ZEC was taken out of the pool; given the
exchange rate at that time of 34~USD per ZEC, this was equivalent to
242,590~USD.  The coins were distributed across 15 t-addresses, which initially
we had not tagged as belonging to any named user.  After running the heuristic
described in Section~\ref{sec:founders}, however, we tagged all of these 
addresses as belonging to founders.  In fact, this was the very first
withdrawal that we identified as being associated with founders.%
	
\item[``Secret Santa''] The second withdrawal spike took place on December 25
2017, at block height 242,642.  In it, 10,000~ZEC was distributed among 10
different t-addresses, each receiving 1,000~ZEC.  None of these t-addresses 
had done a transaction before then, and none have been involved in one
since (i.e., the coins received in this transaction have not yet been spent).
	
\item[``One-man wolf packs''] Both of the deposit spikes in the graph 
correspond to single large deposits from unknown t-addresses that, 
using our analysis 
from Section~\ref{sec:T-to-T}, we identified as residing in single-address
clusters.  For the first spike, however, many of the deposited amounts came 
directly from a founder address identified by our heuristics 
(Heuristic~\ref{heuristic:founder}), so
given our analysis in Section~\ref{sec:founders} we believe this may also be
associated with the founders.

\end{sarahlist}

While this figure already provides some information about how the pool is used
(namely that most of the money put into it is withdrawn almost immediately
afterwards), it does not tell us who is actually using the pool.  For this, we
attempt to associate addresses with the types of participants identified
in Section~\ref{sec:back-participants}: founders, miners, and `other' 
(encompassing both services and individual users).

When considering deposits into the shielded pool, it is easy to 
associate addresses with founders and miners, as the consensus rules dictate 
that they must put their block rewards into the shielded pool before spending 
them further.  

\begin{figure}
\centering
\includegraphics[width=0.75\linewidth]{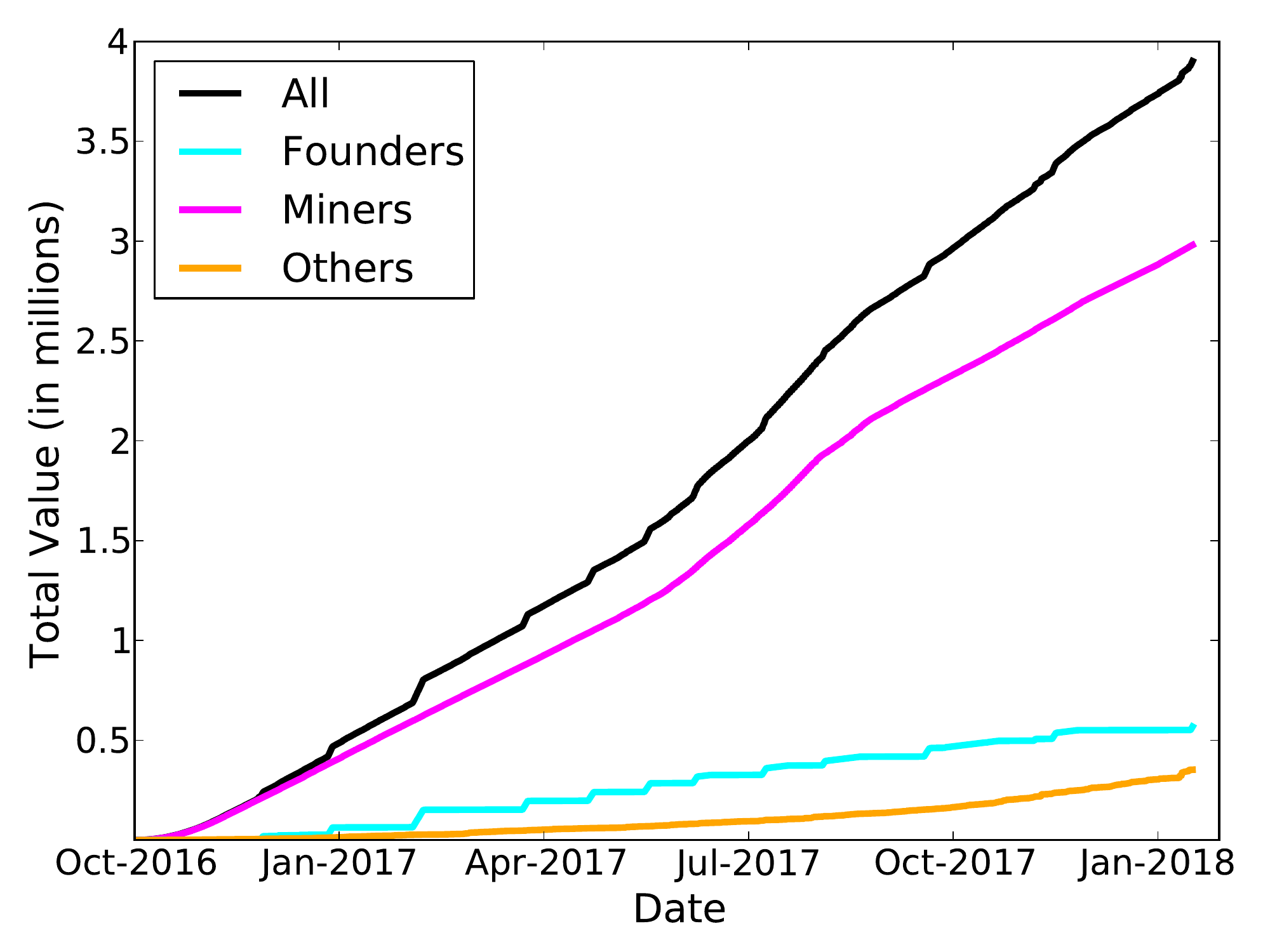}
\caption[Amount of ZEC deposited into the shielded pool by 
miners, founders, and others]{Over time, the amount of ZEC deposited into the shielded pool by 
miners, founders, and others.}
\label{fig:t-to-z}
\end{figure}

As described in Section~\ref{sec:tagging}, we tagged founders
according to the Zcash parameters, and tagged as miners all recipients of
coingen transactions that were not founders.  We then used these tags to
identify a founder deposit as any \ttoz transaction using one or more
founder addresses as input, and a miner deposit as any \ttoz transaction
using one or more miner addresses as input. 
The results are in Figure~\ref{fig:t-to-z}.

Looking at this figure, it is clear that miners are the main participants
putting money into the pool.  This is not particularly surprising, given that
all the coins they receive must be deposited into the pool at least once, so
if we divide that number of coins by the total number deposited we would
expect at least 63.7\% of the deposits to come from miners. (The actual number
is 76.7\%.) Founders, on the other hand, don't put
as much money into the pool (since they don't have as much to begin with), 
but when they do they put in large amounts that cause visible step-like 
fluctuations to the overall line.  

\begin{figure}
\centering
\includegraphics[width=100mm]{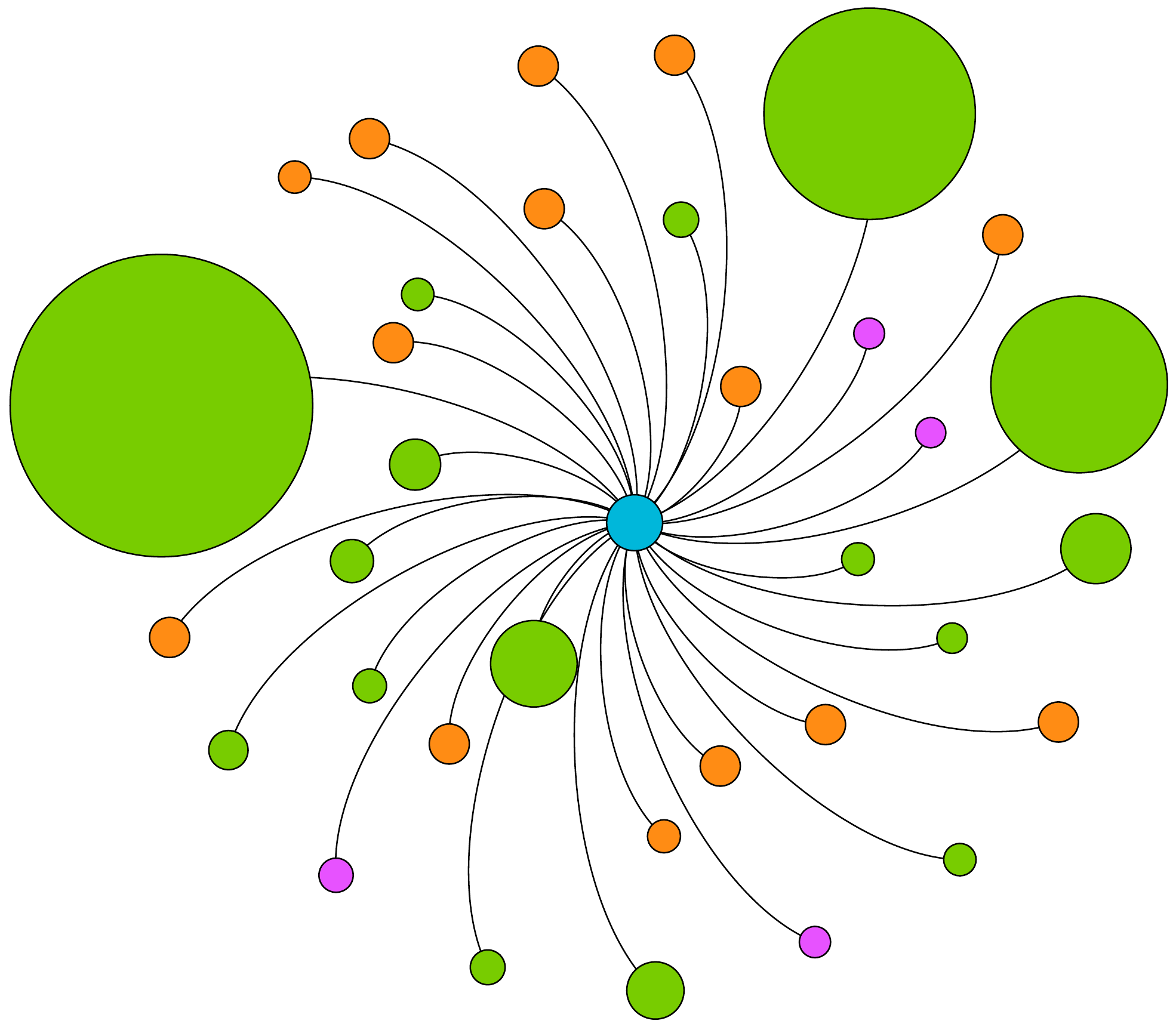}
\caption[Addresses that put more than 10,000 ZEC into the shielded 
pool over time]{The addresses that have put more than 10,000 ZEC into the shielded 
pool over time, where the size of each node is proportional to the
value it has put into the pool.  The addresses of miners are green, of 
founders are orange, and of unknown `other' participants are purple.}\label{fig:TtoZ10k}
\end{figure}

In terms of the heaviest users, we looked at the individual addresses
that had put more than 10,000 ZEC into the pool.  The results are in
Figure~\ref{fig:TtoZ10k}.

In fact, this figure incorporates the heuristics we develop in
Sections~\ref{sec:founders} and~\ref{sec:miners}, although it looked very
similar when we ran it before applying our heuristics (which makes sense,
since our heuristics mainly act to link \ztot transactions).
Nevertheless, it demonstrates again that most of the heavy users of the pool
are miners, with founders also depositing large amounts but spreading them 
over a wider variety of addresses.  Of the four `other' addresses, one of them 
belonged to ShapeShift, and the others belong to untagged clusters.

\begin{figure}
\centering
\begin{subfigure}[b]{0.7\textwidth}
\centering
\includegraphics[width=\linewidth]{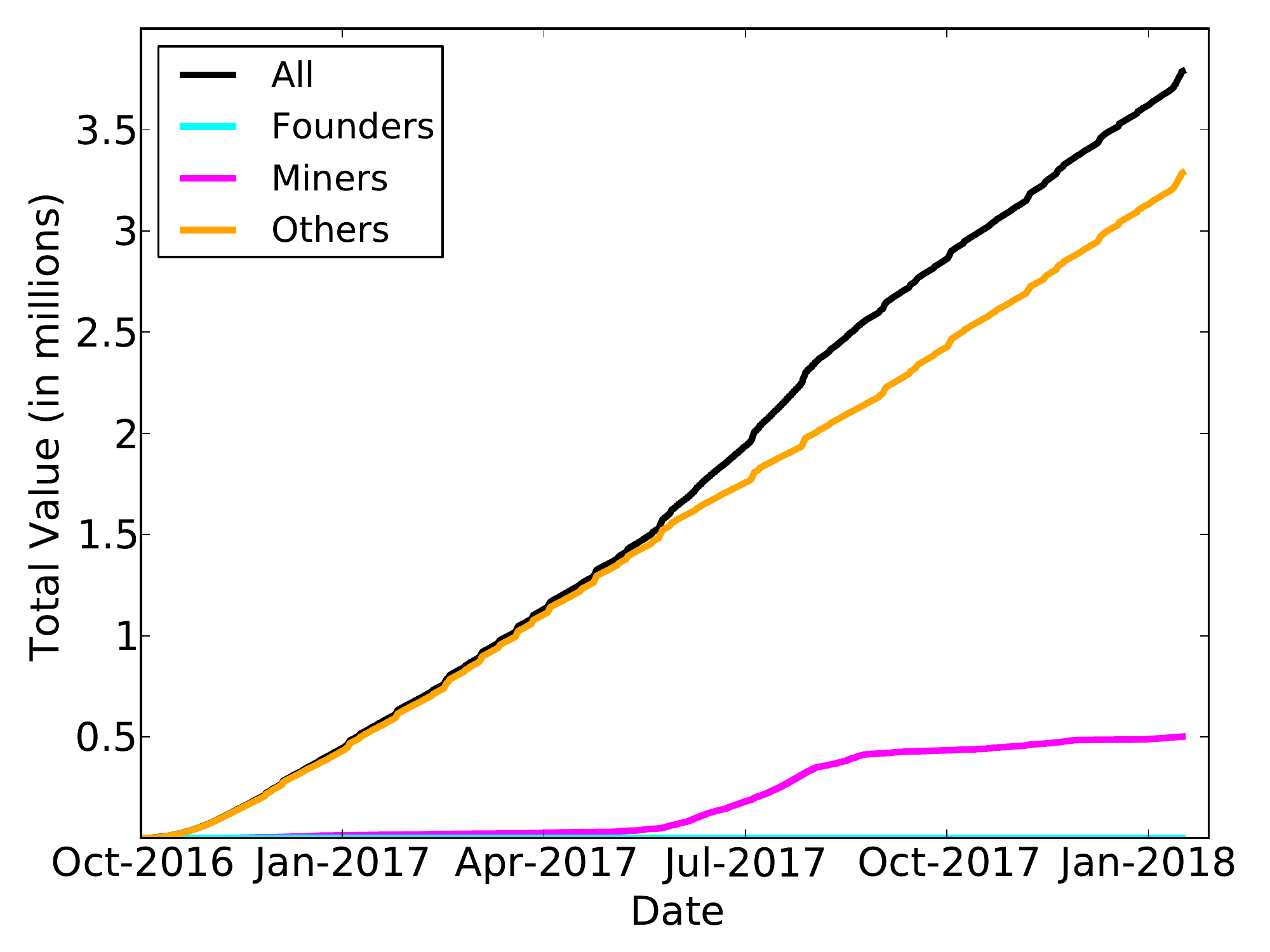}
\caption{No heuristics}
\label{fig:zt-none}
\end{subfigure}
~
\begin{subfigure}[b]{0.7\textwidth}
\includegraphics[width=\linewidth]{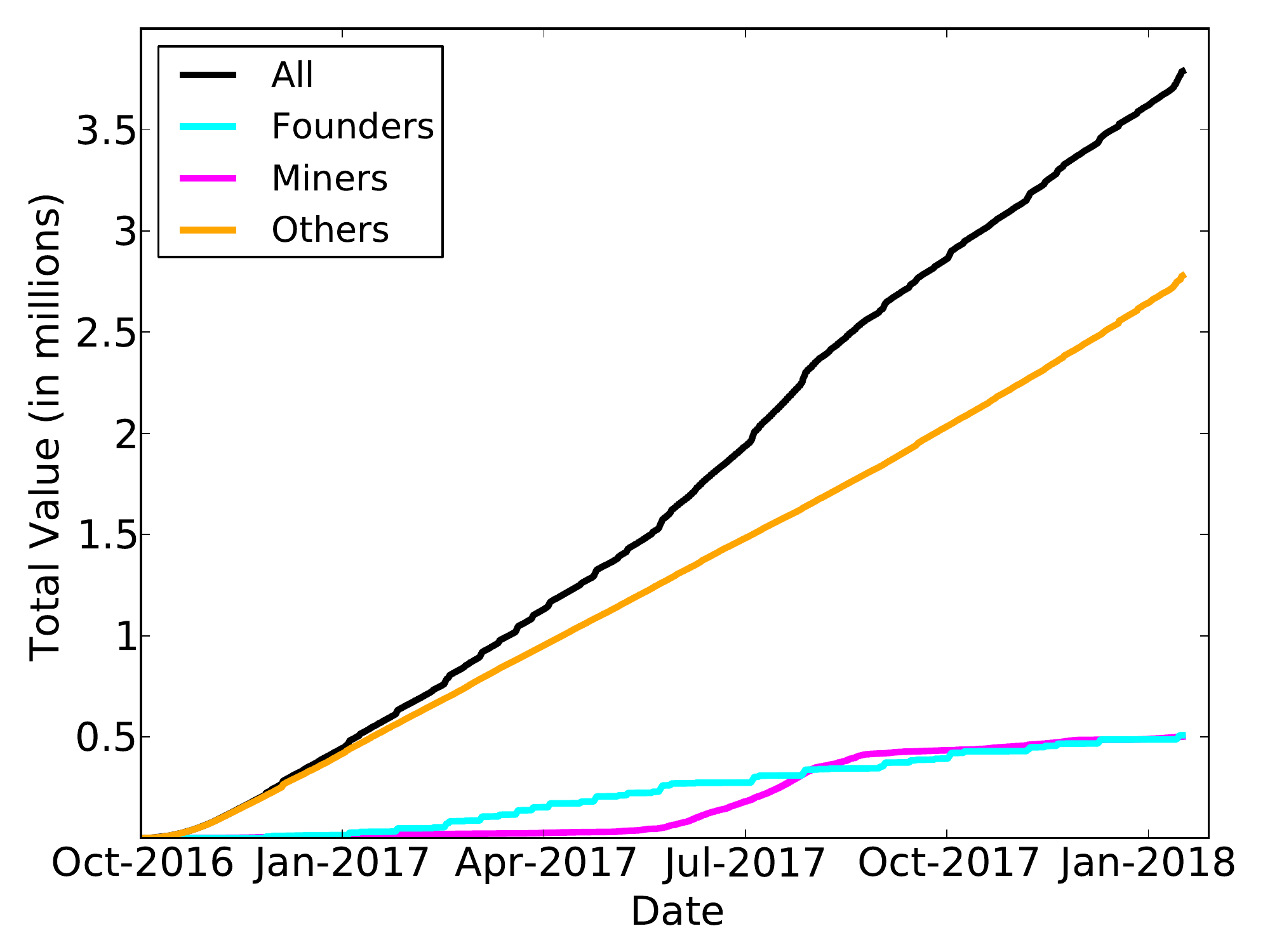}
\caption{Founder heuristic}
\label{fig:zt-founders}
\end{subfigure}
~
\begin{subfigure}[b]{0.7\textwidth}
\includegraphics[width=\linewidth]{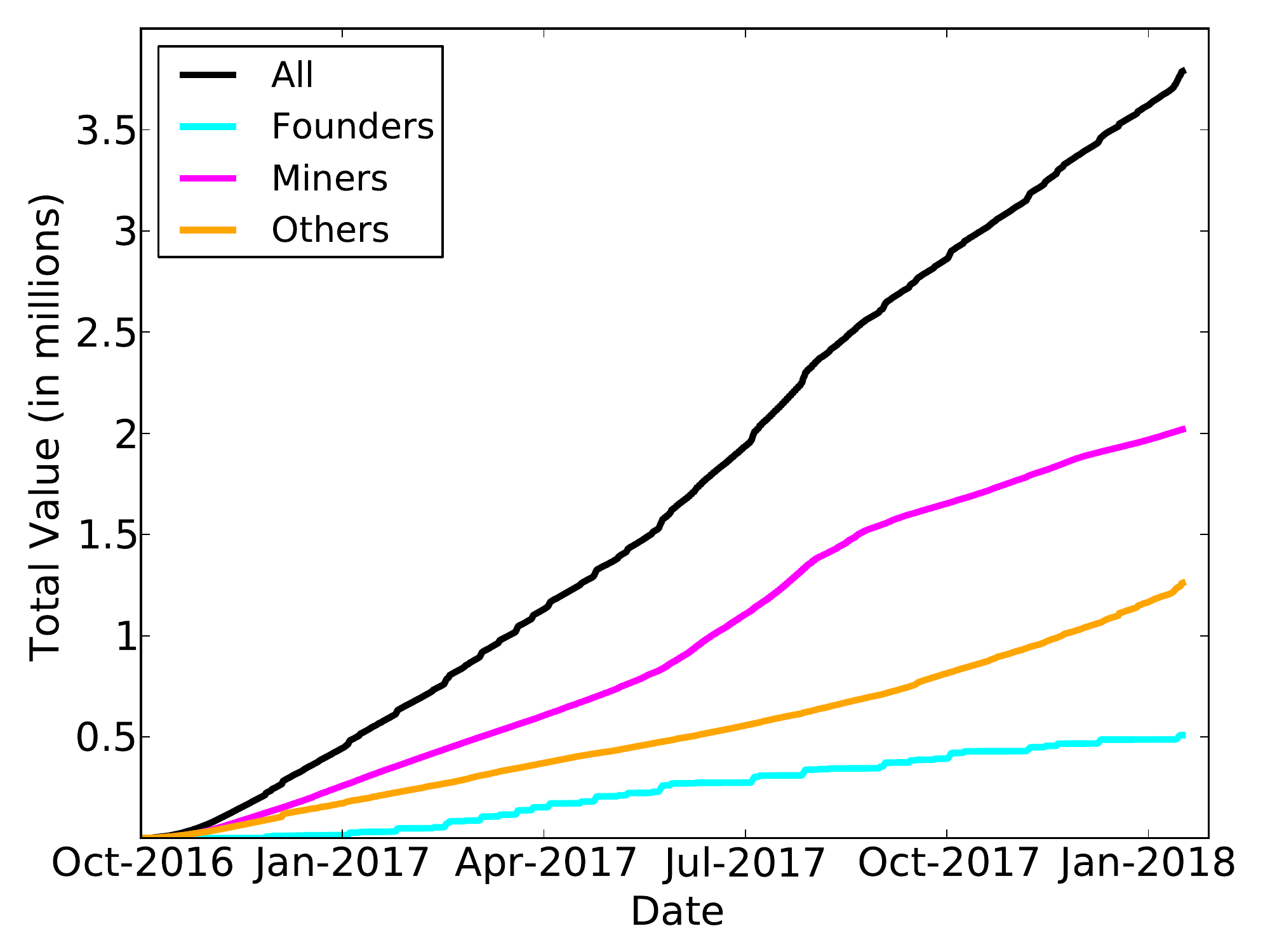}
\caption{Founder and miner heuristics}
\label{fig:zt-both}
\end{subfigure}
\caption[\ztot transactions associated with miners, founders, and
`other', with our heuristics.]{The \ztot transactions we associated with miners, founders, and
`other', after running some combination of our heuristics, in millions of transactions.}
\label{fig:all-heuristics}
\end{figure}

While it is interesting to look at \ttoz transactions on their own, the main
intention of the shielded pool is to provide an anonymity set, so that when
users withdraw their coins it is not clear whose coins they are.  In that
sense, it is much more interesting to link together 
\ttoz and \ztot transactions, which acts to reduce the anonymity set.  More
concretely, if a \ttoz transaction can be linked to a \ztot transaction, then
those coins can be ``ruled out'' of the anonymity set of future users 
withdrawing coins from the pool.  
We thus devote our attention to this type of analysis for the rest of the
section.

The most na{\"i}ve way to link together these transactions would be to see if
the same addresses are used across them; i.e., if a miner uses the same
address to withdraw their coins as it did to deposit them.  By running this
simple form of linking, we see the results in Figure~\ref{fig:zt-none}.
This figure shows that we are not able to identify any withdrawals as
being associated with founders, and only a fairly small number as associated
with miners: 49,280 transactions in total, which account for 13.3\% of the
total value in the pool.  %

Nevertheless, using heuristics that we develop for identifying founders (as
detailed in Section~\ref{sec:founders}) and miners (Section~\ref{sec:miners}),
we are able to positively link most of the \ztot activity with one of
these two categories, as seen in Figures~\ref{fig:zt-founders}
and~\ref{fig:zt-both}.  In the end, of the 177,009 \ztot transactions, we were 
able to tag 120,629 (or 68\%) of them as being associated with miners, 
capturing 52.1\% of the value coming out of the pool, and 2,103 of them as 
being associated with founders (capturing 13.5\% of the value).  We then
examine the remaining 30-35\% of the activity surrounding the shielded pool 
in Section~\ref{sec:others}.

\subsection{Founders}\label{sec:founders}

After comparing the list of founder addresses against the outputs of all 
coingen transactions, we found that 14 of them had been used.  Using 
these addresses, we were able to identify founder deposits into the pool, as 
already shown in Figure~\ref{fig:t-to-z}.  Table~\ref{tab:founders} provides a
closer inspection of the usage of each of these addresses.

\begin{table}[t]
\centering
\begin{tabular}{c S[table-format=4.0]S[table-format=6.1] S[table-format=3.0]} 
\toprule
& {\# Deposits} & {Total value} & {\# Deposits (249)} \\
\midrule
1 & 548 & 19600.4 & 0 \\ 
2 & 252 & 43944.6 & 153 \\
3 & 178 & 44272.5 & 177 \\
4 & 192 & 44272.5 & 176 \\
5 & 178 & 44272.5 & 177 \\ 
6 & 178 & 44272.5 & 177 \\
7 & 178 & 44272.5 & 177 \\
8 & 178 & 44272.5 & 177 \\
9 & 190 & 44272.5 & 176 \\ 
10 & 188 & 44272.5 & 176 \\
11 & 190 & 44272.5 & 176 \\
12 & 178 & 44272.5 & 177 \\
13 & 191 & 44272.5 & 175 \\
14 & 70 & 17500 & 70 \\
\midrule
Total & 2889 & 568042.5 & 2164 \\
\bottomrule
\end{tabular}
\caption[Behaviour of the 14 active Zcash founder addresses]{The behaviour of each of the 14 active founder addresses, in terms of 
the number of deposits into the pool, the total value deposited (in ZEC), and 
the number of deposits carrying exactly 249.9999~ZEC in value.}
\label{tab:founders}
\end{table}

This table shows some quite obvious patterns in the behavior
of the founders.  At any given time, only one address is ``active,'' meaning it
receives rewards and deposits them into the pool.  Once it reaches the limit
of 44,272.5~ZEC, the next address takes its place and it is not used again.
This pattern has held from the third address onwards.  What's more, the amount
deposited was often the same: exactly 249.9999~ZEC, which is roughly the
reward for 100 blocks.  This was true of 74.9\%
of all founder deposits, and 96.2\% of all deposits from the third address
onwards.  There were only ever five other deposits into the pool carrying 
value between 249 and 251~ZEC (i.e., carrying a value close but not equal to
249.9999~ZEC).

Thus, while we were initially unable to identify any withdrawals associated
with the founders (as seen in Figure~\ref{fig:zt-none}), these patterns
indicated an automated use of the shielded pool that might also carry into 
the withdrawals.  Upon examining the withdrawals from the 
pool, we did not find any with a value exactly equal to 249.9999~ZEC.  We did,
however, find 1,953 withdrawals of exactly 250.0001~ZEC (and 1,969 carrying a
value between 249 and 251~ZEC, although we excluded the extra ones from our
analysis).

\begin{figure}[t]
\centering
\includegraphics[width=110mm]{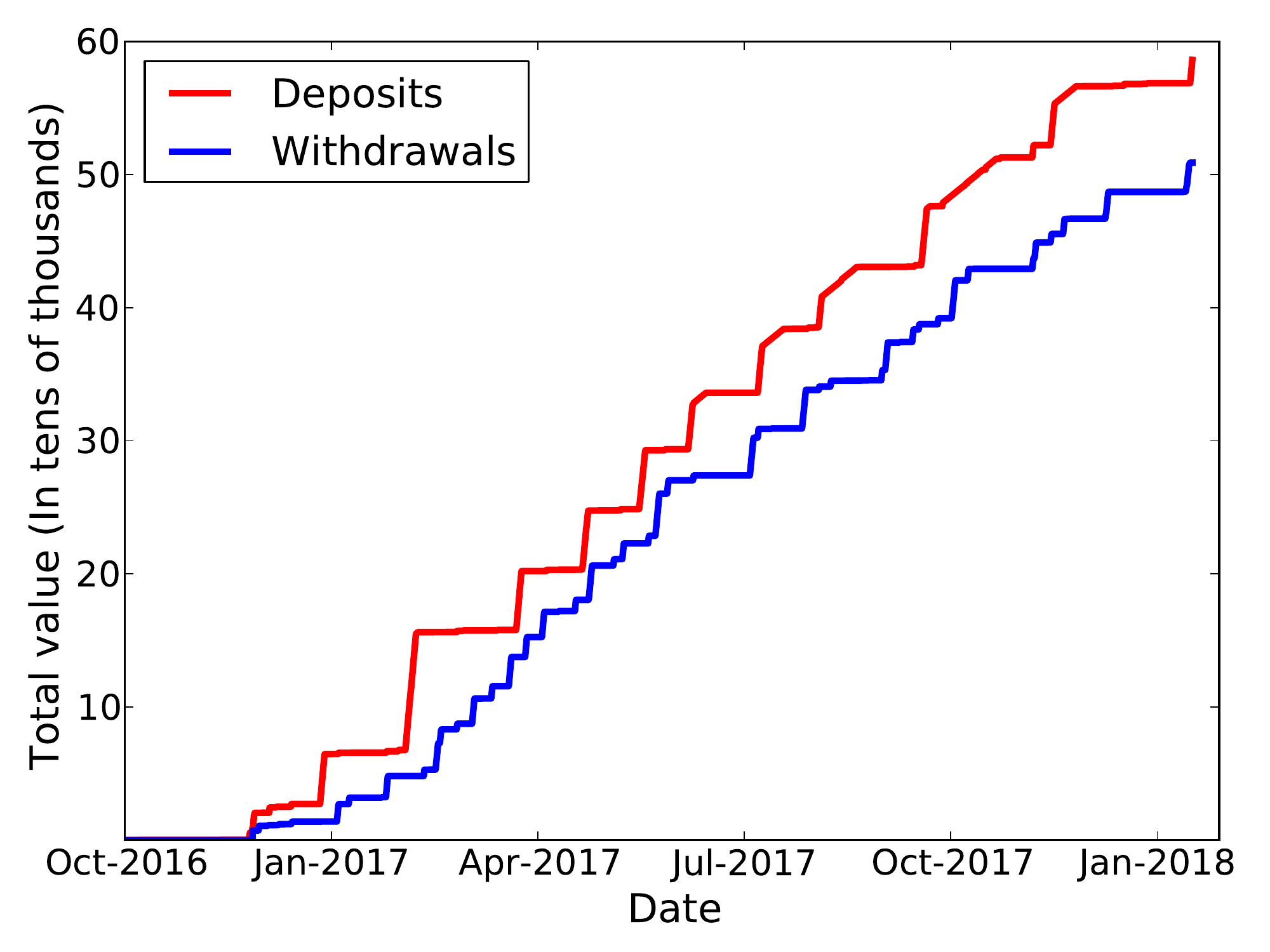}
\caption[Founder deposits and withdrawals into the pool detected using our heuristic]{Over time, the founder deposits into the pool (in red) and 
withdrawals from the pool (in blue), after running
Heuristic, in tens of thousands of transactions~\ref{heuristic:founder}.}
\label{fig:founders_correlation}
\end{figure}

The value alone of these withdrawals thus provides some correlation with the
deposits, but to further explore it we also looked at the timing of
the transactions.  When we examined the intervals between consecutive
deposits of 249.9999~ZEC, we found that 85\% happened within 
6-10 blocks of the previous one.  Similarly, when examining the intervals between
consecutive withdrawals of 250.0001~ZEC, we found that 1,943 of the 1,953
withdrawals also had a proximity of 6-10 blocks.  Indeed, both the deposits
and the withdrawals 
proceeded in step-like patterns, in which many transactions were made within 
a very small number of blocks (resulting in the step up), at which point 
there would be a pause while more block rewards were accumulated (the step
across).  This pattern is visible in Figure~\ref{fig:founders_correlation},
which shows the deposit and withdrawal transactions associated with
the founders.  Deposits are typically made in
few large steps, whereas withdrawals take many smaller ones. 

\begin{heuristic}\label{heuristic:founder}
Any \ztot transaction carrying 250.0001 ZEC in value is done by the founders.
\end{heuristic}

In terms of false positives, we cannot truly know how risky this
heuristic is, short of asking the founders.  This is in contrast to the
t-address clustering heuristics presented in Section~\ref{sec:T-to-T}, in
which we were not attempting to assign addresses to a specific owner, so 
could validate the heuristics in other ways.  Nevertheless, the high
correlation between both the value and timing of the transactions led us to
believe in the reliability of this heuristic.

As a result of running this heuristic, we added 75 more addresses to our
initial list of 48 founder addresses (of which, again, only 14 had been used).
Aside from the correlation showed in Figure~\ref{fig:founders_correlation}, 
the difference in terms of our ability to tag founder withdrawals is seen 
in Figure~\ref{fig:zt-founders}.

\subsection{Miners}\label{sec:miners}

The Zcash protocol specifies that all newly generated coins are required to be
put into the shielded pool before they can be spent further.  As a result, we 
expect that a large quantity of the ZEC being deposited into the pool are from
addresses associated with miners.  

\subsubsection{Deposits}

\begin{figure}[t]
\centering
\includegraphics[width=110mm]{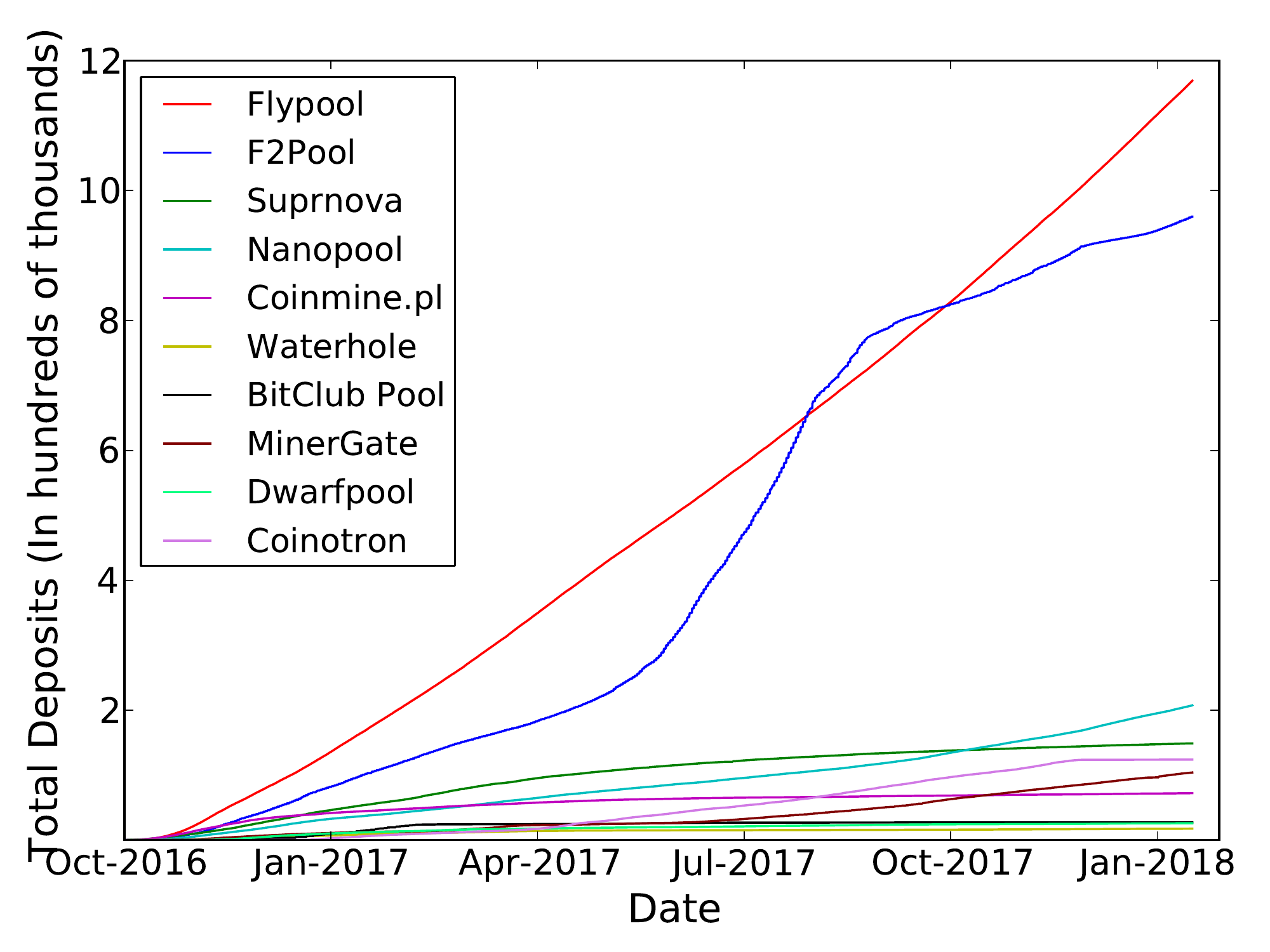}
\caption[Value of deposits made by known mining pools into the 
shielded pool]{Over time, the value of deposits made by known mining pools into the 
shielded pool, in hundreds of thousands of transactions.}
\label{fig:pools}
\end{figure}

As discussed earlier and seen in Figure~\ref{fig:t-to-z}, it is easy to identify 
miner deposits into the pool
due to the fact that they immediately follow a coin generation.  
Before going further, we split the category of miners into individual
miners, who operate on their own, and mining pools, which represent
collectives of potentially many individuals.  In total, we gathered 19 
t-addresses associated with Zcash mining pools, using the scraping methods
described in Section~\ref{sec:tagging}.  Table~\ref{tab:miners} lists these 
mining pools, as well as the number of addresses they control and the number 
of \ttoz transactions we associated with them.  Figure~\ref{fig:pools} plots 
the value of their deposits into the shielded pool over time.

\begin{table}
\centering
\begin{tabular}{lS[table-format=1.0]S[table-format=5.0]S[table-format=4.0]} 
\toprule
Name & {Addresses} & {\ttoz} & {\ztot} \\
\midrule
Flypool & 3 & 65631 & 3 \\ 
F2Pool & 1 & 742 & 720 \\
Nanopool & 2 & 8319 & 4107 \\
Suprnova & 1 & 13361 & 0 \\
Coinmine.pl & 2 & 3211 & 0 \\
Waterhole & 1 & 1439 & 5 \\
BitClub Pool & 1 & 196 & 1516 \\
MiningPoolHub & 1 & 2625 & 0 \\
Dwarfpool & 1 & 2416 & 1 \\
Slushpool & 1 & 941 & 0 \\
Coinotron & 2 & 9726 & 0 \\
Nicehash & 1 & 216 & 0 \\
MinerGate & 1 & 13 & 0 \\
Zecmine.pro & 1 & 6 & 0 \\ 
\bottomrule
\end{tabular}
\caption[Summary of identified Zcash mining pool activity]{A summary of our identified mining pool activity, in terms of the
number of associated addresses used in coingen transactions, and the numbers 
of each type of transaction interacting with the pool.}
\label{tab:miners}
\end{table}

In this figure, we can clearly see that the two dominant mining pools are
Flypool and F2Pool.  Flypool consistently
deposits the same (or similar) amounts, which we can see in their linear
representation.  F2Pool, on the other hand, has bursts of large deposits mixed
with periods during which it is not very active, which we can also see
reflected in the graph.  Despite their different behaviors, the amount
deposited between the two pools is similar.

\subsubsection{Withdrawals}

While the withdrawals from the pool do not solely re-use the small number
of mining addresses identified using deposits (as we saw in our na{\"i}ve 
attempt to link miner \ztot transactions in Figure~\ref{fig:zt-none}), they 
do typically re-use some of them, so can frequently be identified anyway.

In particular, mining pool payouts in Zcash are similar to how many of them 
are in Bitcoin~\cite{Meiklejohn2013a,SP:Eyal15}.  The block reward is 
often paid into a single address, controlled by
the operator of the pool, and the pool operator then deposits some set of 
aggregated block rewards into the shielded pool.  They then pay the individual 
reward to each of the individual miners as a way of ``sharing the pie,'' 
which results in \ztot transactions with many outputs. (In Bitcoin, some
pools opt for this approach while some form a ``peeling chain'' in which they
pay each individual miner in a separate transaction, sending the change back
to themselves each time.) In the payouts
for some of the mining pools, the list of output t-addresses sometimes 
includes one of the t-addresses known to be associated with the mining pool 
already.  We thus tag these types of payouts as belonging to the mining pool,
according to the following heuristic:

\begin{heuristic}\label{heuristic:miners}
If a \ztot transaction has over 100 output t-addresses, one of which belongs
to a known mining pool, then we label the transaction as a mining withdrawal
(associated with that pool), and label all non-pool output t-addresses as 
belonging to miners.
\end{heuristic}

As with Heuristic~\ref{heuristic:founder}, short of asking the mining pool
operators directly it is impossible to validate this heuristic.  Nevertheless,
given the known operating structure of Bitcoin mining pools and the way this
closely mirrors that structure, we again believe it to be relatively safe.

As a result of running this heuristic, we tagged 110,918 addresses as
belonging to miners, and linked a much more significant portion of the \ztot
transactions, as seen in Figure~\ref{fig:zt-both}.  As the last column in
Table~\ref{tab:miners} shows, however, this heuristic captured the activity 
of only a small number of the mining pools, and the large jump in linked
activity is mostly due to the high coverage with F2Pool (one of the two
richest pools).  This implies that further heuristics developed specifically
for other pools, such as Flypool, would increase the linkability even more.
Furthermore, a more active strategy in which we mined with the pools to
receive payouts would reveal their structure, at which point
(according to the 1.1M deposited by Flypool shown in Figure~\ref{fig:pools}
and the remaining value of 1.2M attributed to the `other' category shown in
Figure~\ref{fig:zt-both}) we would shrink the anonymity set even
further.\footnote{It is possible that we have already captured some of the
    Flypool activity, as many of the miners receive payouts from multiple pools.
    We thus are not claiming that all remaining activity could be attributed to
    Flypool, but potentially some substantial portion.}

\subsection{Other Entities}\label{sec:others}

Once the miners and founders have been identified, we can assume the remaining
transactions belong to more general entities.  In this section we look into 
different means of categorizing these entities in order to identify how the 
shielded pool is being used.

In particular, we ran the heuristic due to Quesnelle~\cite{zcash-anon}, which
said that if a unique value (i.e., a value never seen in the blockchain before
or since) is deposited into the pool and then, after some short
period of time, the exact same value is withdrawn from the pool, the deposit
and the withdrawal are linked in what he calls a 
\emph{round-trip transaction}.

\begin{heuristic}
{\textbf{\cite{zcash-anon}}}
\label{heuristic:bonus}
For a value $v$, if there exists exactly one \ttoz transaction carrying value 
$v$ and one \ztot transaction carrying value $v$, where the 
\ztot transaction happened after the \ttoz one and within some small number of
blocks, then these transactions are linked.
\end{heuristic}

In terms of false positives, the fact that the value is unique in the
blockchain means that the only possibility of a false positive is if some of
the \ztoz transactions split or aggregated coins in such a way that another 
deposit (or several other deposits) of a different amount were altered within 
the pool to yield an amount identical to the initial deposit.  While this is
possible in theory, we observe that of the 12,841 unique values we identified,
9,487 of them had eight decimal places (the maximum number 
in Zcash), and 98.9\% of them had more than three decimal
places.  We thus view it as highly unlikely that these exact values were
achieved via manipulations in \ztoz transactions.

\begin{figure}[t]
\centering
\includegraphics[width=110mm]{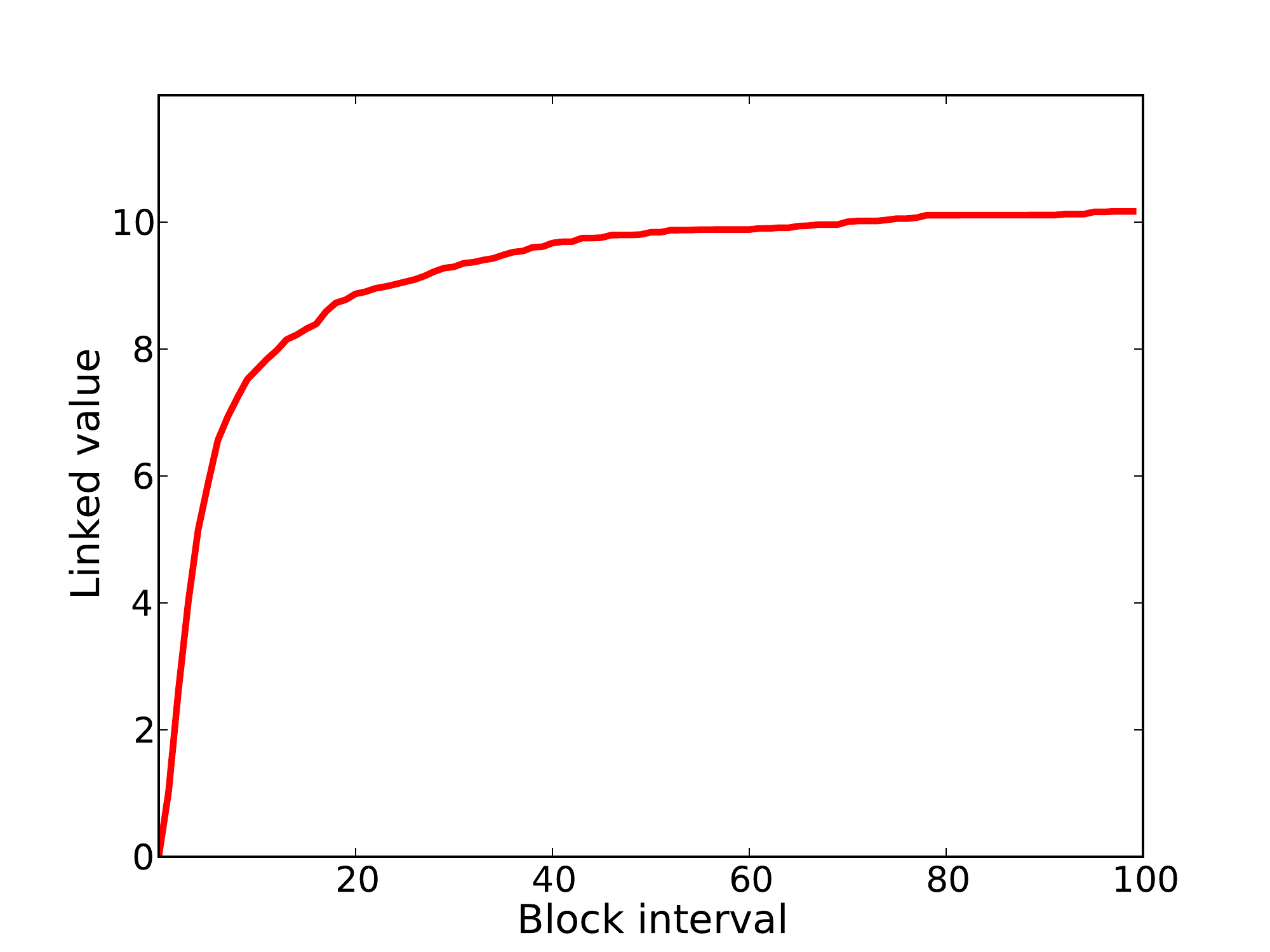}
\caption[Value linked by Heuristic~\ref{heuristic:bonus}]{The value linked by Heuristic~\ref{heuristic:bonus}, as a function of
the block interval required between the deposit and withdrawal transactions.}
\label{fig:bonus}
\end{figure}

By running this heuristic, we identified 12,841 unique values, which means 
we linked 12,841 transactions.  The values total 
1,094,513.23684~ZEC and represent 28.5\% of all coins 
ever deposited in the pool.  Interestingly, most (87\%) of the linked coins 
were in transactions attributed to the founders and miners, so had already 
been linked by our previous heuristics.  We believe this lends further 
credence to their soundness.  In terms of the block interval, we 
ran Heuristic~\ref{heuristic:bonus} for every interval between $1$ and $100$ 
blocks; the results are in Figure~\ref{fig:bonus}.

As this figure shows, even if we assume a conservative block interval of
$10$ (meaning the withdrawal took place 25 minutes after the deposit), we
still capture 70\% of the total value, or over 700K~ZEC.  If we require the
withdrawal to have taken place within an hour of the deposit, we get 83\%.

\section{Interactions within the Shielded Pool}\label{sec:z-to-z}

In this section we consider private transactions; i.e., \ztoz transactions
that interact solely with the shielded pool.  As seen in
Section~\ref{sec:tx-usage}, these transactions form a small percentage of the
overall transactions.  However, \ztoz transactions form a crucial part of the
anonymity core of Zcash.  In particular, they make it difficult to identify the
round-trip transactions from Heuristic~\ref{heuristic:bonus}.  

Our analysis identified 6,934 \ztoz transactions, with 8,444 \vjoinsplits.  As 
discussed in Section~\ref{sec:back-zcash}, the only information revealed by
\ztoz transactions is the miner's fee, the time of the transaction, and the 
number of \vjoinsplits used as input.  Of these, we looked at the time of
transactions and the number of \vjoinsplits in order to gain some insight as
to the use of these operations.

We found that 93\% of \ztoz transactions took just one \vjoinsplit as input.  
Since each \vjoinsplit can have at most two shielded outputs as its input, 
the majority of \ztoz transactions thus take no more than two shielded outputs 
as their input.  This increases the difficulty of categorizing \ztoz 
transactions, because we cannot know if a small number of users are making 
many transactions, or many users are making one transaction.

\begin{figure}[t]
\centering
\includegraphics[width=110mm]{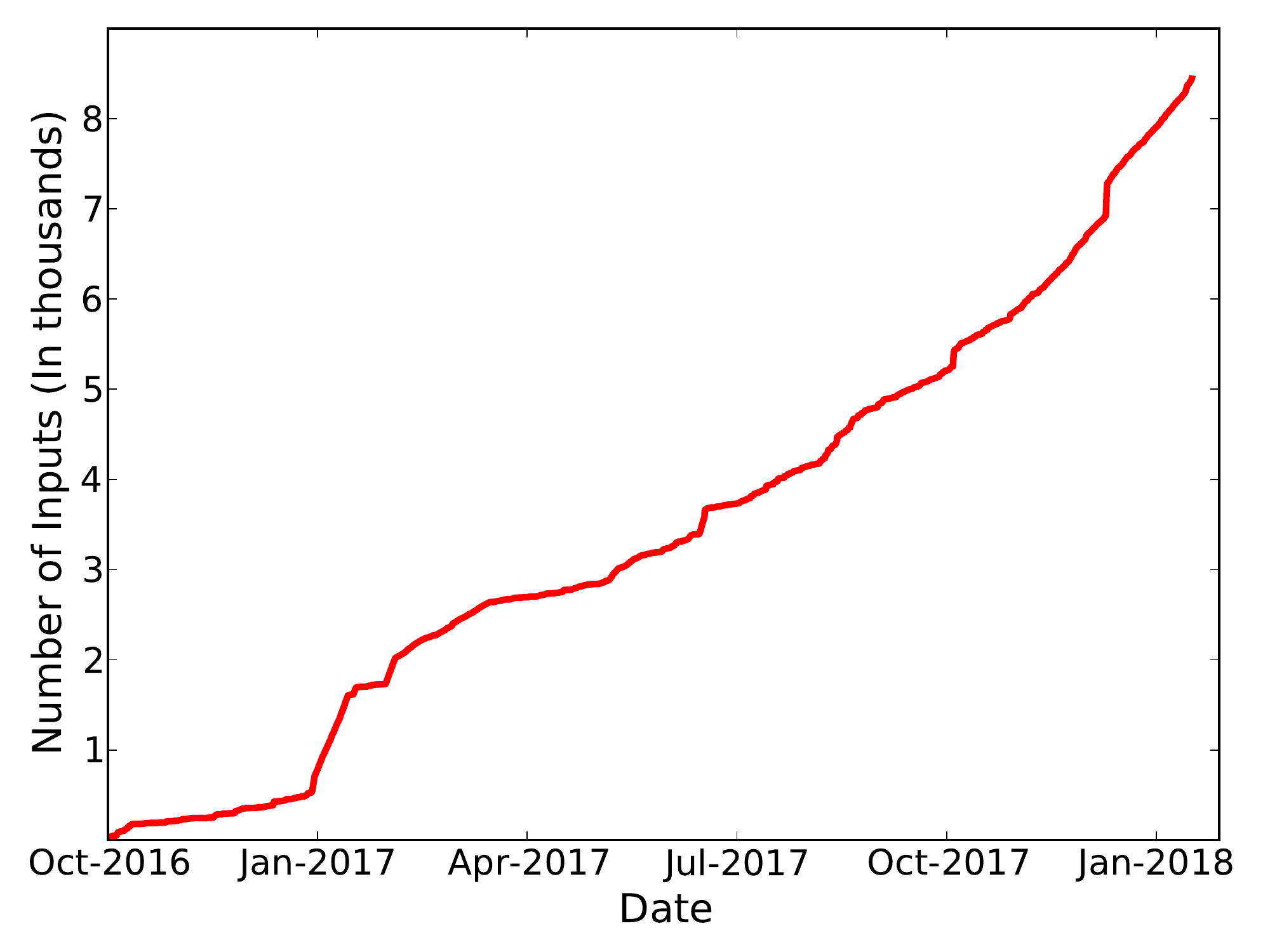}
\caption[Count of \ztoz \vjoinsplits over time]{The number of \ztoz \vjoinsplit transactions over time.}
\label{fig:zz2}
\end{figure}

In looking at the timing of \ztoz transactions, however, we conclude that it 
is likely that a small number of users were making many transactions.
Figure~\ref{fig:zz2} plots the cumulative number of \vjoinsplits over time.
The occurrences of \vjoinsplits are somewhat irregular, with 17\%
of all \vjoinsplits occurring in January 2017.
There are four other occasions when a sufficient number of \vjoinsplits occur
within a sufficiently short period of time as to be visibly noticeable.  It 
seems likely that these occurrences belong to the same group of users, or at
least by users interacting with the same service.

Finally, looking back at the number of \ttoz and \ztot transactions identified
with mining pools in Table~\ref{tab:miners}, it is possible that BitClub Pool
is responsible for up to 1,300 of the \ztoz transactions, as it had 196
deposits into the pool and 1,516 withdrawals.  This can happen only because
either (1) the pool made extra \ztoz transactions, or (2) it sent change from 
its \ztot
transactions back into the shielded pool.  As most of BitClub Pool's \ztot
transactions had over 200 output t-addresses, however, we conclude that the
former explanation is more likely.

\section{Case Study: The Shadow Brokers}\label{sec:shadow-brokers}

The Shadow Brokers (TSB) are a hacker collective that has been active since
the summer of 2016, and that leaks tools supposedly created by the
NSA.  Some of these leaks are released as free samples, but many are sold
via auctions and as monthly bundles.  Initially, TSB accepted payment only 
using Bitcoin.  Later, however, they began to
accept Zcash for their monthly dump service.  In this section we discuss
how we identified \ttoz transactions that could represent payments to TSB.
We identified twenty-four clusters (created using our analysis in
Section~\ref{sec:T-to-T}) matching our criteria for potential TSB 
customers, one of which could be a regular customer.

\subsection{Techniques}

In order to identify the transactions that are most likely to be associated
with TSB, we started by looking at their blog~\cite{TSB}.  In May 2017, 
TSB announced that they would be accepting Zcash for their monthly dump 
service.
Throughout the summer (June through August) 
they accepted both Zcash and Monero, but in September they announced that they 
would accept only Zcash.
Table~\ref{table:TSB-blog-amounts} summarizes the amount they were requesting
in each of these months.  The last blog post was made in October 2017, when 
they stated that all subsequent dumps would cost 500~ZEC.

\begin{table}
\centering
\small
\begin{tabular}{ccccc}
\toprule
May/June & July & August & September & October \\
\midrule
\begin{tabular}[t]{l}
100 
\end{tabular}
&
\begin{tabular}[t]{l}
200 \\
400 \\
\end{tabular}
&
\begin{tabular}[t]{l}
500 \\ 
\end{tabular}
&
\begin{tabular}[t]{l}
100 \\
200 \\
500  
\end{tabular}
&
\begin{tabular}[t]{l}
500 \\
\end{tabular}
\\
\bottomrule
\end{tabular}
\caption[Zcash amounts charged for TheShadowBrokers monthly dumps]{Amounts charged for TSB monthly dumps, in ZEC. In July and September 
TSB offered different prices depending on which exploits were being purchased.}
\label{table:TSB-blog-amounts}
\end{table}

To identify potential TSB transactions, we thus looked at all \ttoz 
transactions not associated with miners or founders that deposited either 
100, 200, 400, or 500 ZEC $\pm~5$ ZEC.  
Our assumption was that users paying TSB were not likely to be regular Zcash
users, but rather were using it with the main purpose of making the payment.  
On this basis, addresses making \ttoz transactions of the above values were 
flagged as a potential TSB customer if the following conditions held:

\begin{enumerate}

\item They did not get their funds from the pool; i.e., there were no \ztot
transactions with this address as an output.  Again, if this were a user 
mainly engaging with Zcash as a way to pay TSB, they would need to to buy
their funds from an exchange, which engage only with t-addresses.

\item They were not a frequent user, in the sense that they had not made 
or received more than 250 transactions (ever).

\item In the larger cluster in which this address belonged, the total amount
deposited by the entire cluster into the pool within one month was 
within 1~ZEC of the amounts
requested by TSB.  Here, because the resulting clusters were small enough to
treat manually, we applied not only Heuristic~\ref{heuristic:inputclusters}
but also Heuristic~\ref{heuristic:others} (clustering by change), making sure
to weed out false positives.   Again, the idea was that suspected TSB customers
would not be frequent users of the pool.

\end{enumerate}

As with our previous heuristics, there is no way to quantify the
false-positive risks associated with this set of criteria, although we see
below that many of the transactions matching it did occur in the time period
associated with TSB acceptance of Zcash.  Regardless, given this limitation we
are not claiming that our results are definitive, but do believe this 
to be a realistic set of criteria that might be applied in the
context of a law enforcement investigation attempting to narrow down potential
suspects.  %

\subsection{Results}

Our results, in terms of the number of transactions matching our requirements
above up until 17 January 2018, are summarized in 
Table~\ref{table:number-of-suspicious-transactions-per-month}.  Before the
first TSB blog post in May, we found only a single matching transaction.  
This is very likely a false positive, but demonstrates that the types of 
transactions we were seeking were not common
before TSB went live with Zcash.  After the blog post, we flagged five clusters
in May and June for the requested amount of 100~ZEC.    There were only two 
clusters that was flagged for 500~ZEC, one of which was from August. 
No transactions of any of the required quantities were flagged in September,
despite the fact that TSB switched to accepting only Zcash in September.  This 
is possible for a number of reasons: our criteria may have caused us to miss 
transactions, or maybe there were no takers.  From October onwards we flagged 
between 1-6 transactions per month.  It is hard to know if these represent users 
paying for old data dumps or are simply false positives.  

\begin{table}
\centering
\begin{tabular}{lcccc}
\toprule
Month & 100 & 200 & 400 & 500 \\
\midrule
October (2016)   & 0   &  0  &  0  &  0  \\
November         & 0   &  0  &  0  &  0  \\
December         & 0   &  0  &  0  &  0  \\
January (2017)   & 1   &  0  &  0  &  0  \\
February         & 0   &  0  &  0  &  0  \\
March            & 0   &  0  &  0  &  0  \\
April            & 0   &  0  &  0  &  0  \\
May (before) & 0   &  0  &  0  &  0  \\
May (after)  & 3   &  1  &  0  &  0  \\
June       & 2   &  1  &  1  &  0  \\
July       & 1   &  2  &  0  &  0  \\
August     & 1   &  0  &  0  &  1  \\
September  & 0   &  0  &  0  &  0  \\
October    & 2   &  0  &  0  &  0  \\
November   & 1   &  0  &  0  &  0  \\
December   & 2   &  3  &  0  &  1  \\
January (2018)   & 0   &  1  &  0  &  0 \\
\bottomrule
\end{tabular}
\caption[Number of clusters that put the required Shadow Broker amounts into the pool]{Number of clusters that put the required amounts ($\pm 1$~ZEC) 
into the shielded pool.}
\label{table:number-of-suspicious-transactions-per-month}
\end{table}

Four out of the 24 transactions in 
Table~\ref{table:number-of-suspicious-transactions-per-month} are highly 
likely to be false positives.
First, there is the deposit of 100~ZEC into the pool in January, before
TSB announced their first blog post.  This cluster put an additional 252~ZEC
into the pool in March, so is likely just some user of the pool.
Second and third, there are two deposits of 200~ZEC into the pool in 
June, before TSB announced that one of the July dump prices would cost 
200~ZEC.  %
Finally, there is a deposit of 400~ZEC into the pool in June before TSB 
announced that one of the July dump prices would cost 400~ZEC.

Of the remaining clusters, there is one whose activity is worth discussing.
From this cluster, there was one deposit into the pool in
June for 100~ZEC, one in July for 200~ZEC, and one in August for 500~ZEC,
matching TSB prices exactly.  The cluster belonged to a new user, and most of
the money in this user's cluster came directly from Bitfinex (Cluster~3).

\section{Discussion and Future work}  

Our heuristics would have been significantly less effective if the founders
interacting with the pool behaved in a less regular fashion.  In particular,
by always withdrawing the same amount in the same time interval, it became
possible to distinguish founders withdrawing funds from other users.  Given
that the founders are both highly invested in the currency and knowledgeable
about how to use it in a secure fashion, they are in the best place to ensure
the anonymity set is large.

Ultimately, the only way for Zcash to truly ensure the
size of its anonymity set is to require all transactions to take place within
the shielded pool, or otherwise significantly expand the usage of it.  This 
may soon be computationally feasible given emerging advances in the 
underlying cryptographic techniques~\cite{jubjub}, or even if more mainstream
wallet providers like Jaxx roll out support for z-addresses.  More broadly, 
we view it as an interesting 
regulatory question whether or not mainstream exchanges would continue to 
transact with Zcash if it switched to supporting only z-addresses.

Our study was an initial exploration, and thus left many avenues open for
further exploration.  For example, it may be possible to classify more \ztoz 
transactions by analyzing the time intervals between the transactions in more 
detail, or by examining other metadata such as the miner's fee or even the
size (in bytes) of the transaction.  Additionally, the
behavior of mining pools could be further identified by a study that actively
interacts with them.

\section{Conclusions}\label{sec:zec:conclusions}

We present the first in-depth exploration of Zcash, with a
particular focus on its anonymity guarantees.  To achieve this, we applied
both well-known clustering heuristics that have been developed for Bitcoin
and attribution heuristics we developed ourselves that take into account 
Zcash's shielded pool and its unique cast of characters.  As with previous 
empirical analyses of other cryptocurrencies, our study has shown that most 
users are not taking advantage of the main privacy feature of Zcash at all.
Furthermore, the participants who do engage with the shielded pool do so in a
way that is identifiable, which has the effect of significantly eroding the 
anonymity of other users by shrinking the overall anonymity set.

\chapter{Tracing Transactions Across Cryptocurrency Ledgers}
\label{chapterlabelTracing}

\newcommand{\curin}{\mathsf{curIn}}
\newcommand{\curout}{\mathsf{curOut}}
\newcommand{\rate}{\mathsf{rate}}
\newcommand{\userdest}{\mathsf{addr_u}}
\newcommand{\servdest}{\mathsf{addr_s}}
\newcommand{\amt}{\mathsf{amt}}
\newcommand{\tx}{\mathsf{tx}}
\newcommand{\minerfee}{\mathsf{fee}}
\newcommand{\blockbefore}{\delta_b}
\newcommand{\blockafter}{\delta_a}
\newcommand{\status}{\mathsf{status}}
\newcommand{\address}{\mathsf{address}}
\newcommand{\withdraw}{\mathsf{withdraw}}
\newcommand{\incomingCoin}{\mathsf{inCoin}}
\newcommand{\incomingType}{\mathsf{inType}}
\newcommand{\outgoingCoin}{\mathsf{outCoin}}
\newcommand{\outgoingType}{\mathsf{outType}}
\newcommand{\transaction}{\mathsf{tx}}
\newcommand{\transactionURL}{\mathsf{txURL}}
\newcommand{\error}{\mathsf{error}}

\section{Overview}

For the past decade, cryptocurrencies such as Bitcoin have been touted for
their transformative potential, both as a new form of electronic cash and as 
a platform to ``re-decentralize'' aspects of the Internet and computing in
general.  
Traditionally, criminals attempting to cash out illicit funds would have
to use exchanges;
indeed, most tracking techniques rely on identifying the addresses associated
with these exchanges as a way to observe when these deposits
happen~\cite{sarah-fistfulofbitcoins}.  Nowadays, however, exchanges 
typically implement strict Know Your Customer/Anti-Money Laundering 
(KYC/AML) policies to comply with regulatory requirements, meaning 
criminals (and indeed all users) risk revealing their real identities when 
using them.  Users also run risks when storing their coins in accounts at 
custodial 
exchanges, as exchanges may be hacked or their coins may otherwise become
inaccessible~\cite{mtgoxwired,quadriga}.  As an alternative, there
have emerged in the past few years frictionless trading platforms such 
as ShapeShift\footnote{\url{https://shapeshift.io}} and
Changelly,\footnote{\url{https://changelly.com}} in which users are 
able to trade between cryptocurrencies without having to store their coins
with the platform provider.  Furthermore, while ShapeShift now requires users
to have verified accounts~\cite{ss-id}, this was not the case before October
2018.

Part of the reason for these trading platforms to exist is the sheer rise in
the number of different cryptocurrencies: according to the popular
cryptocurrency data tracker CoinMarketCap there were 36
cryptocurrencies in September 2013, only 7 of which had a stated market
capitalization of over 1 million
USD,\footnote{\url{https://coinmarketcap.com/historical/20130721/}} whereas
in January 2019 there were 2117 cryptocurrencies, of which the top 10 had a market
capitalization of over 100 million USD.  
Given this proliferation of new cryptocurrencies and platforms that make it
easy to transact across them, it becomes important to 
consider not just whether or not flows of coins can be tracked within the
transaction ledger of a given currency, but also if they can be tracked as
coins move across their respective ledgers as well.  This 
is especially important given that there are documented cases of criminals
attempting to use these cross-currency trades to obscure the flow of their 
coins: the WannaCry ransomware operators, for example, 
were observed using ShapeShift to convert their ransomed bitcoins into
Monero~\cite{wannacry}.  More generally, these services have the potential to 
offer an insight into the broader cryptocurrency ecosystem and the thousands
of currencies it now contains.

In this Chapter, we initiate an exploration of the usage of these 
cross-currency trading platforms, and the potential 
they offer in terms of the ability to track flows 
of coins as they move across different transaction ledgers.  Here we rely on 
three distinct sources of data: the cryptocurrency blockchains, 
the data collected via our own interactions with these trading platforms, 
and\dash as we describe in Section~\ref{sec:data}\dash the information offered 
by the platforms themselves via their public APIs. 

We begin in Section~\ref{sec:phases} by identifying the specific on-chain 
transactions associated with an advertised ShapeShift transaction, which we 
are able to do with a relatively high degree of success (identifying both the 
deposit and withdrawal transactions 81.91\% of the time, on average).  We 
then describe in
Section~\ref{sec:tracking} the different transactional patterns that can be 
traced by identifying the relevant on-chain transactions, focusing
specifically on patterns that may be indicative of trading or money
laundering, and on the ability to link addresses across different currency
ledgers.  We then move in
Section~\ref{sec:clusters} to consider both old and new heuristics for
clustering together addresses associated with ShapeShift, with particular
attention paid to our new heuristic concerning the common social relationships
revealed by the usage of ShapeShift.  Finally, we bring all the analysis
together by applying it to several case studies in
Section~\ref{sec:case-studies}.  Again, our particular focus in this last 
section is on the phenomenon of trading and other profit-driven activity, 
and the extent to which usage of the 
ShapeShift platform seems to be motivated by criminal activity or a more 
general desire for anonymity.

\section{Background}\label{sec:back}

\subsection{Digital asset trading platforms}\label{sec:back-shapeshift}

In contrast to a traditional exchange, a digital asset trading 
platform allows users to move between different cryptocurrencies without
needing to set up an account, and thus without needing to following KYC/AML 
regulations. Instead, a user 
approaches the service and selects a supported input currency $\curin$ (i.e., 
the currency from which they would like to move money) and a supported output 
currency $\curout$ (the currency which they would like to obtain).  
A user additionally specifies a destination
address $\userdest$ in the $\curout$ blockchain, which is the address to which 
the output currency will be sent.  The service then presents the user with 
an exchange rate $\rate$ and an address $\servdest$ in the $\curin$ blockchain 
to which to send money.  The user then sends this address the amount $\amt$ in 
$\curin$ they wish to convert, and after some delay the service sends the 
appropriate amount of the output currency to the specified destination 
address.  This means that an interaction with either of these services
results in two transactions: one on the $\curin$ blockchain sending $\amt$ to
$\servdest$, and one on the $\curout$ blockchain sending (roughly)
$\rate\cdot\amt$ to $\userdest$. 

This describes an interaction with an abstracted platform; today, the two 
best-known examples are ShapeShift and Changelly, although Changelly does
require account creation.  Each platform supports
dozens of cryptocurrencies, ranging from better-known ones such as Bitcoin and
Ethereum to lesser-known ones such as FirstBlood and Clams.  Many of the
supported cryptocurrencies actually operate as ERC20 or BTC tokens, meaning
they run as contracts on top of the Ethereum and Bitcoin blockchains, 
respectively, rather than as their own standalone platforms.  
In Section~\ref{sec:data},
we describe in more depth the operations of these concrete platforms and our
own interactions with them.

\section{Data Collection and Statistics}\label{sec:data}

In this section, we describe our data sources, as well as some preliminary
statistics about the collected data.  We begin in
Section~\ref{sec:changelly-data} by describing our own 
interactions with Changelly, a trading platform with a
limited personal API.  We then describe in Section~\ref{sec:shapeshift-data} 
both our own interactions with
ShapeShift, and the data we were able to scrape from their public API, which
provided us with significant insight into their overall set of transactions.
Finally, we describe in Section~\ref{sec:blockchain-data} our collection of
the data backing eight different cryptocurrencies.

\subsection{Changelly}\label{sec:changelly-data}

Changelly offers a simple API\footnote{\url{https://api-docs.changelly.com/}}
that allows registered users to carry out transactions with
the service.  Using this API, we engaged in 22 transactions, using the most
popular ShapeShift currencies (Table~\ref{tab:popular}) to guide our choices 
for $\curin$ and $\curout$. 

While doing these transactions, we observed that they would sometimes take 
up to an hour to complete.  This is because Changelly attempts to minimize 
double-spending risk by requiring users to wait for a set number 
of confirmations (shown to the user at the time of their transaction) in the 
$\curin$ blockchain before executing the transfer on the $\curout$ blockchain.  
We used this observation to
guide our choice of parameters in our identification of on-chain transactions
in Section~\ref{sec:phases}.

\subsection{ShapeShift}\label{sec:shapeshift-data}

ShapeShift's API\footnote{\url{https://info.shapeshift.io/api}} allows 
users to execute their own transactions, of which we did 18 in total.  %
As with Changelly, we were able to gain some valuable insights about the
operation of the platform via these personal interactions.  
Whereas ShapeShift did not disclose the number of confirmations they
waited for on the $\curin$ blockchain, we again observed long delays,
indicating that they were also waiting for a sufficient number.

Beyond these personal interactions, the API provides information on 
the operation of the service as 
a whole. Most notably, it provides three separate pieces of information: (1)
the current trading rate between any pair of cryptocurrencies, (2) 
a list of up to 50 of the most recent transactions that have taken place 
(across all users), and (3) full details of a specific ShapeShift transaction 
given the address $\servdest$ in the $\curin$ blockchain (i.e., the address to
which the user sent their coins).

For the trading rates, ShapeShift provides the following information for all
cryptocurrency pairs $(\curin,\curout)$: the rate, the limit (i.e., the 
maximum that can be exchanged), the minimum that can be exchanged, and the 
miner fee (denominated in $\curout$).  For the 50 most recent transactions, 
information is provided in the form:
$(\curin,\curout,\amt,t,\mathsf{id})$, where the first three of these are as
discussed in Section~\ref{sec:back-shapeshift}, $t$ is a UNIX timestamp, and
$\mathsf{id}$ is an internal identifier for this transaction.  For the
transaction information, when provided with a specific $\servdest$ ShapeShift
provides the tuple $(\status,\address,\withdraw,\incomingCoin,
\incomingType,\outgoingCoin,\allowbreak \outgoingType,\transaction,\transactionURL,\error)$.
The $\status$ field is a flag that is either \verb#complete#, to mean the
transaction was successful; \verb#error#, to mean an issue occurred with the
transaction or the queried address was not a ShapeShift address; or
\verb#no_deposits#, to mean a user initiated a transaction but did not send
any coins.  The $\error$ field appears when an error is returned and gives a
reason for the error.  The $\address$ field is the same address $\servdest$
used by ShapeShift, and $\withdraw$ is the address $\userdest$ (i.e., the
user's recipient address in the $\curout$ blockchain).  $\incomingType$ and
$\outgoingType$ are the respective $\curin$ and $\curout$ currencies and 
$\incomingCoin$ is the $\amt$ received.  $\outgoingCoin$ is the amount sent in
the $\curout$ blockchain.  Finally, $\transaction$ is the transaction hash in
the $\curout$ blockchain and $\transactionURL$ is a link to this transaction 
in an online explorer. 

Using a simple Web scraper, we downloaded the transactions and rates every
five seconds for close to thirteen months: from November 27 
2017 until December 23 2018.  This resulted in a set of 
2,843,238 distinct transactions. Interestingly, we noticed that several 
earlier test transactions we did with the platform did
not show up in their list of recent transactions, which suggests that their
published transactions may in fact underestimate their overall activity.

\subsubsection{ShapeShift currencies}

\begin{figure}
\centering
\includegraphics[width=\linewidth]{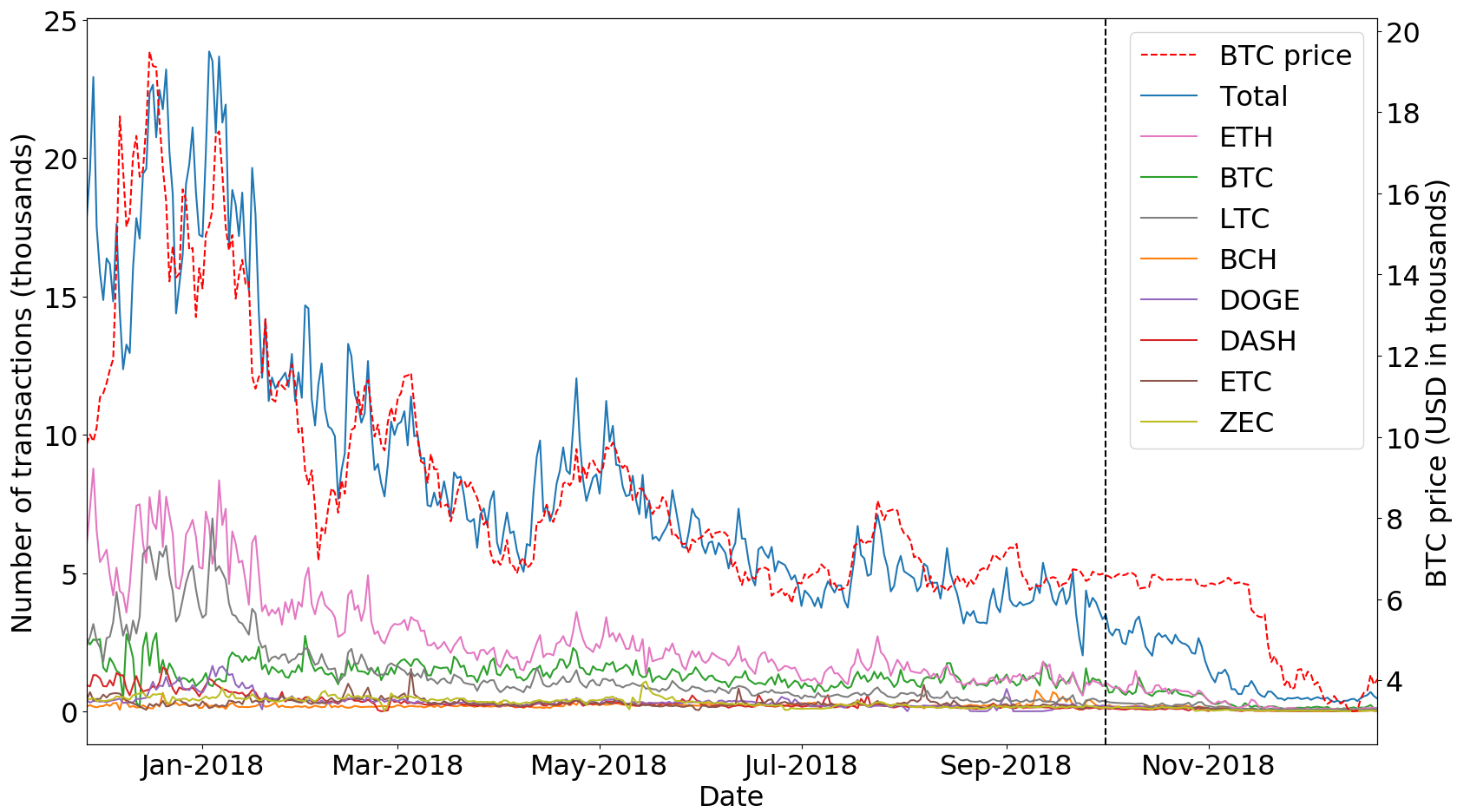}
\caption[Total number of transactions per day reported via ShapeShift's API,
broken down by cryptocurrency]{The total number of transactions per day reported via ShapeShift's API,
and the numbers broken down by cryptocurrency (where a transaction is attributed 
to a coin if it is used as either $\curin$ or $\curout$).  The dotted red line
indicates the BTC-USD exchange rate, and the horizontal dotted black line
indicates when KYC was introduced into ShapeShift.}
\label{fig:ss-time-distro}
\end{figure}

In terms of the different cryptocurrencies used in ShapeShift transactions, 
their popularity was distributed as seen in
Figure~\ref{fig:ss-time-distro}.  As this figure depicts, the overall
activity of ShapeShift is (perhaps unsurprisingly) correlated with the price
of Bitcoin in the same time period.  At the same time, there is a 
decline in the number of transactions after KYC was introduced that is not
clearly correlated with the price of Bitcoin (which is largely steady and
declines only several months later).

ShapeShift supports dozens of cryptocurrencies, and in our data we observed
the use of 65 different ones.   The most commonly
used coins are shown in Table~\ref{tab:popular}.  %
It is clear that Bitcoin and Ethereum are 
the most heavily used currencies, which is perhaps not surprising given the 
relative ease with which they can be exchanged with fiat currencies on more 
traditional exchanges, and their rank in terms of market capitalization.  

\begin{table}[t]
\centering
{\footnotesize
\begin{tabular}{lcS[table-format=7.0]S[table-format=6.0]S[table-format=6.0]}
\toprule
Currency & Abbr. & Total & $\curin$ & $\curout$ \\
\midrule
Ethereum & ETH & 1385509 & 892971 & 492538 \\
Bitcoin  & BTC & 1286772 & 456703 & 830069 \\
Litecoin & LTC & 720047 & 459042 & 261005 \\
Bitcoin Cash & BCH & 284514 & 75774 & 208740 \\
Dogecoin & DOGE & 245255 & 119532 & 125723 \\
Dash & DASH & 187869 & 113272 & 74597 \\
Ethereum Classic & ETC & 179998 & 103177 & 76821 \\
Zcash & ZEC & 154142 & 111041 & 43101 \\
\bottomrule
\end{tabular}
}
\caption[Eight most traded coins used on ShapeShift]{The eight most popular coins used on ShapeShift, in terms of the total 
units traded, and the respective units traded with that coin as $\curin$ and 
$\curout$.}
\label{tab:popular}
\end{table}

\subsection{Blockchain data}\label{sec:blockchain-data}

For the cryptocurrencies we were interested in exploring further, it was also
necessary to download and parse the respective blockchains, in order to 
identify the on-chain transactional behavior of ShapeShift and Changelly.  
It was not feasible to do this for all 65 currencies used on ShapeShift
(not to mention that given the low volume of transactions for many of them, 
it would likely not yield additional insights anyway), so we chose to focus 
instead on just the top 8, as seen in Table~\ref{tab:popular}.  
Together, these 
account for 95.7\% of all ShapeShift transactions if only one of $\curin$ 
and/or $\curout$ is one of the eight, and 60.5\% if both are.  

For each of these currencies, we ran a full node in order to download the
entire blockchain.  For the ones supported by the BlockSci tool~\cite{blocksci} 
(Bitcoin, Dash and Zcash), 
we used it to parse and analyze their blockchains.  BlockSci does not,
however, support the remaining five currencies.  For these we
thus parsed the blockchains using Python scripts, stored the data as Apache 
Spark parquet files, and analyzed them using custom scripts.  In total, we
ended up working with 654~GB of raw blockchain data and 434~GB of 
parsed blockchain data.

\section{Identifying Blockchain Transactions}\label{sec:phases}

In order to gain deeper 
insights about the way these trading platforms are used, it is necessary to 
identify 
not just their internal transactions but also the transactions that appear on
the blockchains of the traded currencies.  This section presents heuristics
for identifying these on-chain transactions, and the next section explores
the additional insights these transactions can offer.

Recall from Section~\ref{sec:back-shapeshift} that an interaction with
ShapeShift
results in the deposit of coins from the user to the service on the $\curin$ 
blockchain (which we refer to as ``Phase~1''), and the withdrawal of coins
from the service to the user on the $\curout$ blockchain
(``Phase~2'').  
To start with Phase~1, we thus seek to identify the deposit transaction on 
the input ($\curin$) blockchain.  Similarly to Portnoff et
al.~\cite{portnoff-kdd}, we consider two main requirements for
identifying the correct on-chain transaction: (1) that it 
occurred reasonably close in time to the point at which it was
advertised via the API, and (2) that the value it carried was identical to the 
advertised amount.

For this first requirement, we look for candidate transactions as follows.  
Given a ShapeShift transaction with timestamp $t$, we first find the block 
$b$ (at some height 
$h$) on the $\curin$ blockchain that was mined at the time closest to $t$.  We 
then look at the transactions in all blocks with height in the range 
$[h-\blockbefore,h+\blockafter]$, where $\blockbefore$ and $\blockafter$ 
are parameters specific to $\curin$.  We looked at both earlier and
later blocks based on the observation in our own interactions that the
timestamp published by ShapeShift would sometimes be earlier and sometimes be
later than the on-chain transaction.

For each of our eight currencies, we ran this heuristic for every ShapeShift 
transaction using $\curin$ as the currency in question, with every possible
combination of $\blockbefore$ and $\blockafter$ ranging from $0$ to $30$.
This resulted in a set of candidate transactions with zero hits (meaning no
matching transactions were found), a single hit, or multiple hits.  To rule out
false positives, we initially considered as successful only ShapeShift 
transactions with a single candidate on-chain transaction, although we
describe below an augmented heuristic that is able to tolerate multiple hits.  
We then used the values of $\blockbefore$ and $\blockafter$ that maximized 
the number of single-hit transactions for each currency.  
As seen in Table~\ref{tab:phase12}, the optimal choice of these parameters 
varies significantly across currencies, according to their different block 
rates; typically we needed to look further before or after for currencies in
which blocks were produced more frequently.  

In order to validate the results
of our heuristic for Phase~1, we use the additional capability of the ShapeShift 
API described in Section~\ref{sec:shapeshift-data}. In particular, we queried the 
API on the recipient address of every transaction identified by our heuristic 
for Phase~1.  If the response of the API was affirmative, we flagged the recipient 
address as belonging to ShapeShift and we identified the transaction in which it
received coins as the $\curin$ transaction.  This also provided a way to
identify the corresponding Phase~2 transaction on the $\curout$ blockchain,
as it is just the $\transaction$ field returned by the API.  As we
proceed only in the case that the API returns a valid result, we gain
ground-truth data in both Phase~1 and Phase~2.  In other words, this method
serves to not only validate our results in Phase~1 but also provides a way to
identify Phase~2 transactions.

The heuristic described above is able to handle only single hits; i.e., the
case in which there is only a single candidate transaction.  Luckily, it is
easy to augment this heuristic by again using the API.  For example, assume we 
examine
a BTC-ETH ShapeShift transaction and we find three candidate transactions in
the Bitcoin blockchain after applying the basic heuristic described above.  To 
identify which of
these transactions is the right one, we simply query the API on all three
recipient addresses and check that the $\status$ field is affirmative (meaning
ShapeShift recognizes this address) and that the $\outgoingType$ field is ETH.
In the vast majority of cases this uniquely identifies the correct transaction
out of the candidate set, meaning we can use the API to both validate our
results (i.e., we use it to eliminate potential false positives, as described
above) and to augment the heuristic by being able to tolerate multiple
candidate transactions.  The augmented results for Phase~1 can be found in the
last column of Table~\ref{tab:phase12} and clearly demonstrate the benefit
of this extra usage of the API.  In the most dramatic example, we were able to
go from identifying the on-chain transactions for ShapeShift transactions
involving Bitcoin 65.75\% of the time with the basic heuristic to
identifying them 76.86\% of the time with the augmented heuristic.

\begin{table}[t]
\centering
\begin{tabular}{lccS[table-format=2.2]S[table-format=2.2]}
\toprule
Currency & \multicolumn{2}{c}{Parameters} & {Basic \%} & {Augmented \%} \\
\cmidrule(lr){2-3}
& {$\blockbefore$} & {$\blockafter$} && \\
\midrule
BTC & 0 & 1 & 65.76 & 76.86 \\
BCH & 9 & 4 & 76.96 & 80.23 \\
DASH & 5 & 5 & 84.77 & 88.65  \\
DOGE & 1 & 4 & 76.94 & 81.69  \\
ETH & 5 & 0 & 72.15 & 81.63 \\
ETC & 5 & 0 & 76.61 & 78.67  \\
LTC & 1 & 2 & 71.61 & 76.97  \\
ZEC & 1 & 3 & 86.94 & 90.54 \\
\bottomrule
\end{tabular}
\caption[Percentage of ShapeShift transactions found matching on-chain transactions for both the basic and augmented heuristics]{For the selected (optimal) parameters and for a given currency used
as $\curin$, the percentage of ShapeShift transactions for which we found 
matching on-chain transactions for both the basic (time- and value-based) 
and the augmented (API-based) Phase~1
heuristic.  The augmented heuristic uses the API and thus also represents 
our success in identifying Phase~2 transactions.}
\label{tab:phase12}
\end{table}

\subsection{Accuracy of our heuristics}

False negatives can occur for both of our heuristics when there are either too 
many or too few matching transactions in the searched block interval.  These 
are more common for the basic heuristic, as described above and seen in 
Table~\ref{tab:phase12}, because it is conservative in identifying an
on-chain transaction only when there is one candidate.  This rate could be
improved by increasing the searched block radius, at the expense of adding
more computation and potentially increasing the false positive rate.

False positives can occur for both of our heuristics if someone sends the same 
amount as the ShapeShift transaction at roughly the same time, but this 
transaction falls within our searched interval whereas the ShapeShift one 
doesn't.  In theory, this
should not be an issue for our augmented heuristic, since the API
will make it clear that the candidate transaction is not in fact associated
with ShapeShift.  In a small number of cases (fewer than 1\% of all ShapeShift 
transactions), however, the API returned details of a transaction with different
characteristics than the one we were attempting to identify; e.g., it had a
different pair of currencies or a different value being sent.  This happened 
because ShapeShift allows users to re-use an existing deposit address, and the 
API returns only the latest transaction using a given address. 

If we blindly took the results of the API, then this would lead to false 
positives in our augmented heuristic for both Phase~1 and Phase~2.  We thus 
ensured that the transaction returned by the
API had three things in common with the ShapeShift transaction:
(1) the pair of currencies, (2) the amount being sent, and (3) the timing,
within the interval specified in Table~\ref{tab:phase12}.  If there was any 
mismatch, we discarded the transaction.  For example, given a ShapeShift
transaction indicating an ETH-BTC shift carrying 1~ETH and occurring at 
time $t$, we looked for all addresses that received 1~ETH at time $t$ or up to 
$5$ blocks earlier.  We then queried the API on these addresses and kept only
those transactions which reported shifting 1~ETH to BTC.  
While our augmented heuristic still might produce false positives in the
case that a user quickly makes two different transactions using the same 
currency pair, value, and deposit address, we view this as unlikely, 
especially given the relatively long wait times we observed ourselves when
using the service (as mentioned in Section~\ref{sec:shapeshift-data}).

\subsection{Alternative Phase~2 identification}

Given that our heuristic for Phase~2 involved just querying the API for the
corresponding Phase~1 transaction, it is natural to wonder what would be
possible without this feature of the API, or indeed if there are any
alternative strategies for identifying Phase~2 transactions.  Indeed, it is
possible to use a similar heuristic for identifying Phase~1 transactions, by
first looking for transactions in blocks that were mined close to the 
advertised transaction time, and then looking for ones in which the amount was
close to the expected amount.  Here the amount must be estimated according to
the advertised $\amt$, $\rate$, and $\minerfee$.  In theory, the amount sent 
should be $\amt\cdot\rate - \minerfee$, although in practice the rate can
fluctuate so it is important to look for transactions carrying a total value 
within a reasonable error rate of this amount.

When we implemented and applied this heuristic, we found that our accuracy 
in identifying Phase~2 transactions decreased
significantly, due to the larger set of transactions that carried an amount
within a wider range (as opposed to an exact amount, as in Phase~1) and the
inability of this type of heuristic to handle multiple candidate transactions.
More importantly, this approach provides no ground-truth information at all:
by choosing conservative parameters it is possible to limit the number of
false positives, but this is at the expense of the false negative rate (as,
again, we observed in our own application of this heuristic) and in general
it is not guaranteed that the final set of transactions really are associated
with ShapeShift.  As this is the exact guarantee we can get by using the API,
we continue in the rest of the paper with the results we obtained there, but
nevertheless mention this alternative approach in case this feature of the API
is discontinued or otherwise made unavailable.

\section{Tracking Cross-Currency Activity}\label{sec:tracking}

In the previous section, we saw that it was possible in many cases to identify
the on-chain transactions, in both the $\curin$ and $\curout$ blockchains, 
associated with the transactions advertised by
ShapeShift.  In this section, we take this a step further and show how
linking these transactions can be used to identify more complex patterns 
of behavior.

\begin{figure*}[t]
\centering
\begin{subfigure}[b]{0.34\textwidth}
\centering
\includegraphics[width=\linewidth]{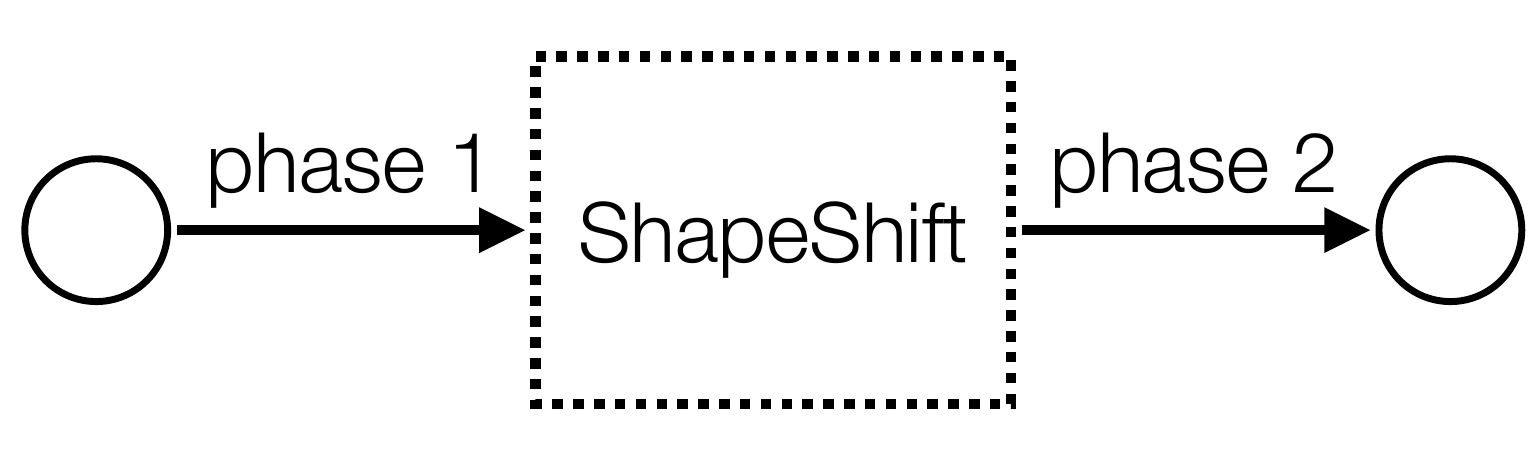}
\caption{Pass-through}
\label{fig:passthrough}
\end{subfigure}
~
\begin{subfigure}[b]{0.24\textwidth}
\centering
\includegraphics[width=\linewidth]{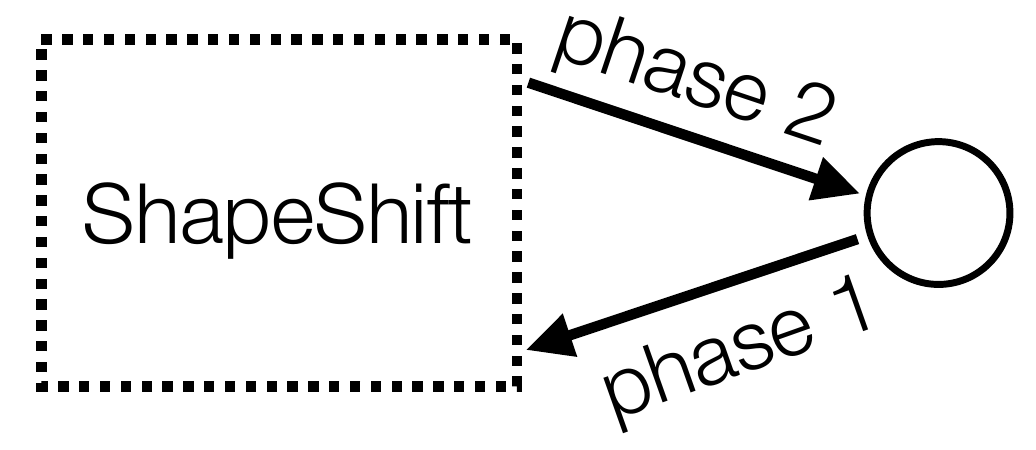}
\caption{U-turn}
\label{fig:uturn}
\end{subfigure}
~
\begin{subfigure}[b]{0.34\textwidth}
\centering
\includegraphics[width=\linewidth]{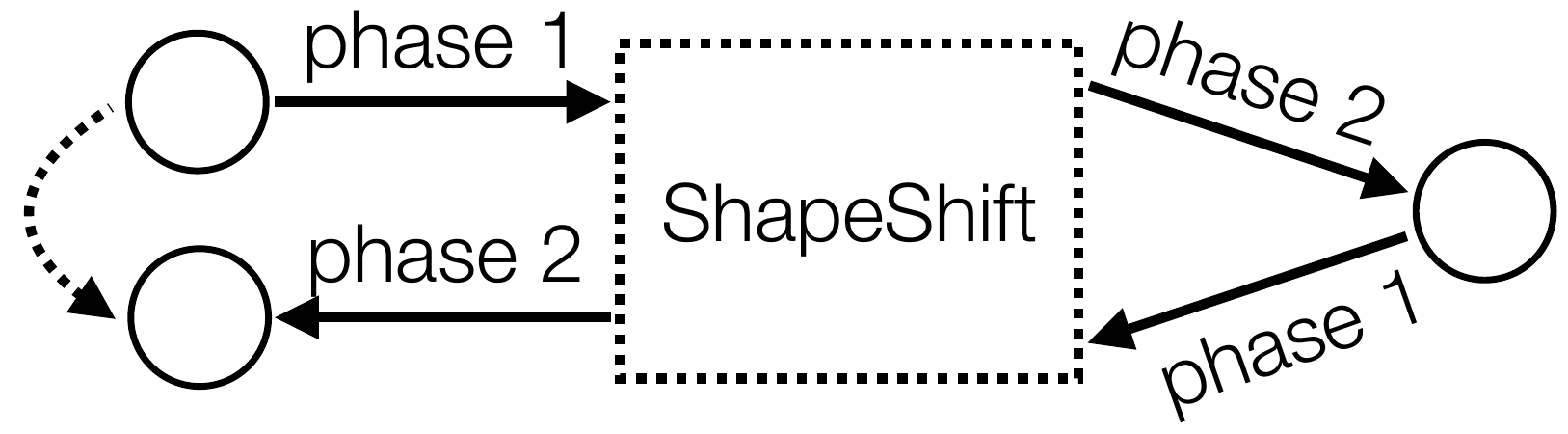}
\caption{Round-trip}
\label{fig:roundtrip}
\end{subfigure}
\caption[Transactional patterns analysed in ShapeShift]{The different transactional patterns, according to how
they interact with ShapeShift and which phases are required to identify them.}
\label{fig:tx-types}
\end{figure*}

As shown in Figure~\ref{fig:tx-types}, we consider these for three main types
of transactions.  In particular, we look at (1) \emph{pass-through}
transactions, which represent the full flow of money as it moves from one
currency to the other via the deposit and withdrawal transactions; 
(2) \emph{U-turns}, in which a user who has shifted into one currency 
immediately shifts back; and (3) \emph{round-trip} transactions, which are
essentially a combination of the first two and follow a user's flow of money
as it moves from one currency to another and then back to the original one.
Our interest in these particular patterns of behavior is largely based on the 
role they play in tracking money as it moves across the ledgers of different 
cryptocurrencies.  In particular, our goal is to test the validity of the
implicit assumption made by criminal usage of the platform\dash such as we
examine further in Section~\ref{sec:case-studies}\dash that ShapeShift 
provides additional anonymity beyond simply transacting in a given currency.

In more detail, identifying pass-through transactions allows us to create a 
link between the input address(es) in the deposit on the $\curin$ 
blockchain and the output address(es) in the withdrawal on the 
$\curout$ blockchain.

Identifying U-turns allows us to see when a user has interacted with
ShapeShift not because they are interested in holding units of the $\curout$
cryptocurrency, but because they see other benefits in shifting coins back and
forth.  There are several possible motivations for this: for example, traders 
may quickly shift back and forth between two different cryptocurrencies in 
order to profit from differences in their price.  We investigate this
possibility in Section~\ref{sec:trading-bots}.  Similarly, people performing 
money laundering or 
otherwise holding ``dirty'' money may engage in such behavior under the belief 
that once the coins are moved back into the $\curin$ blockchain, they are 
``clean'' after moving through ShapeShift regardless of what happened with the 
coins in the $\curout$ blockchain.

Finally, identifying round-trip transactions allows us to create a link
between the input address(es) in the deposit on the $\curin$ 
blockchain with the output address(es) in the later withdrawal on 
the $\curin$ blockchain.  %
Again, there are many reasons why users might engage in such
behavior, including the trading and money laundering examples given above.  As
another example, if a $\curin$ user wanted to make an anonymous payment to 
another $\curin$ user, they might attempt to do so via a round-trip
transaction (using the address of the other user in the second pass-through
transaction), under the same assumption that ShapeShift would sever the link
between their two addresses.

\subsection{Pass-through transactions}\label{sec:pass}

\begin{figure}[t]
\centering
\includegraphics[width=0.7\linewidth]{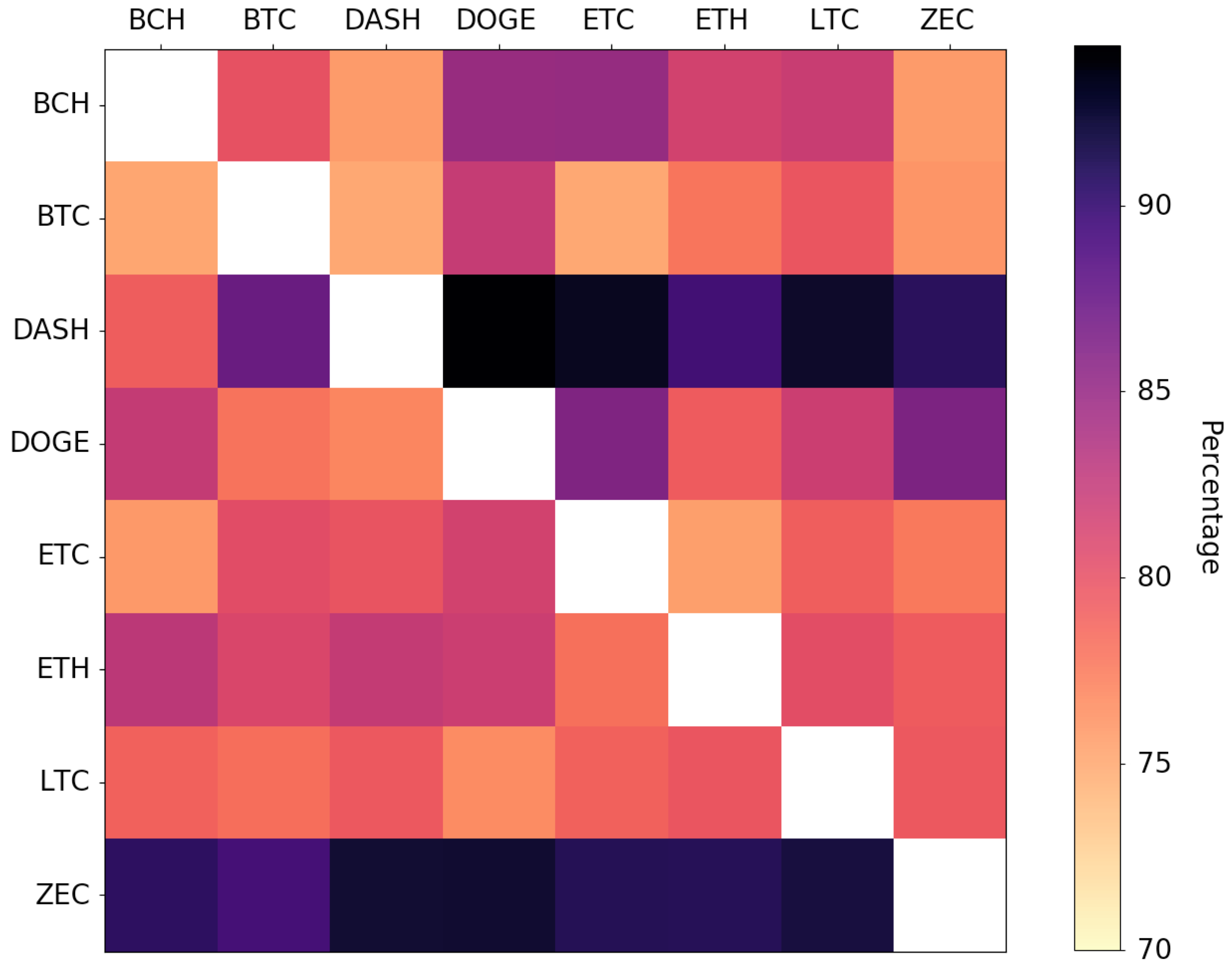}
\caption[Number of transactions identified as being a pass-through]{For each pair of currencies, the number of transactions we identified 
as being a pass-through from one to the other, as a percentage of 
the total number of transactions between those two currencies.}
\label{fig:passthrough-heatmap}
\end{figure}

Given a ShapeShift transaction from $\curin$ to $\curout$, the methods from
Section~\ref{sec:phases} already provide a way to identify pass-through
transactions, as depicted in Figure~\ref{fig:passthrough}.  In particular, 
running the augmented heuristic for
Phase~1 transactions identifies not only the deposit transaction in the
$\curin$ blockchain but also the Phase~2 transaction (i.e., the withdrawal 
transaction in the $\curout$
blockchain), as this is exactly what is returned by the API.
As discussed above, this has the effect on anonymity of tracing the flow of 
funds across this ShapeShift transaction and linking its two 
endpoints; i.e., the input address(es) in 
the $\curin$ blockchain with the output address(es) in the $\curout$ 
blockchain.  The results, in terms of the percentages of all possible 
transactions between a pair $(\curin,\curout)$ for which we found the
corresponding on-chain transactions, are in Figure~\ref{fig:passthrough-heatmap}.

The figure demonstrates that our success in identifying these types of
transactions varied somewhat, and depended\dash not unsurprisingly\dash on
our success in identifying transactions in the $\curin$ blockchain.  This
means that we were typically least successful with $\curin$ blockchains with
higher transaction volumes, such as Bitcoin, because  
we frequently ended up with multiple
hits (although here we were still able to identify more than 74\% of
transactions).  
In contrast, the dark stripes for Dash and
Zcash demonstrate our high level of success in identifying pass-through 
transactions with those currencies as $\curin$, due to our high level of
success in their Phase~1 analysis in general (89\% and 91\% respectively).  
In total, across all eight currencies we were
able to identify 1,383,666 pass-through transactions. 

\subsection{U-turns}\label{sec:uturn}

As depicted in Figure~\ref{fig:uturn}, we consider a U-turn to be a pattern 
in which a user has just sent money from 
$\curin$ to $\curout$, only to turn around and go immediately back to 
$\curin$.  This means linking two transactions:  the Phase~2 
transaction used to send money to $\curout$ and the Phase~1 transaction
used to send money back to $\curin$.  In terms of timing and amount, we require 
that the second transaction happens within 30 minutes of the first, and that 
it carries within 1\% of the value that was generated by the first Phase~2
transaction. This value is returned by the ShapeShift API in the 
$\outgoingCoin$ field.

While the close timing and amount already give some indication that these two
transactions are linked, it is of course possible that this is a coincidence 
and they were in fact carried out by different users.  In order to gain 
additional confidence that it was the same user, we have two options.
In UTXO-based cryptocurrencies (see Section~\ref{sec:back-accounting}), we could
see if the input is the same UTXO that was created in the Phase~2 transaction,
and thus see if a user is spending the coin immediately.
In cryptocurrencies based instead on accounts, such as Ethereum, 
we have no choice but to look just at the addresses.  Here we thus define a
U-turn as seeing if the address that was used as the output in the Phase~2
transaction is used as the input in the later Phase~1 transaction.

Once we identified such candidate pairs of transactions $(\tx_1,\tx_2)$, 
we then ran the augmented heuristic from Section~\ref{sec:phases} to
identify the relevant output address in the $\curout$ blockchain, according
to $\tx_1$.  We then ran the same heuristic to identify the relevant input 
address in the $\curout$ blockchain, this time according to $\tx_2$.

\begin{figure}[t]
\centering
\includegraphics[width=0.7\linewidth]{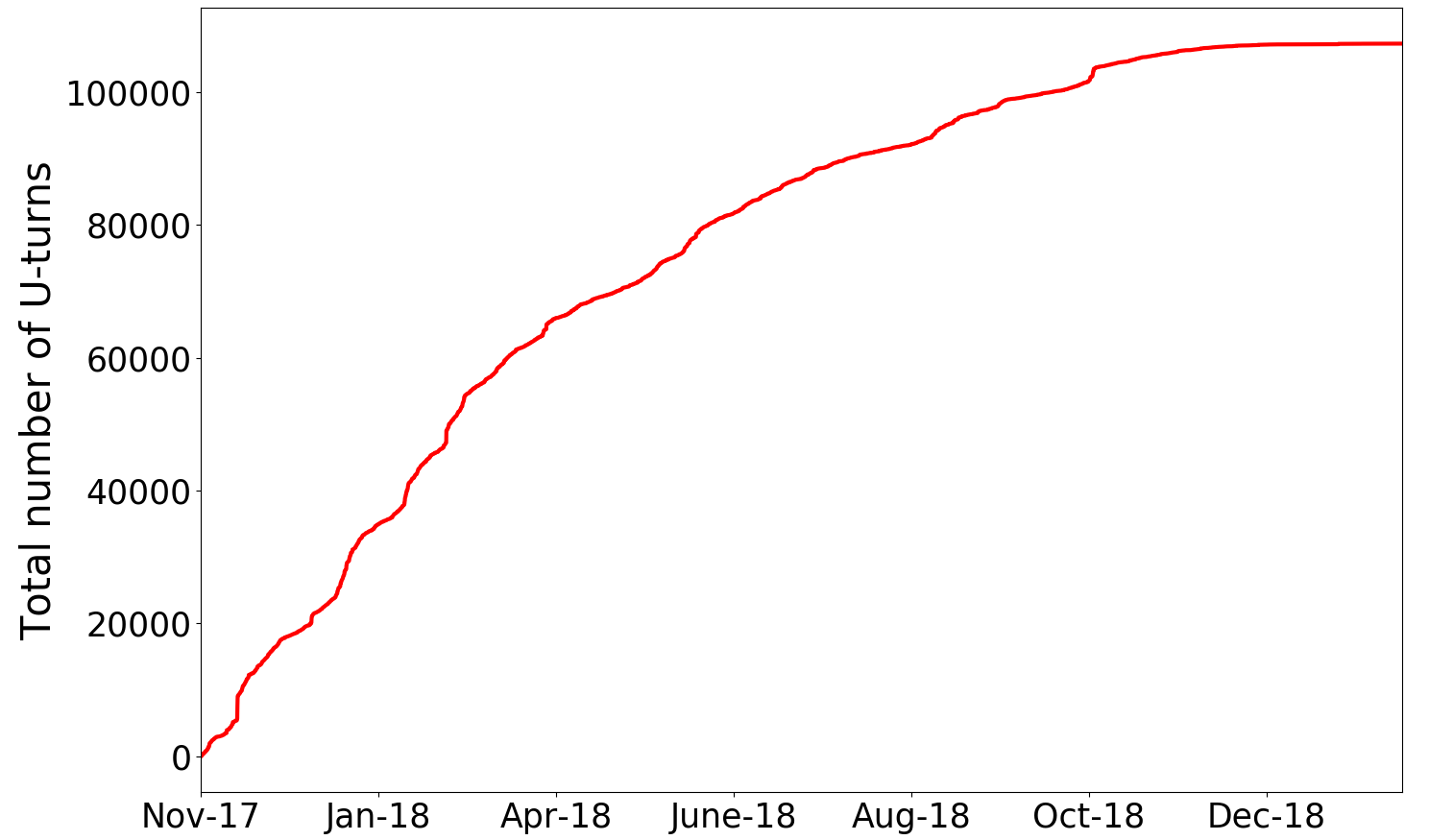}
\caption[Total number of U-turns over time]{The total number of U-turns over time, as identified by our basic 
heuristic.}
\label{fig:AllUturns}
\end{figure}

\begin{figure}[t]
\centering
\includegraphics[width=0.7\linewidth]{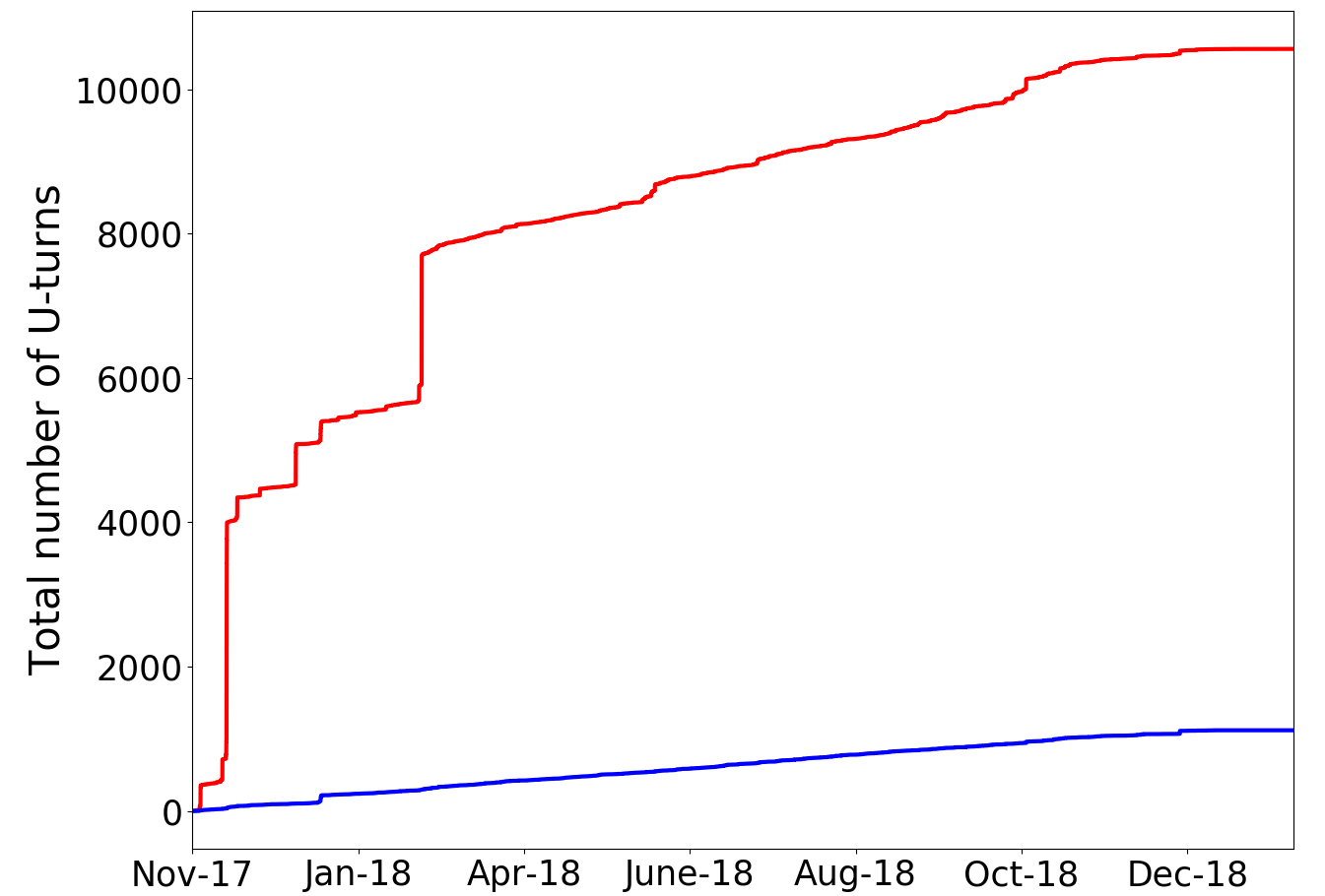}
\caption[Total number of U-turns over time identified by our heuristics]{The total number of U-turns over time, as identified by our 
address-based (in red) and UTXO-based (in blue) heuristics.}
\label{fig:VerUturns}
\end{figure}

In fact though, what we really identified in Phase~2 was not just an address but, 
as described above, a newly created UTXO.  If the input used in $\tx_2$ was this 
same UTXO, then we found a U-turn according to the first heuristic.  If instead 
it corresponded just to the same address, then we found a U-turn according
to the second heuristic.  The results of both of these heuristics, in addition
to the basic identification of U-turns according to the timing and amount, can 
be found in Table~\ref{tab:UTurnsPerCurrency}, and plots showing their
cumulative number over time can be found in Figures~\ref{fig:AllUturns}
and~\ref{fig:VerUturns}.  In total, we identified 107,267 U-turns according 
to our basic heuristic,
10,566 U-turns according to our address-based heuristic, and 
1,120 U-turns according to our UTXO-based heuristic. 

\begin{table}[t]
\centering
\begin{tabular}{lS[table-format=5]S[table-format=3]S[table-format=3]}
\toprule
Currency & {\# (basic)} & {\# (addr)} & {\# (utxo)}\\
\midrule
BTC & 36666 & 565 & 314 \\
BCH & 2864 & 196 & 81 \\
DASH & 3234 & 2091 & 184 \\
DOGE & 546 & 75 & 75 \\
ETH & 53518 & 5248 & {-} \\
ETC & 1397 & 543 & {-} \\
LTC & 8270 & 1429 & 244 \\
ZEC & 772 & 419 & 222\\
\bottomrule
\end{tabular}
\caption[Number of U-turns identified for each cryptocurrency]{The number of U-turns identified for each cryptocurrency, according
to our basic heuristic concerning timing and value, and both the address-based 
and UTXO-based heuristics concerning identical ownership.  Since Ethereum and
Ethereum Classic are account-based, the UTXO heuristic cannot be applied to
them.}
\label{tab:UTurnsPerCurrency}
\end{table}

While the dominance of both Bitcoin and Ethereum should be expected given
their overall trading dominance, we also observe that both Dash and Zcash 
have been used extensively as
``mixer coins'' in U-turns, and are in fact more popular for this purpose than
they are overall.
Despite this indication that users may prefer to use privacy coins as the
mixing intermediary, Zcash has the highest percentage of identified UTXO-based 
U-turn transactions.  Thus, these users not only do not gain
extra anonymity by using it, but in fact are easily identifiable given that 
they did not change the address used in 419 out of 772 (54.24\%) cases, 
or\dash even worse\dash immediately shifted back the exact same coin they 
received in 222 (28.75\%) cases.  
In the case of Dash, the results suggest something a bit different.  Once
more, the usage of a privacy coin was not very
successful since in 2091 out of the 3234 cases the address that received the
fresh coins was the same as the one that shifted it back.  It was the exact
same coin in only 184 cases, however, which suggests that although the user is
the same, there is a local Dash transaction between the two ShapeShift
transactions.  We defer a further discussion of this asymmetry to
Section~\ref{sec:zcash-dash}, where we also discuss more generally the use of 
anonymity features in both Zcash and Dash.

Looking at Figure~\ref{fig:VerUturns}, we can see a steep rise in the number
of U-turns that used the same address in December 2017, which is not
true of the ones that used the same UTXO or in the overall number of
U-turns in Figure~\ref{fig:AllUturns}.  Looking into this further, we observed
that the number of U-turns was particularly elevated during this period for
four specific pairs of currencies: DASH-ETH, DASH-LTC, ETH-DASH, and 
LTC-ETH.  This thus affected primarily the address-based heuristic due to the
fact that (1) Ethereum is account-based so the UTXO-based heuristic does not
apply, and (2) Dash has a high percentage of U-turns 
using the same address, but a much smaller percentage using the same UTXO.  
The amount of money shifted in these U-turns varied significantly in terms of
the units of the input currency, but all carried between 115K and 138K in USD.  
Although the ShapeShift transactions that were
involved in these U-turns had hundreds of different addresses in the $\curin$ 
blockchain, they used only a small number of addresses in the $\curout$
blockchain: $4$ addresses in Ethereum, 13 in Dash, and 9 in Litecoin.
As we discuss further in Section~\ref{sec:new-heuristic}, the re-use of 
addresses and the fact that the total amount of money (in
USD) carried by the transactions was roughly the same indicates that perhaps a
small group of people was responsible for creating this spike in the graph.

\subsection{Round-trip transactions}\label{sec:rtt}

As depicted in Figure~\ref{fig:roundtrip}, a round-trip transaction requires
performing two ShapeShift transactions: one out of
the initial currency and one back into it.  
To identify round-trip transactions, we effectively combine the results of the
pass-through and U-turn transactions; i.e., we tagged something as a
round-trip transaction if the output of a pass-through
transaction from X to Y was identified as being involved in a U-turn 
transaction, which was itself linked to a later pass-through transaction from
Y to X (of roughly the same amount).  As described at the beginning of the
section, this has the powerful effect of creating a link between the sender
and recipient within a single currency, despite the fact that money flowed
into a different currency in between.

In more detail, we looked for consecutive ShapeShift transactions 
where for a given pair of cryptocurrencies X and Y: 
(1) the first transaction was of the form X-Y; (2) the second transaction 
was of the form Y-X; (3) the second transaction happened relatively
soon after the first one; and (4) the value carried by the two 
transaction was approximately the same.  For the third property, we required
that the second transaction happened within 30 minutes of the first.  For the
fourth property, we
required that if the first transaction carried $x$ units of $\curin$ then the
second transaction carried within $0.5\%$ of the value in the (on-chain) 
Phase~2 transaction, according to the $\outgoingCoin$ field provided by the
API.  

As with U-turns, we considered an additional restriction to capture the case
in which the user in the $\curin$ blockchain stayed the same, meaning money 
clearly did not change hands.  Unlike with U-turns, however, this restriction 
is less to provide accuracy for the basic heuristic and more to isolate the 
behavior of people engaged in day trading or money laundering (as opposed to
those meaningfully sending money to other users).  For 
this pattern, we identify the input addresses used in Phase~1 for the first
transaction, which represent the user who initiated the round-trip transaction 
in the $\curin$ blockchain.  We then identify the output addresses used in
Phase~2 for the second transaction, which represent the user who was the
final recipient of the funds.  If the address was the same, then it is clear
that money has not changed hands.  Otherwise, the round-trip transaction acts
as a heuristic for linking together the input and output addresses.

\begin{table}[t]
\centering
\begin{tabular}{lS[table-format=5]S[table-format=2]S[table-format=2]}
\toprule
Currency & {\# (regular)} & {\# (same addr)} \\
\midrule
BTC & 35019 & 437  \\
BCH & 1780 & 84  \\
DASH & 3253 & 2353  \\
DOGE & 378 & 0 \\
ETH & 45611 & 4085 \\
ETC & 1122 & 626  \\
LTC & 6912 & 2733  \\
ZEC & 472 & 172 \\
\bottomrule
\end{tabular}
\caption[Number of regular round-trip transactions identified for each 
cryptocurrency]{The number of regular round-trip transactions identified for each 
cryptocurrency, and the number that use the same initial and final address.}
\label{tab:RTTsPerCurrency}
\end{table}

The results of running this heuristic (with and without the extra restriction) 
are in Table~\ref{tab:RTTsPerCurrency}.  In total, we identified 95,547
round-trip transactions according to our regular heuristic, and identified 
10,490 transactions where
the input and output addresses were the same.  Across different currencies,
however, there was a high level of variance in the results.  While this could 
be a result of the different levels of accuracy in Phase~1 for 
different currencies, the 
more likely explanation is that users indeed engage in different patterns of
behavior with different currencies.  For Bitcoin, for example, there was a 
very small percentage (1.2\%) of round-trip transactions that used the
same address.  This suggests that either users are aware of the
general lack of anonymity in the basic Bitcoin protocol and use ShapeShift to
make anonymous payments, or that if they do use round-trip transactions as a 
form of money laundering they are at least careful enough to change their
addresses.  More simply, it may just be the case that generating new addresses
is more of a default in Bitcoin than it is in other currencies.

In other currencies, however, such as Dash, Ethereum Classic, Litecoin, and
Zcash, there were relatively high percentages of round-trip transactions that 
used the same 
input and output address: 72\%, 56\%, 40\%, and 36\% respectively.  In Ethereum
Classic, this may be explained by the account-based nature of the currency, 
which means that it is common for one entity to use only one address, although 
the percentage for Ethereum is much lower (9\%).  In Dash and Zcash, as we 
have already seen in Section~\ref{sec:uturn} and explore further in
Section~\ref{sec:zcash-dash}, it may simply be the case that users assume they
achieve anonymity just through the use of a privacy coin, so do not take extra
measures to hide their identity.

\section{Clustering Analysis}\label{sec:clusters}

\subsection{Shared ownership heuristic}\label{sec:multi-input}

As described in Sections~\ref{sec:changelly-data}
and~\ref{sec:shapeshift-data}, we engaged in transactions with both ShapeShift 
and Changelly, which provided us with some ground-truth evidence of addresses
that were owned by them.  We also collected three sets of
tagging data (i.e., tags associated with addresses that describe their
real-world owner): for Bitcoin we used the data available from
WalletExplorer,\footnote{\url{https://www.walletexplorer.com/}} which covers a
wide variety of different Bitcoin-based services; for Zcash we used
hand-collected data from Kappos et al.~\cite{217535}, which covers only
exchanges; and for Ethereum we used the data available from
Etherscan,\footnote{\url{https://etherscan.io/}} which covers a variety of
services and contracts.

In order to understand the behavior of these trading platforms and the
interaction they had with other blockchain-based services such as exchanges,
our first instinct was to combine these tags with the now-standard 
``multi-input'' clustering heuristic for 
cryptocurrencies~\cite{reid2013analysis,Meiklejohn2013a}, which states that 
in a transaction with multiple input addresses, all inputs belong to the same
entity.  %
When we applied this clustering heuristic to an earlier version of our 
dataset~\cite{anon-us}, however, the results were fairly uneven.  For
Dogecoin, for example, the three ShapeShift transactions we performed revealed 
only three addresses, which each had done a very small number of
transactions.  By clustering the addresses we sent coins to and received from the three Changelly transactions we performed, we identified 24,893 addresses, which in total had received over 67
trillion DOGE.  These results suggest that the trading platforms operate a
number of different clusters in each cryptocurrency, and perhaps even change
their behavior depending on the currency, which in turns makes it clear that
we did not capture a comprehensive view of the activity of either.

More worrying, in one of our Changelly transactions, we received coins from a 
Ethereum address that had been tagged as belonging to HitBTC, a prominent exchange.
This suggests that Changelly may occasionally operate using exchange accounts,
which would completely invalidate the results of the clustering heuristic, as
their individually operated addresses would end up in the same cluster as all
of the ones operated by HitBTC.  We thus decided not to use this type of
clustering, and to instead focus on a new clustering heuristic geared at
identifying common social relationships.

\subsection{Common relationship heuristic}\label{sec:new-heuristic}

As it was clear that the multi-input heuristic would not yield meaningful
information about shared ownership, we chose to switch our focus away from the
interactions ShapeShift had on the blockchain and look instead at
the relationships between individual ShapeShift users.  In particular, we
defined the following heuristic:

\begin{heuristic}\label{heuristic:clusters}
If two or more addresses send coins to the same address in the $\curout$
blockchain, or if two or more addresses receive coins from the same address 
in the $\curin$ blockchain, then these addresses have some common social 
relationship.
\end{heuristic} 

The definition of a common social relationship is (intentionally) vague, and 
the implications of this heuristic are indeed less clear-cut than those of
heuristics around shared ownership.
Nevertheless, we consider what it means for two different addresses, in 
potentially two different blockchains, to have sent coins to the same address;
we refer to these addresses as belonging in the \emph{input} cluster of the
output address (and analogously refer to the \emph{output} cluster for an
address sending to multiple other addresses).
In the case in which the addresses are most closely linked, it could represent 
the same user consolidating money held across different currencies into a 
single one.  It could also represent different users interacting with a common 
service, such as an exchange.  Finally, it could simply be two users who do 
not know each other directly but happen to be sending money 
to the same individual.  What cannot be the case, however, is that the
addresses are not related in any way.

To implement this heuristic, we parsed transactions into a graph where we
defined a node as an address and a directed edge $(u,v)$ as existing when 
one address $u$ initiated a ShapeShift transaction sending coins to $v$, which
we identified using the results of our pass-through analysis from 
Section~\ref{sec:phases}. (This means that the inputs in our graph are
restricted to those for which we ran Phase~1 to find the address, and thus 
that our input clusters contain only the top 8 currencies.  
In the other direction, however, we obtain the address directly from the API, 
which means output clusters can contain all currencies.)
Edges are further weighted by the number of transactions sent from $u$ to $v$.  
For each node, the cluster centered on that address was then defined as all
nodes adjacent to it (i.e., pointing towards it).

Performing this clustering generated a graph with 2,895,445
nodes (distinct addresses) and 2,244,459 edges.  
Sorting the clusters by in-degree reveals the entities that received
the highest number of ShapeShift transactions (from the top 8
currencies, per our caveat above).
The largest cluster had 12,868 addresses\dash many of them belonging to
Ethereum, Litecoin, and Dash\dash and was centered on a Bitcoin 
address belonging to CoinPayments.net, a multi-coin payment processing
gateway.  Of the ten largest clusters, three others (one associated with 
Ripple and two with Bitcoin addresses) are also connected with CoinPayments, 
which suggests that ShapeShift is a popular platform amongst its users.  

Sorting the individual clusters by out-degree reveals instead the users who 
initiated the highest number of ShapeShift transactions.
Here the largest cluster (consisting of 2314 addresses) was centered on a 
Litecoin address, and the second largest cluster was centered on an Ethereum 
address that belonged to Binance (a popular exchange).  Of the
ten largest clusters, two others were centered on Binance-tagged addresses,
and three were centered on other exchanges (Freewallet, Gemini, and Bittrex).
While it makes sense that exchanges typically dominate on-chain activity in
many cryptocurrencies, it is somewhat surprising to also observe that
dominance here, given that these exchanges already allow users to shift between 
many different cryptocurrencies.  Aside from the potential for better rates or
the perception of increased anonymity, it is thus unclear why a user wanting to
shift from one currency to another would do so using ShapeShift
as opposed to using the same service with which they have already stored 
their coins.

Beyond these basic statistics, we apply this heuristic to several of the 
case studies we investigate in the next section.  We also revisit here
the large spike in the number of U-turns that we observed in
Section~\ref{sec:uturn}.  Our hypothesis then was that this spike was caused
by a small number of parties, due to the similar USD value carried by the
transactions and by the re-use of a small number of addresses across Dash,
Ethereum, and Litecoin.  Here we briefly investigate this further by 
examining the clusters centered on these addresses.  

Of the 13 Dash addresses, 
all but one of them formed small input and output clusters that were comprised 
of addresses solely from Litecoin and Ethereum.  
Of the 9 Litecoin addresses, 6 had input clusters consisting solely
of Dash and Ethereum addresses, with two of them consisting solely of Dash 
addresses.  
Finally, of the 4
Ethereum addresses, all of them had input clusters consisting solely of Dash
and Litecoin addresses.  One of them, however, had a diverse set of addresses 
in its output cluster, belonging to Bitcoin, Bitcoin Cash, and a number of 
Ethereum-based tokens.  These results thus still suggest a small number of 
parties, due to the tight connection between the three currencies in the 
clusters, although of course further investigation would be needed to get a
more complete picture.

\section{Patterns of ShapeShift Usage}\label{sec:case-studies}

In this section, we examine potential applications of the analysis developed
in previous sections, in terms of identifying specific usages of ShapeShift.
As before, our focus is on anonymity, and the potential that such platforms
may offer for money laundering or other illicit purposes, as well as for
trading.  To this end, we
begin by looking at two case studies associated with explicitly criminal
activity and examine the interactions these criminals had with the ShapeShift
platform.  We then switch in Section~\ref{sec:trading-bots} to look at
non-criminal activity, by attempting to identify trading bots that use
ShapeShift and the patterns they may create.  Finally, in
Section~\ref{sec:zcash-dash} we look at the role
that privacy coins (Monero, Zcash, and Dash) play, in order to 
identify the extent to
which the usage of these coins in ShapeShift is motivated by a desire for
anonymity.

\subsection{Starscape Capital}

In January 2018, an investment firm called Starscape Capital raised over
2,000~ETH (worth 2.2M~USD at the time) during their Initial Coin Offering,
after promising users a 50\% return in exchange for investing in their
cryptocurrency arbitrage fund.  Shortly afterwards, all of their social media 
accounts disappeared, and it was reported that an amount of ETH worth
517,000~USD was sent from their wallet to ShapeShift, where it was shifted into 
Monero~\cite{scheck_shifflett_2018}.  

We confirmed this for ourselves by observing that the address known to be 
owned by Starscape Capital participated in 192 Ethereum transactions across a 
three-day span (January 19-21), during which it received and sent 
2,038 ETH; in total it sent money in 133 transactions.  We found that 109 
of these transactions sent money to 
ShapeShift, and of these 103 were shifts to Monero conducted on January 21 
(the remaining 6 were shifts to Ethereum). The total amount shifted into 
Monero was 465.61~ETH (1388.39~XMR), and all
of the money was shifted into only three different Monero addresses, of which
one received 70\% of the resulting XMR.  Using the clusters defined in 
Section~\ref{sec:new-heuristic}, we did not find evidence of any other 
addresses (in any other currencies)
interacting with either the ETH or XMR addresses associated with Starscape
Capital.  %

\subsection{Ethereum-based scams}

EtherScamDB\footnote{https://etherscamdb.info/} is a website that, based on 
user reports that are manually investigated by its operators, collects and 
lists Ethereum addresses that have been involved in scams.
As of January 30 2019, they had a total
of 6374 scams listed, with 1973 associated addresses.  
We found that 194 of these addresses (9\% of those listed) had been involved 
in 853 transactions to ShapeShift, of which 688 had a $\status$ field of 
\verb#complete#.
Across these successful transactions, 1797~ETH was shifted to other
currencies: 74\% to Bitcoin, 19\% to Monero, 3\% to Bitcoin Cash,
and 1\% to Zcash.

The scams which successfully shifted the highest volumes belonged to so-called 
trust-trading and MyEtherWallet scams. Trust-trading is a scam based on
the premise that users who send coins prove the legitimacy of their
addresses, after which the traders ``trust'' their address and send 
back higher amounts (whereas in fact most users send money and simply receive
nothing in return).  
This type of scam shifted over 918~ETH, the majority of which was converted 
to Bitcoin (691~ETH, or 75\%). 
A MyEtherWallet scam is a phishing/typosquatting scam where scammers
operate a service with a similar name to the popular online wallet
MyEtherWallet,\footnote{https://www.myetherwallet.com/} in order to trick
users into giving them their account details.
These scammers shifted the majority of the stolen ETH to Bitcoin
(207~ETH) and to Monero (151~ETH).

Given that the majority of the overall stolen coins was shifted to Bitcoin, we 
next investigated whether or not these stolen coins could be tracked further
using our analysis.  %
In particular, we looked to see if they performed a U-turn or a round-trip
transaction, as discussed in Section~\ref{sec:tracking}. We identified one 
address, associated with a trust-trading scam, that participated in 34 
distinct round-trip transactions, all coming back to a different address from
the original one.  All
these transactions used Bitcoin as $\curout$ and used the same address in
Bitcoin to both receive and send coins; i.e., we identified the U-turns in
Bitcoin according to our address-based heuristic.  In total, more than 70~ETH
were circulated across these round-trip transactions.

\subsection{Trading bots}\label{sec:trading-bots}

ShapeShift, like any other cryptocurrency exchange, can be used by 
traders who wish to take advantage of the volatility in cryptocurrency prices.
The potential advantages of doing this via ShapeShift, 
as compared with other platforms that focus more on the exchange between
cryptocurrencies and fiat currencies, are that (1) ShapeShift transactions can
be easily automated via their API, and (2) a single ShapeShift transaction
acts to both purchase desired coins and dump unwanted ones.  
Such trading usually requires large volumes of transactions and high precision 
on their the timing, due to the constant fluctuation in cryptocurrency prices.  
We thus looked for activity that involved large numbers of similar
transactions in a small time period, on the theory that it would be
associated primarily with trading bots.

We started by searching for sets of consecutive ShapeShift
transactions that carried approximately the same value in $\curin$ (with an error
rate of 1\%) and involved the same currencies.  When we did this, however, we 
found thousands of such sets.  We thus added the extra conditions that there
must be at least 15 transactions in the set that took place in a
span of five minutes; i.e., that within a five-minute block of ShapeShift
transactions there were at least 15 involving the same currencies and carrying
the same approximate USD value.  This resulted in 107 such sets.  

After obtaining our 107 trading clusters, we removed transactions that 
we believed were false positives in that they happened to have a similar 
value but were clearly the odd one out.  For example, in a cluster of 20 
transactions with 19 ETH-BTC transactions and one LTC-ZEC transaction, 
we removed the latter.  We were thus left with clusters of either a 
particular pair (e.g., ETH-BTC) or two pairs where the $\curout$ or the 
$\curin$ was the same (e.g., ETH-BTC and ZEC-BTC), which suggests either 
the purchase of a rising coin or the dump of a declining one.  We sought to
further validate these clusters by using our heuristic from
Section~\ref{sec:new-heuristic} to see if the clusters shared common
addresses.  While we typically did not find this in UTXO-based currencies (as
most entities operate using many addresses), in account-based currencies 
we found that in almost every case there was one particular address that was 
involved in the trading cluster. 

\begin{figure}[t]
\centering
\includegraphics[width=0.8\linewidth]{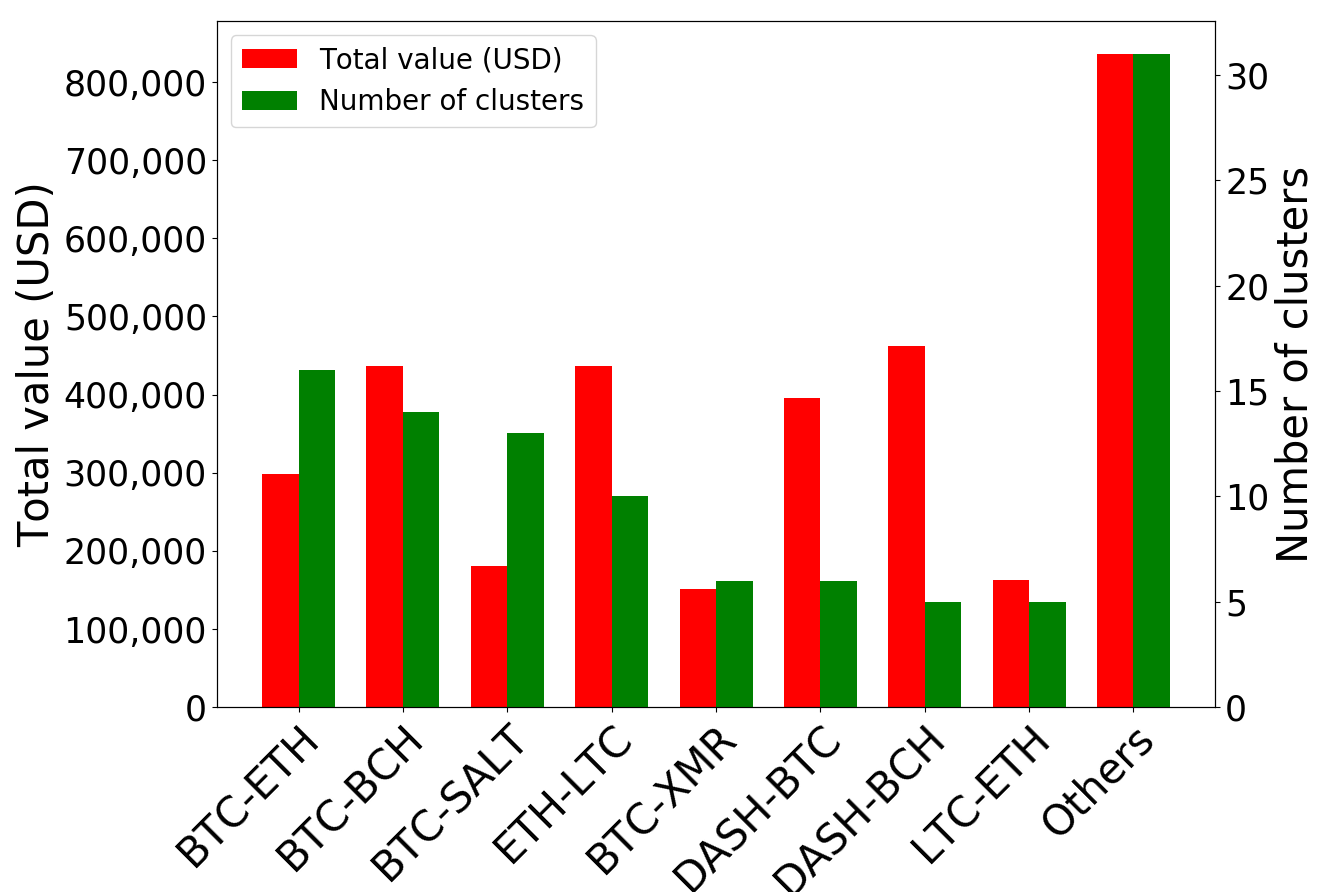}
\caption[Trading bot clusters categorized by traded currencies]{Our 107 clusters of likely trading bots, categorized by the
pair of currencies they trade between and the total amount transacted by 
those clusters (in USD).}
\label{fig:barchart-bots}
\end{figure}

We summarize our results in Figure~\ref{fig:barchart-bots}, in terms of the 
most common pairs of currencies and the total money exchanged by trading 
clusters using those currencies.  It is clear that 
the most common interactions are performed between the most popular currencies
overall, with the exception of Monero (XMR) and SALT.  In particular, we found 
six clusters consisting of 17-20 transactions that exchanged BTC 
for XMR, and 13 clusters that exchanged BTC for SALT, an Ethereum-based 
token.
The sizes of each trading cluster varied between 
16 and 33 transactions and in total comprise 258 transactions, each of which
shifted exactly 0.1~BTC.  In total they originated from 514 
different Bitcoin addresses, which may make it appear as though 
different people carried out these transactions.  After applying our 
pass-through heuristic, however, we
found that across all the transactions there were only two distinct SALT
addresses used to receive the output.  It is thus instead likely that this
represents trading activity involving one or two entities.

\subsection{Usage of anonymity tools}\label{sec:zcash-dash}

Given the potential usage of ShapeShift for money laundering or other criminal
activities, we sought to understand the extent to which its users seemed
motivated to hide the source of their funds.  While using ShapeShift is
already one attempt at doing this, we focus here on the combination of using
ShapeShift and so-called ``privacy coins'' (Dash, Monero, and Zcash) that are 
designed to offer improved anonymity guarantees.  

In terms of the effect of the introduction of KYC into ShapeShift, the
number of transactions using Zcash as $\curin$ averaged 164 per day 
the month before, and averaged 116 per day the month after.  We also saw a 
small decline with Zcash as $\curout$: 69 per day before and 43 per day 
after.  Monero and Dash, however, saw much higher declines, and
in fact saw the largest declines across all eight cryptocurrencies.  The 
daily average the month before was $136$ using Monero as $\curin$, whereas it 
was $47$ after.  Similarly, the daily average using it as $\curout$ was 
$316$ before and $62$ after. For Dash, the daily average as $\curin$ 
was $128$ before and $81$ after, and the daily average as $\curout$ 
was $103$ before and $42$ after.

In terms of the blockchain data we had (according to the most popular 
currencies), our analysis in what follows is restricted to Dash and Zcash,
although we leave an exploration of Monero as interesting future work.

\subsubsection{Zcash}

The main anonymity feature in Zcash is known as the \emph{shielded pool}.
Briefly, transparent Zcash transactions behave just like Bitcoin transactions
in that they reveal in the clear the sender and recipient (according to
so-called \emph{t-addresses}), as well as the value being sent.  This
information is hidden to various degrees, however, when interacting with
the pool.  In particular, when putting money into the pool the recipient is
specified using a so-called \emph{z-address}, which hides the recipient but still
reveals the sender, and taking money out of the pool hides the sender (through
the use of zero-knowledge proofs~\cite{SP:BCGGMT14}) but reveals the
recipient.  Finally, Zcash is designed to provide privacy mainly in the case
in which users transact \emph{within} the shielded pool, which hides the
sender, recipient, and the value being sent.

\begin{figure}
\centering
\includegraphics[width=0.6\linewidth]{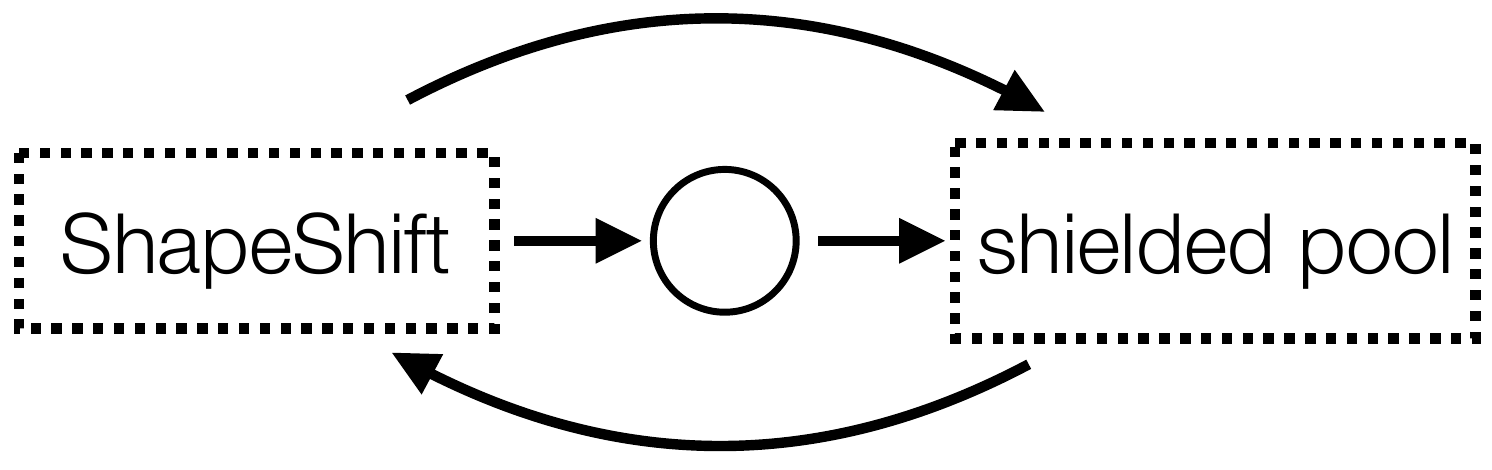}
\caption[Types of transactions investigated between ShapeShift and Zcash]{The three types of interactions we investigated between ShapeShift and
the shielded pool in Zcash.}
\label{fig:nugget}
\end{figure}

We considered three possible interactions between ShapeShift and the shielded
pool, as depicted in
Figure~\ref{fig:nugget}: (1) a user shifts coins directly from ShapeShift into
the shielded pool, (2) a user shifts to a t-address but then uses that
t-address to put money into the pool, and (3) a user sends money directly from
the pool to ShapeShift.

For the first type of interaction, we found 29,003 transactions that used ZEC
as $\curout$.  Of these, 758 had a z-address as the output address, meaning
coins were sent directly to the shielded pool.  The total value put into the
pool in these transactions was 6,707.86~ZEC, which is 4.3\% of all the ZEC
received in pass-through transactions.  When attempting to use
z-addresses in our own interactions with ShapeShift, however, we consistently 
encountered
errors or were told to contact customer service.  It is thus not clear if
usage of this feature is supported at the time of writing.

For the second type of interaction, 
there were 1309 where the next transaction (i.e., the transaction in which this
UTXO spent its contents) involved putting money into the pool.  The total
value put into the pool in these transactions was 12,534~ZEC, which is
8.2\% of all the ZEC received in pass-through transactions.

For the third type of interaction, we found 111,041 pass-through transactions
that used ZEC as $\curin$.  Of these, 3808 came directly
from the pool, with a total value of 22,490~ZEC (14\% of all the
ZEC sent in pass-through transactions).

Thus, while the usage of the anonymity features in Zcash was not
necessarily a large fraction of the overall usage of Zcash in ShapeShift,
there is clear potential to move large amounts of Zcash (representing over 10
million USD at the time it was transacted) by combining ShapeShift with the
shielded pool.  

\subsubsection{Dash}
Our parameters for identifying a 
CoinJoin were thus that (1) the transaction must have at least three inputs, 
(2) the outputs must consist solely of values from the list of possible 
denominations (modulo the fees), and (3) and all output values must be 
the same.  
In fact, given how Dash operates there is always one output with a
non-standard value, so it was further necessary to relax the second and third
requirements to allow there to be at most one address that does not carry the
specified value.

We first looked to see how often the DASH sent to ShapeShift had 
originated from a CoinJoin, which meant identifying if the inputs of a 
Phase~1 transaction were outputs from a CoinJoin.  
Out of 100,410 candidate transactions, we found 2,068 that came from a 
CoinJoin, carrying a total of 11,929~DASH in value (6.5\% of the total value
across transactions with Dash as $\curin$).
Next, we looked at whether or not users performed a CoinJoin after 
receiving coins from ShapeShift, which meant identifying if the outputs of a
Phase~2 transaction had been spent in a CoinJoin.  
Out of 50,545 candidate 
transactions, we found only 33 CoinJoin transactions, carrying a total of
187~DASH in value (0.1\% of the total value across transactions using Dash as
$\curout$).

If we revisit our results concerning the use of U-turns in Dash 
from Section~\ref{sec:uturn}, we recall that there was a large asymmetry in
terms of the results of our two heuristics: only 5.6\% of the U-turns used the
same UTXO, but 64.6\% of U-turns used the same address.  This suggests that some
additional on-chain transaction took place between the two ShapeShift 
transactions, and indeed upon further inspection we identified many cases
where this transaction was a CoinJoin.  There thus appears to have
been a genuine attempt to take advantage of the privacy that Dash offers, but
this was completely ineffective due to the use of the same address that both 
sent and received the mixed coins.

\section{Conclusions}\label{sec:conclusions}

In this study, we presented a characterization of the usage of the ShapeShift
trading platform over a thirteen-month period, focusing on the ability to link
together the ledgers of multiple different
cryptocurrencies.  To accomplish this task, we looked at these trading 
platforms from several different perspectives, ranging from the correlations 
between the transactions they produce in the cryptocurrency ledgers to the 
relationships they reveal between seemingly distinct users.  The 
techniques we develop demonstrate that it is possible to
capture complex transactional behaviors and trace their activity even as it 
moves across ledgers, which has implications for any
criminals attempting to use these platforms to obscure their flow of money.

\chapter{Forsage: An Anatomy of a Cryptocurrency Pyramid Scheme}
\label{chapterlabelForsage}

\section{Overview}

Cryptocurrencies and smart contracts are new and powerful technologies that promise a range of benefits, including faster monetary transactions, innovative financial instruments, and global financial inclusion for the world’s unbanked. Conversely, though, these same technologies have fuelled new forms of fraud and theft~\cite{zhao02017, Ferreira2019} and new ways of perpetrating existing types of crime~\cite{ransomwarejusticedept,phillips2020tracing}. %
 
{\em Pyramid schemes}, for example, are a prevalent type of scam in which top-tier participants in a hierarchical network recruit and profit at the expense of an expanding base of new participants. They have existed for more than a century, but have recently emerged in a new form: as smart contracts on blockchains such as Ethereum. 

Smart contracts are in some ways an ideal medium for pyramid schemes and other scams. Because they run in decentralized systems, they cannot easily be dismantled by law enforcement agencies. They can instantaneously ingest payments from victims across the globe. They provide privacy protection for their creators in the form of pseudonymous addresses. Finally, as so-called ``trustless’’ applications---with world-readable (byte)code---they present a veneer of trustworthiness to unsuspecting users. 

The flip side of such transparency is that smart contracts offer researchers a degree of visibility into the mechanics of online (and offline) scams that is without historical precedent. Not only is the (byte)code specifying the scam’s mechanics visible on chain, but so is every transaction performed by every participant. 

In this chapter, we take advantage of this newfound visibility to conduct an in-depth measurement study of the largest smart contract-based pyramid scheme to date, called {\em Forsage Smartway} or {\em Forsage} for short. 

Forsage came into existence in late January 2020. It was at one point the second most active contract in Ethereum by daily transaction count, and remains in the top twenty at the time of writing.  As we show throughout this paper, it is a classic pyramid scheme, defined by the SEC as ``a type of fraud in which participants profit almost exclusively through recruiting other people to participate in the program''~\cite{sec.gov_2013}. The Forsage contract requires players to send currency (Ether) in order to participate. Funds sent by newly recruited users immediately pass through the contract to existing players, with those at the top of the (smart contract-defined) pyramid obtaining the largest returns. 

Understanding the success of Forsage requires study of not just the contract itself, but also its community of hundreds of thousands of users, many of whom have actively discussed and marketed the scam. Consequently, to paint a detailed picture of how Forsage lures and defrauds users, our study combines measurement and analysis of a range of complementary forms of data, including source code, on-chain transaction data, and social media interactions.

Our results come from three basic, mutually illuminating forms of study: smart contract deconstruction (Section~\ref{sec:evaluation}), blockchain analytics (Section~\ref{sec:measurement_study}), and analysis of video and social media interactions (Section~\ref{sec:community}).

We believe that our study's findings are not just relevant to Forsage, but provide durable insights into the conception, mechanics, and evolution of smart-contract scams and financial scams more generally. They also point to effective strategies that  
government authorities and the cryptocurrency community can use to combat pyramid schemes and other scams, as we discuss in Section~\ref{sec:solutions}.

We emphasize that our results, which reveal a combination of classic and smart contract-specific scam characteristics, offer insights not just into Forsage, but into both blockchain and non-blockchain scams more generally. 
\section{Background}

\subsection{Smart contracts}

Forsage is realized as a {\em smart contract}. Smart contracts are applications that execute on {\em blockchains}, decentralized systems that indelibly and immutably record transactions in an authoritative sequence and are best known as the platforms that realize cryptocurrencies such as Bitcoin.   

The most popular public (permissionless) blockchain for smart contracts today is {\em Ethereum}~\cite{buterin_buterin_2014}, whose native currency is known as {\em Ether} (ETH). Ethereum smart contracts are launched in the form of bytecode that runs in a Turing-complete environment known as the Ethereum Virtual Machine (EVM).  Smart contract creators often also publish corresponding source code, typically written in the Solidity programming language, but such publication is optional. {\em Transactions} sent to smart contracts by users are processed by contract code and are publicly visible on chain. 

Transactions may send money to a contract from user accounts or other contracts and must specify payment of execution fees to miners in the form of {\em gas}, a parallel currency converted into ETH upon transaction execution. 
This conversion is calculated by multiplying the amount of work performed by a transaction (its ``gas consumed") by the price of gas in ETH set by user when submitting the transaction~\cite{yellowpaper}. 

Correctness of contract execution is enforced by the consensus mechanism underlying the Ethereum blockchain, so a miner's execution of contract code in the EVM must be agreed upon by all network participants to be included in a confirmed block. 

Other permissionless blockchains with smart contract functionality are growing in popularity, e.g., Tron~\cite{tronwhitepaper}, to which Forsage has also been ported. Ethereum, however, remains the dominant smart contract platform. 

\subsection{Scams}

Scams, i.e., fraudulent schemes involving financial deception, have been documented for centuries. Many scams involving large populations of victims assume the form of {\em pyramid schemes}. The U.S. Securities and Exchange Commission (SEC) defines a pyramid scheme as ``a type of fraud in which participants profit almost exclusively through recruiting other people to participate in the program''~\cite{sec.gov_2013}. Pyramid schemes, which are illegal in most jurisdictions, come in a number of variants. One variant is a {\em Ponzi scheme}, which specifically involves investment in financial instruments. {\em Multi-level marketing} (MLM) schemes, which involve the sale of a product or service, are related to pyramid schemes. They are legal in the U.S., but outlawed in some jurisdictions (e.g., China)~\cite{chinalawMLM}.

\subsection{Blockchain scams}
A multitude of scams have arisen within the blockchain ecosystem. Some scams have solicited investments from victims in new blockchain technologies. Examples include Onecoin, a Ponzi scheme that involved a fake (centralized) blockchain in which victims invested \$19+ billion~\cite{madeira_2020},  Bitconnect, a token that promised returns of 1\% per day and saw investment of \$3.5 billion from victims, as well as other, related \$1+ billion schemes such as Plustoken and WoToken.pro~\cite{palmer_2020, bel_2020}.

Other scams instead use blockchain technology to realize variants of scams, such as pyramid schemes, that were seen well before the advent of blockchains. Prominent examples are Million.Money\footnote{\url{https://million.money}} and Doubleway.io\footnote{\url{https://doubleway.io/}}, which are both currently active, as well as the defunct scheme Bullrun.live.\footnote{\url{http://bullrun.live}}  All three have similarities with Forsage: they use similar promotional materials, have a similar structure for the user dashboard, and use similar language and terminology (e.g., a referrer to the program is called an ``upline''). We explore Forsage user interactions with multiple scam contracts in ~\ref{subsec:user_behavior}.

\section{Forsage Overview}
\label{sec:overview}

The creators and promoters of Forsage advertise it as a \emph{matrix} MLM scheme, despite the lack of a service or product.  It operates primarily on Ethereum, where its initial Matrix contract has been active since January 31st, 2020. Since then, Forsage creators have also launched a Forsage contract on Tron (TRX) and an additional, followup smart contracts called Forsage xGold on both Tron and Ethereum. At the time of this writing, the Forsage authors have since released contracts (\bscaddr{0x2CAa4694cB7Daf7d49A198dC1103C06d4991ae52
}, \bscaddr{0x98872a66D0749C720D8Dc1A80d496b24B04ff7C5}) on the Binance Smart Chain (BSC).

\noindent \paragraph{The Forsage website:}
Users interact with Forsage using the forsage.io website, which shows how much they have paid into and earned from the contract. %
The website encourages the use of user-friendly cryptocurrency tools. It shows users how to purchase cryptocurrency using Trust Wallet, a user-friendly tool to exchange fiat for cryptocurrency, and how to use MetaMask, a browser extension that allows users to easily transact with cryptocurrency.  
The combination of these tools makes Forsage accessible to novice users who may not previously have used cryptocurrencies or smart contracts.

\newcommand{\subf}[2]{%
  {\small\begin{tabular}[t]{@{}c@{}}
  #1\\#2
  \end{tabular}}%
}

\begin{figure}
\centering
\begin{tabular}{cc}
\subf{\includegraphics[width=140mm]{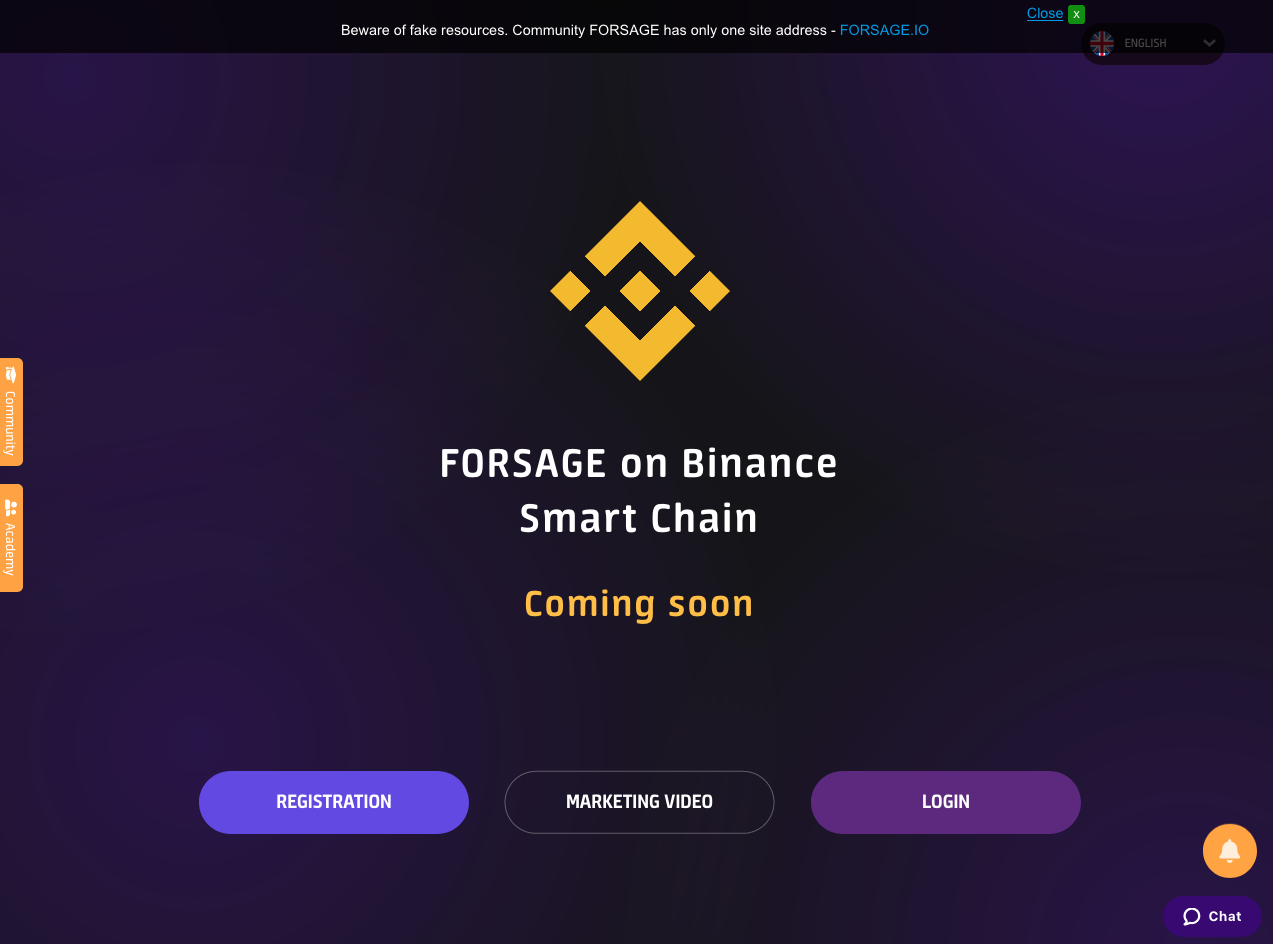}}
     {(a)}
\\
\subf{\includegraphics[width=140mm]{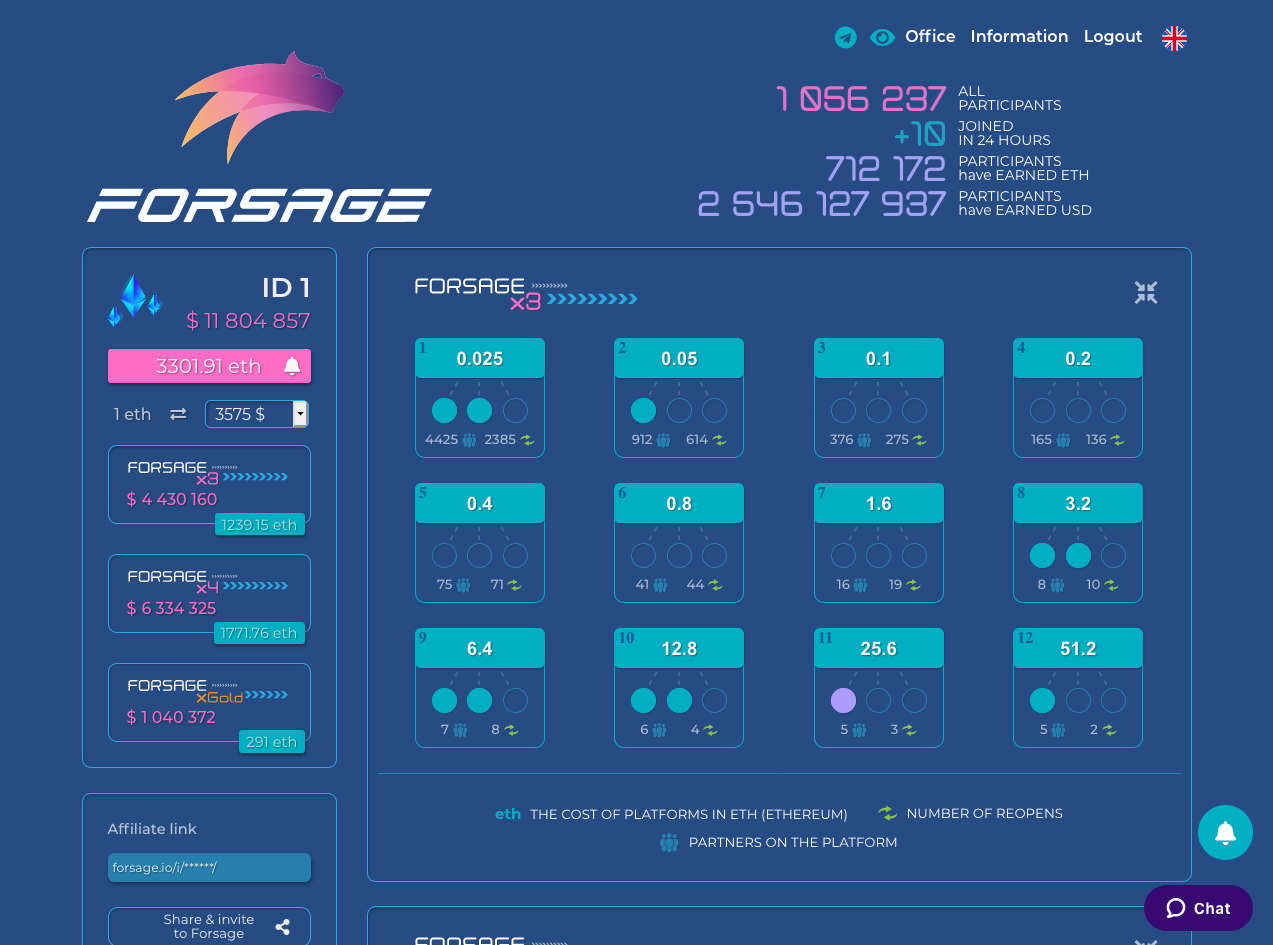}}
     {(b)}
\\
\end{tabular}
\caption[Screenshots of the forsage.io website]{Screenshots of the forsage.io website. (a) Homepage of forsage.io as of May 6th, 2021. The website is marketing the arrival of the new scheme that will be used on the Binance Smart Chain. (b) Landing page of the most profitable user showing the progress page of the X3 matrix and other macro statistics.}
\label{sec:webpagescreenshots}
\end{figure}

\noindent\paragraph{Forsage use and structure:}

A new Forsage user must pay a minimum of 0.05 ETH, which opens up the \emph{slot} at the first \emph{level} in the two matrix systems, called X3 and X4. 
Each matrix consists of 12 slots. To unlock the ability to use the next slot (at level $i+1$), a user must pay twice as much ETH as for their currently highest slot (at level $i$). In both X3 and X4, the first slot costs 0.025~ETH, while the twelfth and final slot costs 51.2~ETH. This means that the total cost to open all slots in either matrix is 102.375~ETH.

Each Forsage user has a \emph{referral code}, created at the time they register. The referral code links a recruited user's account to the account that recruited them, called their \emph{upline}. These referral codes thus organize Forsage users into pyramids, with the oldest accounts at the top. Payments flow upwards within a pyramid as additional users join it. 
The pyramids of users linked by chains of referral code %
are referred to as Forsage \emph{teams}.  It is possible to join Forsage without entering a referral code; users who do so are assigned the referral code of the contract \emph{owner} (the creator).  %

\ref{sec:evaluation} contains an explanation of the logic for payment flow of user funds sent through the Forsage contract.  Briefly, users earn money in the X3 and X4 matrices as follows.

\begin{description}
    \item[X3:] In X3, users earn income by recruiting others into the system. A user must recruit three additional users to recoup their initial investment within each slot.
    Any recruits beyond the first three per slot will generate income for the recruiting user and those further up in their pyramid. Each subsequent slot costs more to open, but its resulting payout if filled with recruits will be higher because the expected payout for each three recruits is equal to the initial cost to open the slot for the recruiter. After a user fills a slot (i.e. recruits 3 users into that slot), Forsage \emph{blocks} the filled slot, causing the user to forfeit future earnings from it until it is unblocked. Unblocking means paying to open the slot at the next level up in the system, at which point this lower-level slot cannot become blocked again.
    
    \item[X4:]  \label{x4reference}  In X4, users can earn both by recruiting other users and by being on an active team whom are opening new slots to ensure minimal blocking. 
    When a user recruits the six additional users necessary to recoup their initial investment in an X4 slot (twice as many as are required in X3), that slot becomes blocked and the user will have received the same amount of money paid to open the slot, with others in their team getting paid as well. X4 also has an element of competition: If a newer user on a team is more active than the user whose referral code they used to join Forsage, that user can switch spots on the team, giving the more active, newer user the profits that would otherwise flow to the older, referring account~\cite{youtube1}. 
\end{description}

\section{Forsage Contract Deconstruction}
\label{sec:evaluation}

Forsage promotional materials imply that the system is trustworthy because its code is open-source, e.g., the promotional materials claim that the contract ``guarantees the purity of conditions.'' 
We took advantage of the availability of the source code to conduct an in-depth analysis of the smart contract's logic and data structures.

\begin{figure}
  \centering
    \includegraphics[width=0.95\linewidth]{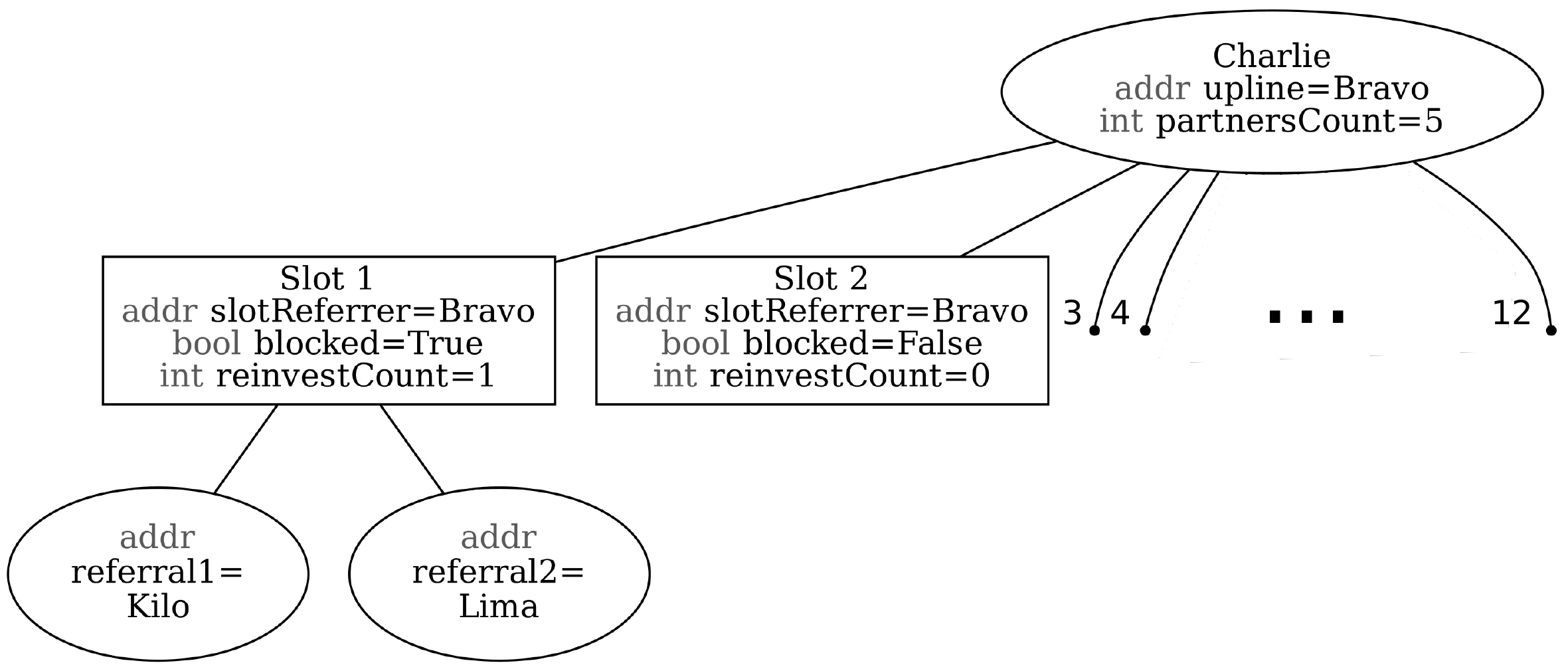}
    \caption[A visualization of the state the Forsage contract keeps for each user in the X3 matrix]{A visualization of the state the contract keeps for each user in the X3 matrix, focusing on a user Charlie. The \texttt{addr} variables point to Ethereum addresses, here given NATO-phonetic names. Matrix slots that have not yet been opened are depicted with a numbered dot, instead of a box.}
    \label{fig:data_structuresx3}
\end{figure}

\paragraph{Methodology and data collection:}
\label{para:data_collection}
The code for the Matrix smart contract is published on Etherscan.\footnote{ \fontsize{7.7}{12}\selectfont{\href{https://etherscan.io/address/0x5acc84a3e955Bdd76467d3348077d003f00fFB97}{etherscan.io/address/0x5acc84a3e955Bdd76467d3348077d003f00fFB97}}}
We first attempted manual source code review, but found the logic too confusing to follow without visualization. We then built a simulator in Python that deployed the contract to a local private test network of Go-Ethereum (Geth) nodes,\footnote{\url{https://github.com/ethereum/go-ethereum}} and used Web3.py\footnote{\url{https://github.com/ethereum/web3.py}} to send sample transactions. We also wrote a visualizer for the contract's state machine using GraphViz~\cite{Ellson01graphviz}. The output of that visualizer assisted in creating Figure~\ref{fig:data_structuresx3}, which depicts the data stored in the contract. Although the open source code is pointed to as a source of legitimacy by Forsage promotional materials, our analysis of the contract took weeks of focused effort by a professional research engineer. Our source code for the visualizer and simulator tools will be released as open source software in the near future.

When the Forsage team launched their Tron implementation of the Matrix smart contract, they also released its source code. We found this Tron code to be nearly identical to the Ethereum original, so we did not specifically analyze it. The latest iteration of Forsage launched on both Ethereum and Tron (as of May 2021), the xGold contract, has no publicly available source code.

The Ethereum and Tron blockchains include the data for all transactions performed by Forsage users.  We mined this publicly available data to perform further analysis.  
To obtain Ethereum data we ran the Go-Ethereum 
(Geth)\footnote{{\url{https://github.com/ethereum/go-ethereum}}} and TurboGeth\footnote{{\url{https://github.com/ledgerwatch/turbo-geth}}} full-node and archive-node software packages, and downloaded 
the entire blockchain up to January 14, 2021.   

We then used the Ethereum-ETL\footnote{{\url{https://github.com/blockchain-etl/ethereum-etl}}} package
to retrieve this data from Geth and store the %
345 million transactions included in the Ethereum blockchain between the launch of Forsage (January 31, 2020) and January 14, 2021.  We wrote custom Python scripts to analyze this data and found 222,516,680 transactions that involved function calls on smart contracts, of which 3,266,722  were to the Forsage smart contract. 
To profile user transactions
outside Forsage, we used the Chainalysis Reactor tool.\footnote{\url{https://www.chainalysis.com/chainalysis-reactor/}}  
Chainalysis Reactor is a web-based investigation platform that connects cryptocurrency transactions to real-world entities, using tags that are either internal to Chainalysis or gathered from public websites and documents. 

To collect Tron transaction data we scraped the TronScan 
API\footnote{{\url{https://tronscan.org/}}} and parsed the results 
directly into CSV form.

\goodbreak

\paragraph{Forsage data structures:}
As discussed in Section~\ref{x4reference}, Forsage consists of two matrix systems, X3 and X4, each consisting of 12 slots.  These two matrices differ in the number of users that act to fill each matrix level (three for X3, six for X4) and the logic for how nodes propagate through them over time. %

The data for each user is stored in a hashtable (Solidity mapping) on the Ethereum blockchain, with the key being the user's address and the value being a Solidity \texttt{struct} with the data for that user's state tree and arrays of pointers to its children.
Figure~\ref{fig:data_structuresx3} visualizes this mapping for a user's X3 tree, with some minor metadata variables omitted.  Each user also has an X4 tree, whose structure is largely similar.  As seen in this figure, each user has an \emph{upline}, which is the user that referred them to the contract. This is distinct from \textsf{slotReferrer}, a variable used per slot as part of the payment logic. The \textsf{slotReferrer} variable is initialized to the upline, but changes over time as users refer each other. The \textsf{reinvestCount} variable keeps track of the number of times a slot has been filled. In our example, Charlie has filled his first matrix slot once already (and then unblocked it by buying a slot at level 2), meaning he has referred $3 \times \mathsf{reinvestCount} + 2 = 5 = \mathsf{partnersCount}$ users.

\paragraph{External API:}
The contract exposes 15 functions to read its state, and two state-changing functions, \texttt{registrationExt} and \texttt{buyNewLevel}. The first registers new users and thus adds them to the contract state. The second changes contract state for an existing user to allow them to continue to gain money from new referrals.

The placement of new users in the contract state depends on the X3 and X4 slots for the user that referred them (their upline). The logic of the contract \emph{scrambles} positions in the upline's matrices and in the matrices of the upline's parent when an upline's slot becomes full, i.e. every time the upline refers a multiple of three users to a given X3 slot ($\mathsf{partnersCount} \bmod 3 = 0$), or a multiple of six users to a given X4 slot. The logic of scrambling leaf nodes in the pyramid depends on the state of the slot \texttt{referrer} variable for the affected matrix slot, as well as the \texttt{blocked} variable for that slot, and in the X4 system an additional \texttt{closedPart} variable for each slot. Scrambling the positions of the existing users in the system helps to make payments through Forsage (falsely!) appear more random. It benefits older users in the pyramid, as users are usually scrambled ``up'' the pyramid to become children of older users rather than newer ones.

\begin{table}
\centering
\small
\begin{tabularx}{0.999\linewidth}{XXXXX}
\toprule
Opcode & \makecell{Avg num\\ per tx\\ (all)} & \makecell{Median\\ (all)} & \makecell{Avg num\\ per tx\\ (Forsage)} & \makecell{Median\\ (Forsage)} \\
\midrule
\normalsize{\texttt{SSTORE}}  & \makecell{4.54\\ $\pm$ 8.10} & \makecell{2} & \makecell{10.76\\ $\pm$ 9.57} & \makecell{6} \\
\\
\normalsize{\texttt{SLOAD}}   & \makecell{17.84\\ $\pm$ 51.6} & \makecell{7} & \makecell{36.86\\ $\pm$ 26.21} & \makecell{29} \\
\bottomrule
\end{tabularx}
\caption[Average number of Ethereum instruction operations per transaction]{Average number of instruction operations per transaction, with standard deviation, for both all transactions and only those that interact with Forsage. Due to the intensive computation required to process this data, this table covers only the thousand blocks between block heights 10,600,000 and 10,601,000 (Roughly 13:00-18:00 UTC on August 5th, 2020) rather than our larger dataset including all transactions from 2020. This smaller dataset still contains 188,920 transactions that interact with smart contracts, 5667 of which interact with Forsage.}
\label{table:number_instructions_per}
\end{table}

\paragraph{Transaction fees:}
\label{subsec:gas}
The fact that Forsage has so much persistent on-chain storage means that its users pay higher gas fees than the average for Ethereum contracts, due to the heavy usage of the (expensive) \texttt{SLOAD} and \texttt{SSTORE} opcodes.
These fees are higher even when comparing Forsage transactions only to other contract function calls in Ethereum (so in particular ignoring simple sends of ETH). In our collected dataset of Ethereum network transactions, we found that the mean transaction fee for all Ethereum transactions that interacted with a contract was 0.00632~ETH with a standard deviation of 0.0618~ETH and a median of 0.00257~ETH.  Forsage transactions paid a higher average transaction fee of 0.0116~ETH with a standard deviation of 0.0108~ETH and a median of 0.00883~ETH. Forsage users pay more than four times as much on average as other smart contract users.

The most gas-expensive EVM operations/opcodes are those that create a new contract (\texttt{CREATE}, \texttt{CREATE2}); store, change, and access data into persistent on-chain state (\texttt{SSTORE}, \texttt{SLOAD}), and call contract functions or send money to other users in the network (\texttt{CALL})~\cite{yellowpaper}. Every transaction that interacts with Forsage through its two main functions, \texttt{registrationExt} and \texttt{buyNewLevel}, uses two of these three most expensive categories, often multiple times: they make use of persistent storage via \texttt{SSTORE} and \texttt{SLOAD} operations, and send money to other users on the network using Solidity operations that compile to the \texttt{CALL} opcode.  Forsage uses an average number of \texttt{CALL} operations, but makes heavy use of \texttt{SSTORE} and \texttt{SLOAD}, as shown in Table~\ref{table:number_instructions_per}. 

\begin{figure}[H]
    \centering
    \includegraphics[width=0.8\linewidth]{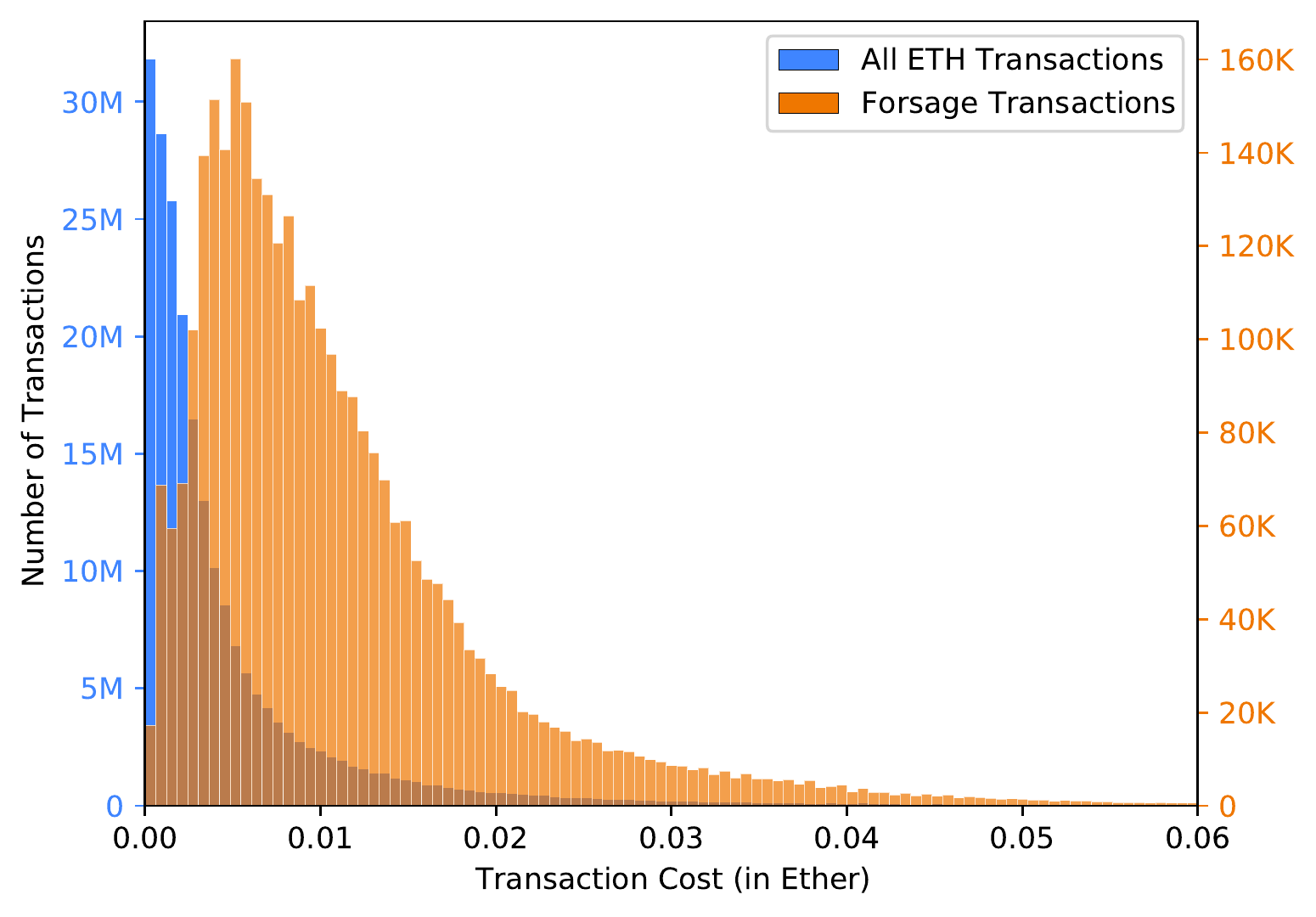}
    \caption[Histogram of transaction costs on the Ethereum blockchain involving smart contract function calls]{Histogram of transaction costs on the Ethereum blockchain---from January 31, 2020 to January 14, 2021---that involve successful smart contract function calls. Blue bars indicate the number of all transactions that paid fees within the given bucket, while orange bars indicate the same data, but only for transactions sent to the Forsage smart contract.  The data excludes outlier transactions with fees above 0.06~ETH, which is above the 99th percentile of all transactions from this time period.}
    \label{fig:tx_fee_histogram}
\end{figure}

Figure~\ref{fig:tx_fee_histogram} shows a superimposed histogram of Forsage transactions relative to all Ethereum transactions. 
The higher gas consumption associated with Forsage results in higher transaction fees overall, as demonstrated by the right-shifted peak in the Forsage curve relative to that of all ETH transactions.

\begin{figure}[H]
\centering
\includegraphics[width=0.9\linewidth]{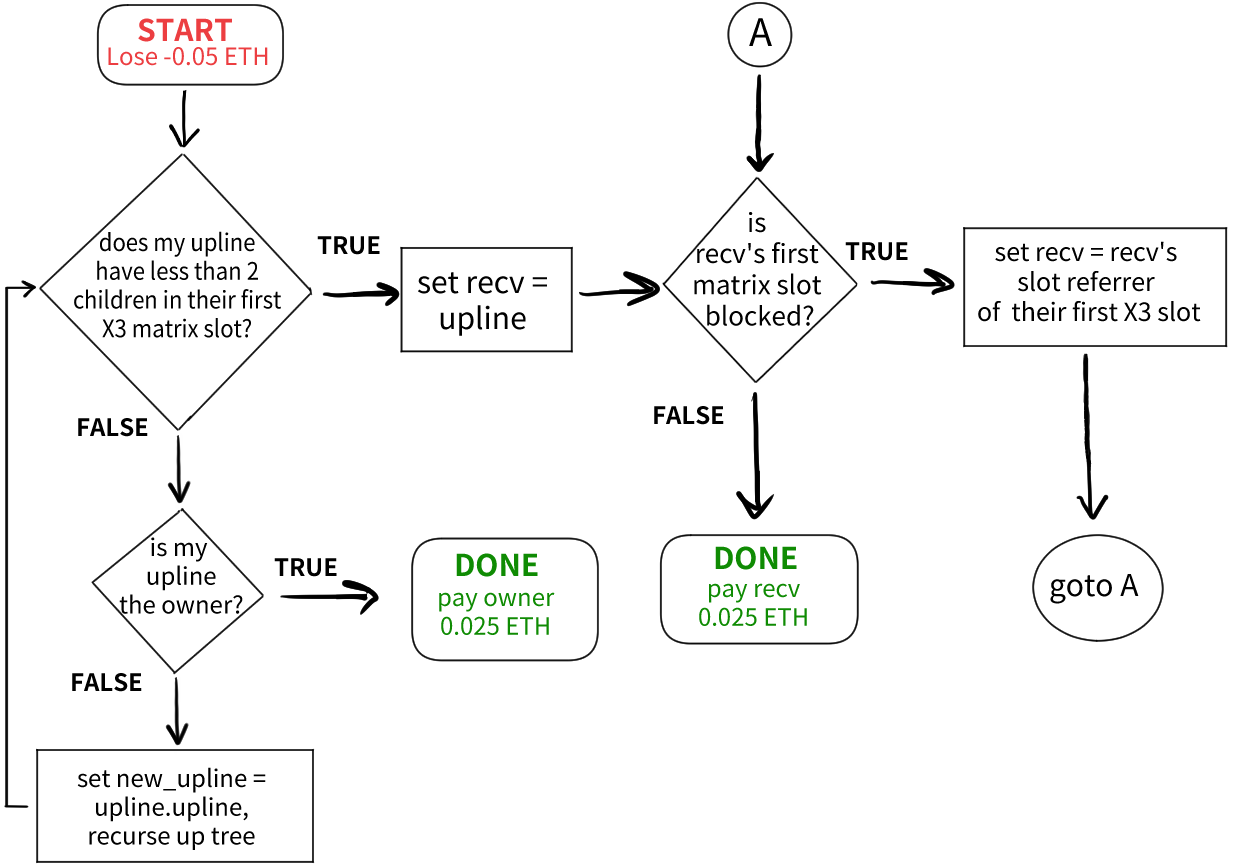}
\caption[Flow chart for the logic of who gets paid when a new user registers in the X3 system]{Flow chart for the logic of who gets paid when a new user registers, in the X3 system. The \texttt{BuyNewLevel} function follows similar logic, but conditioned on the matrix slot being purchased, rather than the first slot.}
\label{forsage-payflow-x3}
\end{figure}

\paragraph{Payment logic:}

There are three ways for a user to get paid in Forsage: (1) by referring new users to the system; (2) when users they have referred in the past buy an additional matrix slot at a level corresponding to one previously purchased by the referrer; and (3) when \emph{spillover} occurs, a condition in the X4 matrix resulting from the slots of another user downstream in the pyramid being blocked. Whenever money is sent to the smart contract by one user, the contract atomically (i.e., in the same transaction) sends those funds to other users based on the logic described below. This allows Forsage promotional materials to claim that the contract ``never stores users' funds.''

When a user buys a new slot, the money they pay typically routes to the first found upline that also has that same slot open. Users are thus incentivized to buy new levels in order to refer users underneath them, which means a user can be generally successful by adding additional matrix slots just before referring additional users, and in general by recruiting as many users as possible.

Figure~\ref{forsage-payflow-x3} show the logic determining who gets paid when a new user registers with the Forsage contract. %
The logic for purchases of new slots (\texttt{buyNewLevel}) is largely similar but depends on the slot purchased rather than the first one (e.g., if a user buys the third slot then the logic is conditioned on the status of their upline's third slot).

\begin{figure}
    \centering
    \includegraphics[width=0.8\linewidth]{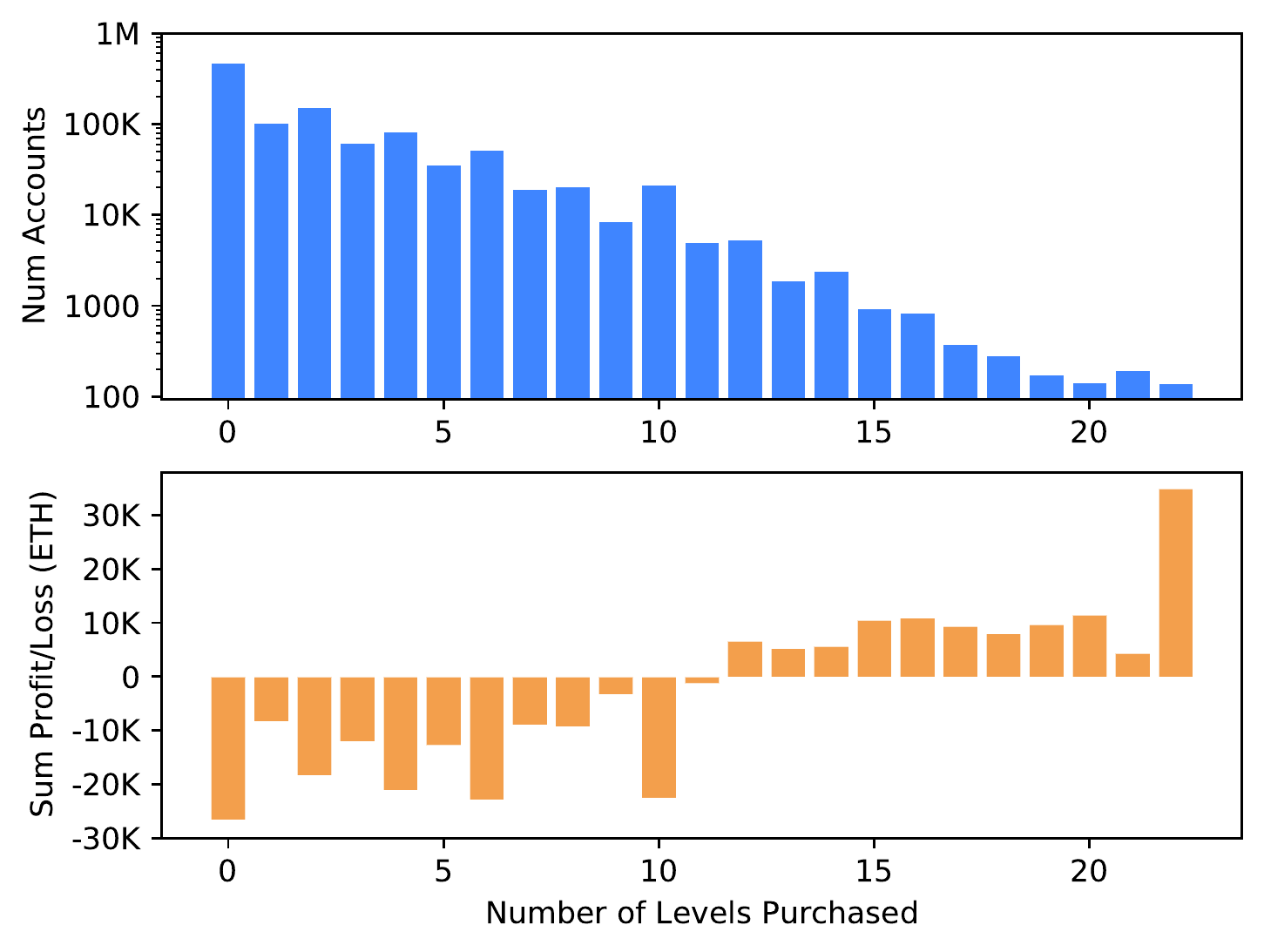}
    \caption[Distribution of users that unlocked a given number of levels in the Forsage contract]{The distribution of how many users had unlocked a given number of levels in the contract (on top, and at log scale), and the collective amount of money gained or lost by the users who had unlocked this number of levels (on bottom, and at linear scale).  Users that bought the most levels were on average the most profitable.}
    \label{fig:num_levels_bought_histogram}
\end{figure}

The flowcharts in these figures show that uplines must keep their slots from becoming blocked, or payments will skip over them.  
To prevent a slot from becoming blocked, a user must buy the slot at the next level. This will also unblock an existing slot if it already has become blocked, and prevent the slot at $level - 1$ from ever becoming blocked again. Figure~\ref{fig:num_levels_bought_histogram} shows the distribution of levels purchased in aggregate for all users in the Forsage contract, as well as the summed profitability for the group of users that purchased that many slot levels. In general users that purchased more levels were also the most profitable users: The average user of the contract purchased 2.13 levels, with a standard deviation of 2.89 and a median of 1 level purchased.

When a new user joins the system, their payment is split into two equal parts and the logic in the flowchart is applied to each half, with one half going through the X3 flowchart and one half through the X4 flowchart, to determine which other user(s) should get each half of the payment.
If the direct upline of this new user is not blocked, then the upline gets the payment.  If the upline has a blocked slot, the contract checks the upline's upline for that matrix slot level to see if it is blocked. This iterates through uplines until the contract finds one that is unblocked, which it then pays.  The contract owner (i.e., the user that created the contract) is always unblocked, so the contract always finds a user to pay. This can sometimes result in the same user being payed twice (once by each half), or uncles and aunts being paid by their nephews and nieces in the tree if it has been previously scrambled.  This condition is called \emph{spillover}.

\begin{figure}
    \centering
    \includegraphics[width=0.8\linewidth]{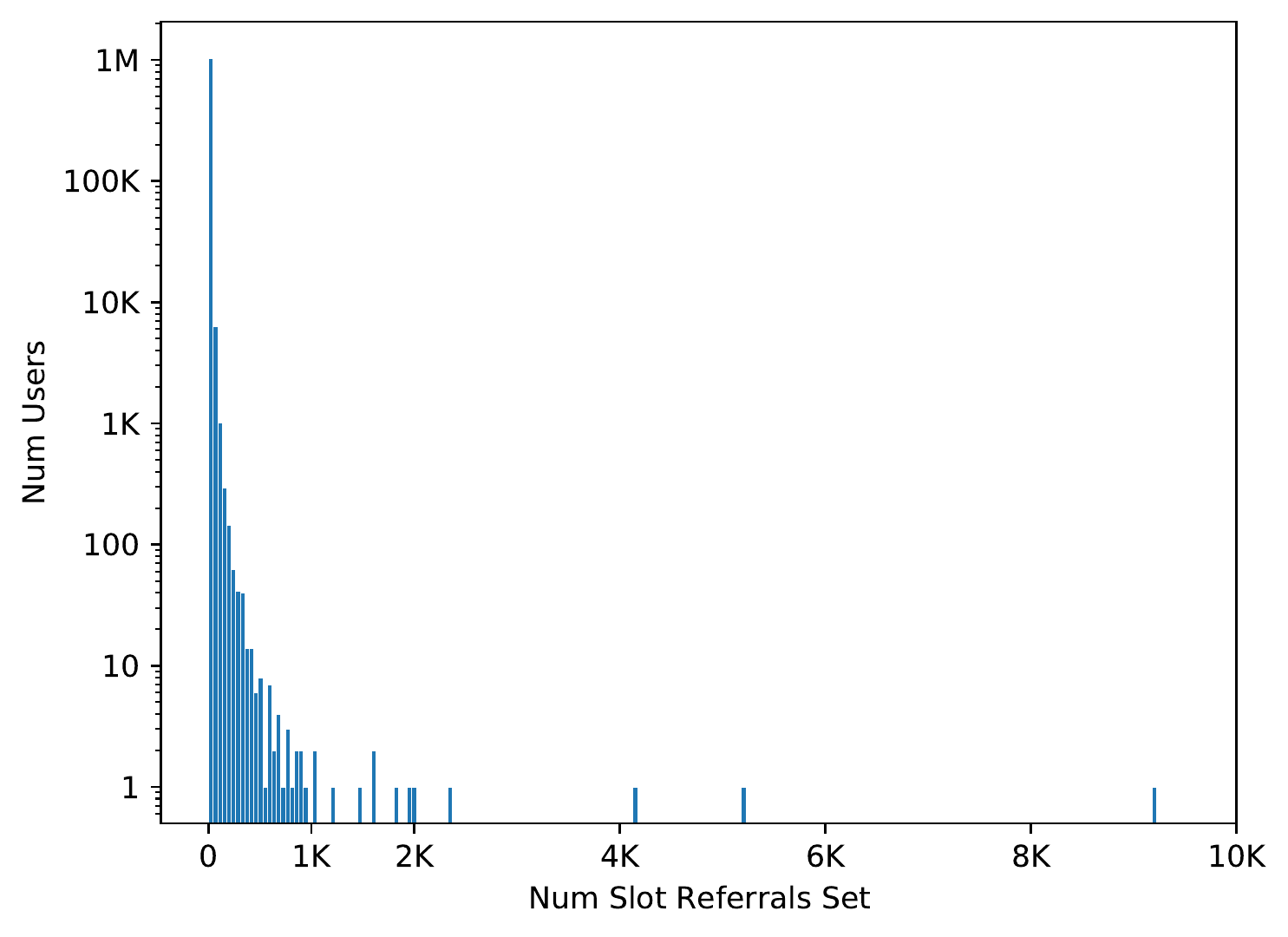}
    \caption[Total number of users acting as slot referrer]{On a log scale, the total number of users (on the y-axis) acting as slot referrer for a given number of \textit{other} users (on the x-axis), for both the X3 and X4 matrices. For example, one user (the contract owner,  \ethaddr{0x81ca1e4de24136ebcf34ca518af87f18fd39d45e}) is slot referrer for 9220 other users.}
    \label{fig:referral_counts}
\end{figure}

Spillover means that it is possible to earn money by receiving payments that should have gone to another user who had blocked slots.
This passive earning is possible only in the X4 system, and only if a spillover recipient's upline is blocked and cannot currently receive payment. 
A given user's chance of spillover is unpredictable, because it depends on the actions of other users. In our analysis of the transactions to Forsage from its conception until January 14th, 2021, we found that 35,251 transactions (only 1.08\%) contained spillover payments. Of those transactions 63\% were registrations, and the remaining 37\% resulted from buying new levels.

Ethereum transaction costs are incurred by each interaction with the Forsage smart contract, and eat into users' profits. Any claims about user profit must thus take gas costs into account.

\paragraph{The privileged role of the owner:}

The Forsage contract is initialized so that the owner account (i.e., the creator of the contract, \ethaddr{81ca1e4de24136ebcf34ca518af87f18fd39d45e
}) has all matrix slots for both X3 and X4 opened for free. Likewise, the owner's slots can never become blocked.  This creates ample opportunities for the owner to profit from the contract, which we confirm empirically in Section~\ref{sec:measurement_study}.

Beyond the ability to earn money by referring users, the owner also has additional opportunities to earn money passively.  If a user sends the contract exactly 0.05 ether for registration without specifically calling the registration function, or calling a function that does not exist, that function call is rerouted to the registration function with the owner set as the user's upline.  Likewise, if the upline gets replaced as the referrer, it is always replaced with a user further up in the pyramid.  Thus, as users refer others and have their slots blocked as a result, the upline for all users eventually converges to the owner of the contract.  Finally, the logic that prevents the owner's slots from becoming blocked also means that the owner's children do not change once set.  This means that the owner maintains the oldest users in the pyramid as children, which results in high spillover in the X4 matrix.

We found that the \textsf{slotReferrer} variable was set to the contract owner for 9220 slots in the Forsage contract. By comparison, the average Forsage user was set as the referrer for 4.14 other accounts (with a standard deviation of 15.92) and the median account was set as the referrer for one other account. Figure~\ref{fig:referral_counts} shows the full distribution of referrers for all accounts.

\section{Contract Measurement Study}
\label{sec:measurement_study}

In this section, we present the results of our measurement study of Forsage contract transactions, which encompasses all monetary transactions in the scheme.  A description of our data collection process is in Section~\ref{para:data_collection}.  We first present statistics capturing the degree of user interaction with the various Forsage contracts on Ethereum and Tron
(\ref{subsec:stats}). We then present an analysis of the account behaviour and profits over the Forsage user population (\ref{subsec:most_profitable}), in particular analyzing where funds are obtained and how funds flow through the five most profitable accounts. 

\subsection{Scheme statistics}\label{subsec:stats}

\begin{table*}[h!]
\centering
\small
\begin{tabular}{m{16mm}m{12mm}m{12mm}m{14mm}m{12mm}m{20mm}m{11mm}}
\toprule
 Contract & Total TXs & Unique sending addresses & Total coins & Total USD & Launch date & Address\\
\midrule
 ETH Matrix  & 3M & 1M & 721k  & 225M & Jan 31, 2020 & \ethaddr{0x5acc84a3e955Bdd76467d3348077d003f00fFB97}  \\ %
  TRX Clone   & 217k &   78k & 537M  & 14M & July 25, 2020 & \trxaddr{TJRv6qukWEz4DKY6gkd3fhX4uahREpTQu6} \\ %
 TRX Matrix  & 1M &  342k &   1B  & 31M & Sept 6, 2020 & \trxaddr{TREbha3Jj6TrpT7e6Z5ukh3NRhyxHsmMug} \\ %
 TRX xGold   & 307k &  105k & 90M  & 2M & Nov 7, 2020 &  \trxaddr{TA6p1BnBf2HJgc77Zk8BHmHoiJzquLCKWb}  \\ %
 ETH xGold   & 37k  &  17k &  8k  & 9M & Jan 4, 2021 & \ethaddr{0x488e3a4bbbb2386ba619eed88319e807c3ddb6c2} \\ %
\bottomrule
\end{tabular}
\caption[Summary statistics of the investigated Forsage smart contracts]{Summary statistics of the four official Forsage smart contracts and one clone. The USD value was calculated by taking a sum of the payments per day and multiplying it by the average of the 24-hour high and low on the respective day.}
\label{table:summary_all_contracts}
\end{table*}

\begin{figure}[t!]
  \centering
    \includegraphics[width=0.8\linewidth]{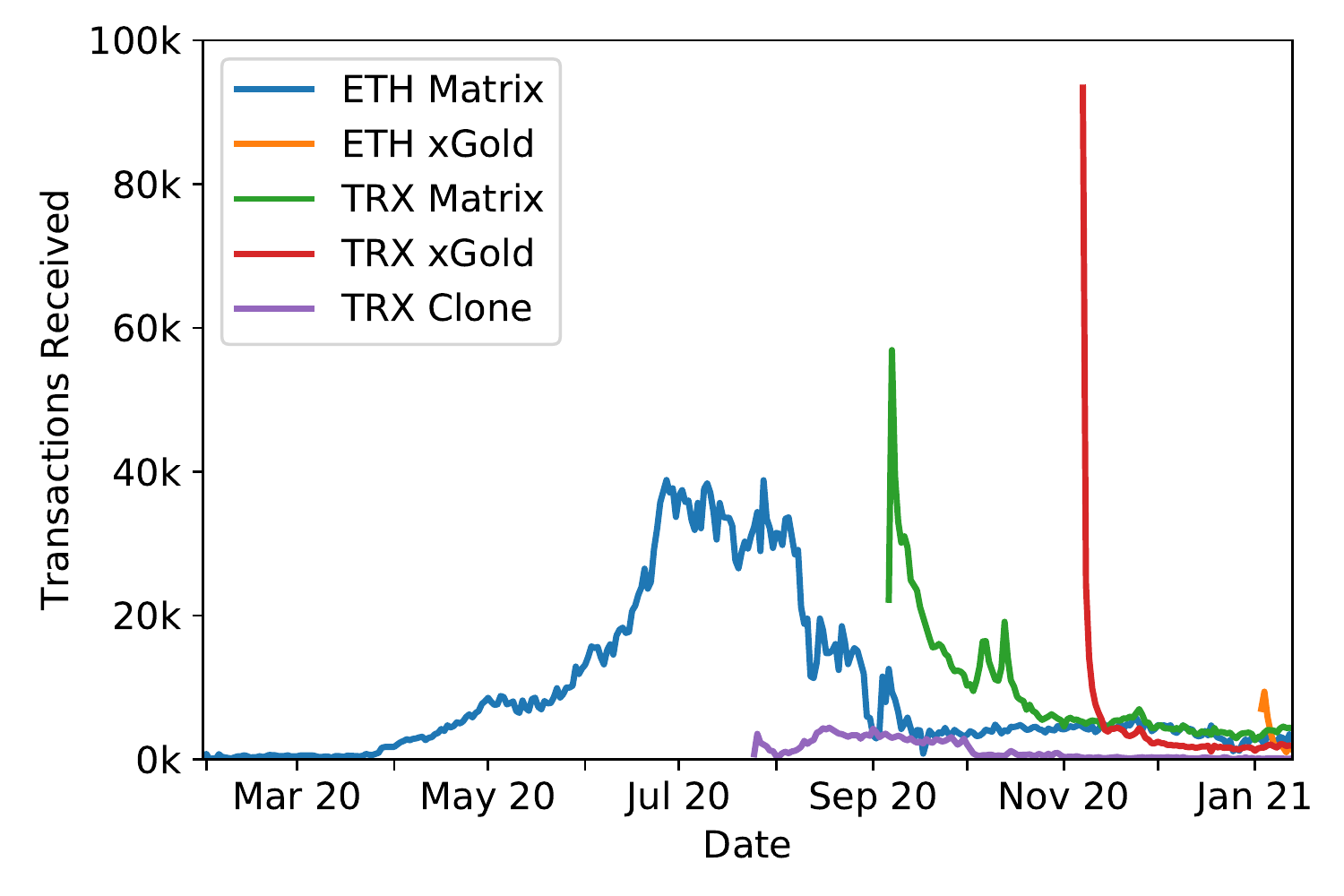}
    \caption[Number of transactions sent from users to the five various scam contracts]{Number of transactions sent from users to the four Forsage contracts across Ethereum and Tron and to an unofficial Tron-based clone.}
    \label{fig:transactionsovertime}
\end{figure}

Table~\ref{table:summary_all_contracts} shows a summary of statistics for the four official Forsage contracts, and one additional contract, TRX Clone, which is a cloned version of the Ethereum Matrix contract operating on Tron.  This clone launched before the official TRX Matrix contract, and has a different domain\footnote{forsagetron.io} but with graphics and style akin to the official website.  The official Forsage website added a warning after the clone's appearance, asking users to ``beware of fake resources'' and stating that the ``forsage.io'' website is the only official domain. 
In total, the table shows that the official Forsage contracts amassed over 267M~USD within the first year of operation. Among all of these contracts, the ETH Matrix contract brought in the most money and raised the highest amount on a single day: 3.7 million USD on August 1, 2020.  
The more recent xGold contracts (deployed on both Ethereum and Tron) were sent a combined 11.53 million USD in ETH and TRX in less than two months.  

Figure~\ref{fig:transactionsovertime} shows the number of transactions received by each contract over time.  For each contract introduced after the original ETH Matrix one, we observe a large number of initial transactions followed by a substantial drop.  We also see a decline in the number of transactions sent to the original ETH Matrix contract after the other contracts become available.  Given the relatively longevity and popularity of the ETH Matrix contract, we focus primarily on it for the remainder of this section.

\begin{figure}
    \centering
    \includegraphics[width=0.7\linewidth]{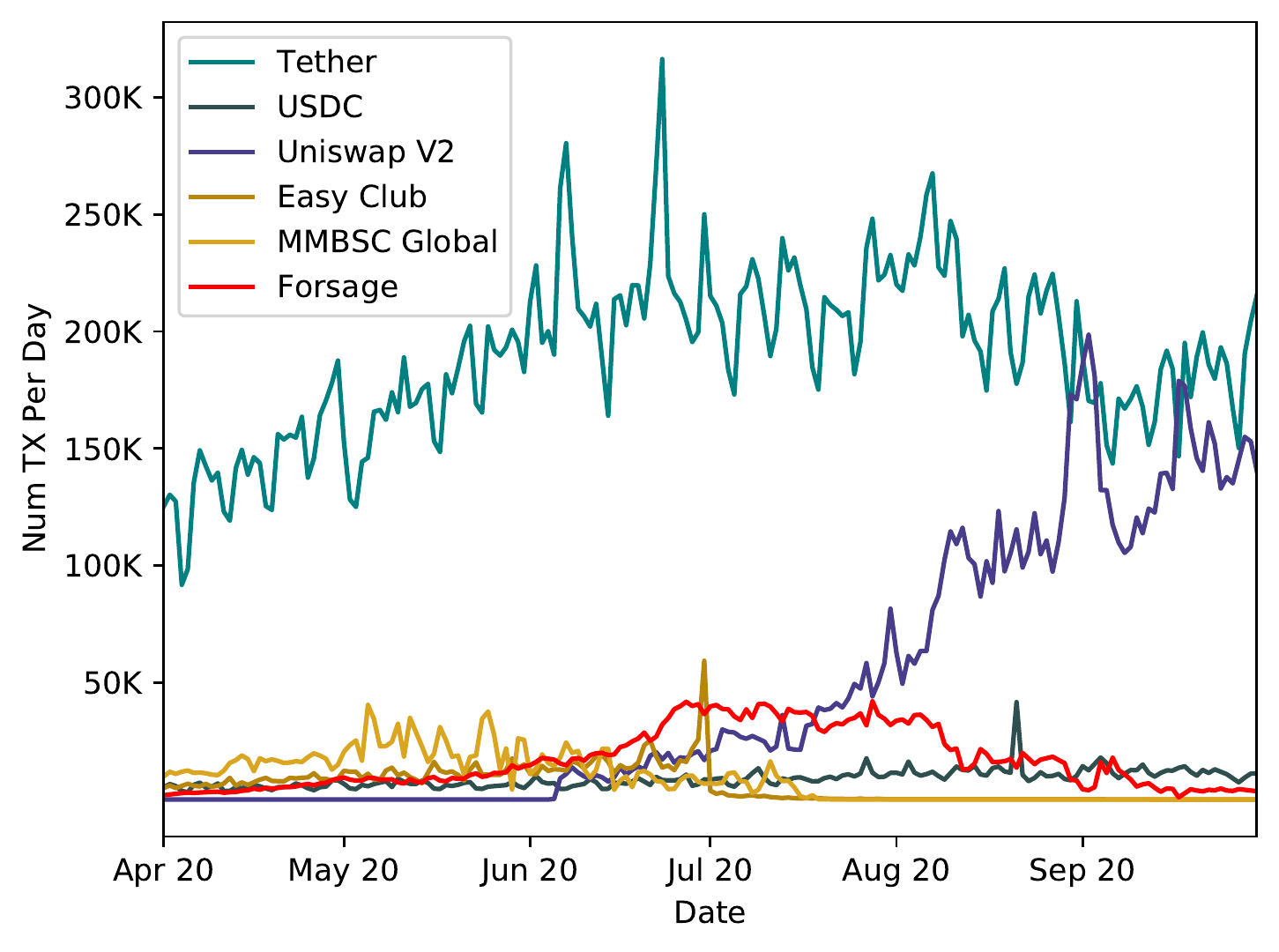}
    \caption[Daily transaction count associated with the six most transacted contracts between April 1 and September 30, 2020]{The daily transaction count associated with the six most transacted contracts between April 1 and September 30, 2020.  Here Forsage refers to the ETH Matrix contract.}
    \label{fig:summer_of_scams}
\end{figure}

To illustrate the popularity of Forsage, Figure~\ref{fig:summer_of_scams} shows the number of daily transactions associated with the six most popular contracts across a six-month period in 2020.  Of these contracts, Tether and USDC are stablecoins; Uniswap is a decentralized exchange; and Easy Club, MMBSC Global, and Forsage are believed to be scams/pyramid schemes. We can see that Tether is consistently the most popular contract and that for most of its peak from June to August, Forsage (as represented by ETH Matrix) had the second highest transaction rate among Ethereum smart contracts. This data is supported by Google Trends results for 2020: From April to August of 2020, Forsage had the highest search traffic globally of any of the smart contracts we studied, including both Tether and Uniswap, the two most heavily used smart contracts on the network as of the time of writing.

\subsection{Account behavior and profitability}
\label{subsec:user_behavior}

\begin{figure}[t!]
	\centering
	\includegraphics[width=0.8\linewidth]{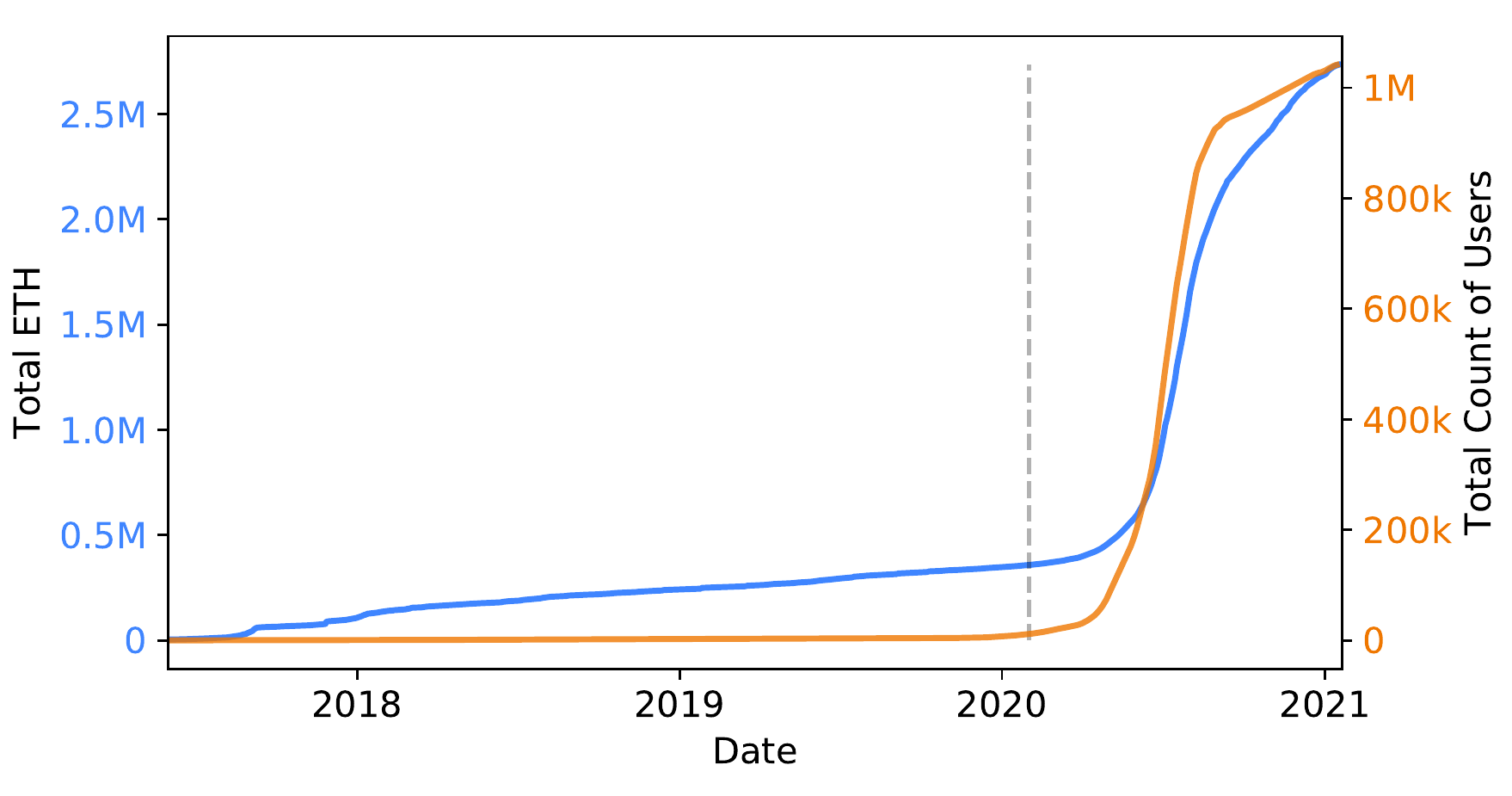}
 	\caption[Total ether received by Forsage users over time and total number of users according to when their accounts were first used]{Total ether received by Forsage users over time and total number of Forsage users according to when their accounts were first used, with a dashed line indicating the Forsage creation date.}
	\label{fig:user_eth_over_time}
\end{figure}

To understand how Forsage users obtained the funds needed to interact with the contract, we looked at the transactions that sent ETH to their accounts, and at when their accounts first became active.  
Figure~\ref{fig:user_eth_over_time} shows the ETH received by Forsage users over time and the cumulative count of active Forsage-related accounts (i.e., the first time an account was used that later interacted with the Forsage contract), with a vertical line indicating when Forsage was deployed.  
It is clear that these accounts became active and began to receive substantially more ether after the deployment of Forsage; in fact, 98.89\% of Forsage users had accounts that did not exist (or at least did not transact) before Forsage. 
We found a similar increase when looking at the number of transactions conducted by these users as well: prior to the deployment of Forsage, 11k accounts were involved in 278k %
transactions, but after Forsage's release this increased to 1.04M users engaging in 
16M transactions. %
While the curve in Figure~\ref{fig:user_eth_over_time} looks quite steep given the timescale, it in fact reflects a steady growth in the first appearance of accounts between April and August 2020, which aligns with the peak of Forsage we saw in Figure~\ref{fig:summer_of_scams}.  Each of these months saw thousands of new accounts appearing per day, on average: 1659 in April, 3653 in May, 8272 in June, 10,798 in July, and 4987 in August.  In contrast there were at most 20 new accounts appearing per day for each month in 2019 (except December, when there were 68).

To identify which types of services were the source of this money, we used tags from Etherscan.  Of the ETH sent to Forsage users, over 56\% (1.5M) came from untagged sources, and only 15\% came from known exchanges, with 5\% of this coming from the decentralized exchange Uniswap. As mentioned in Section~\ref{sec:overview}, Forsage promotional material recommends that users obtain ETH from TrustWallet.  This is a non-custodial service, which means accounts are associated with individual users rather than with the exchange.  Thus, if most users followed this advice, we would expect to see that most of the ETH came from untagged sources.

\begin{figure}
    \centering
    \includegraphics[width=0.65\linewidth]{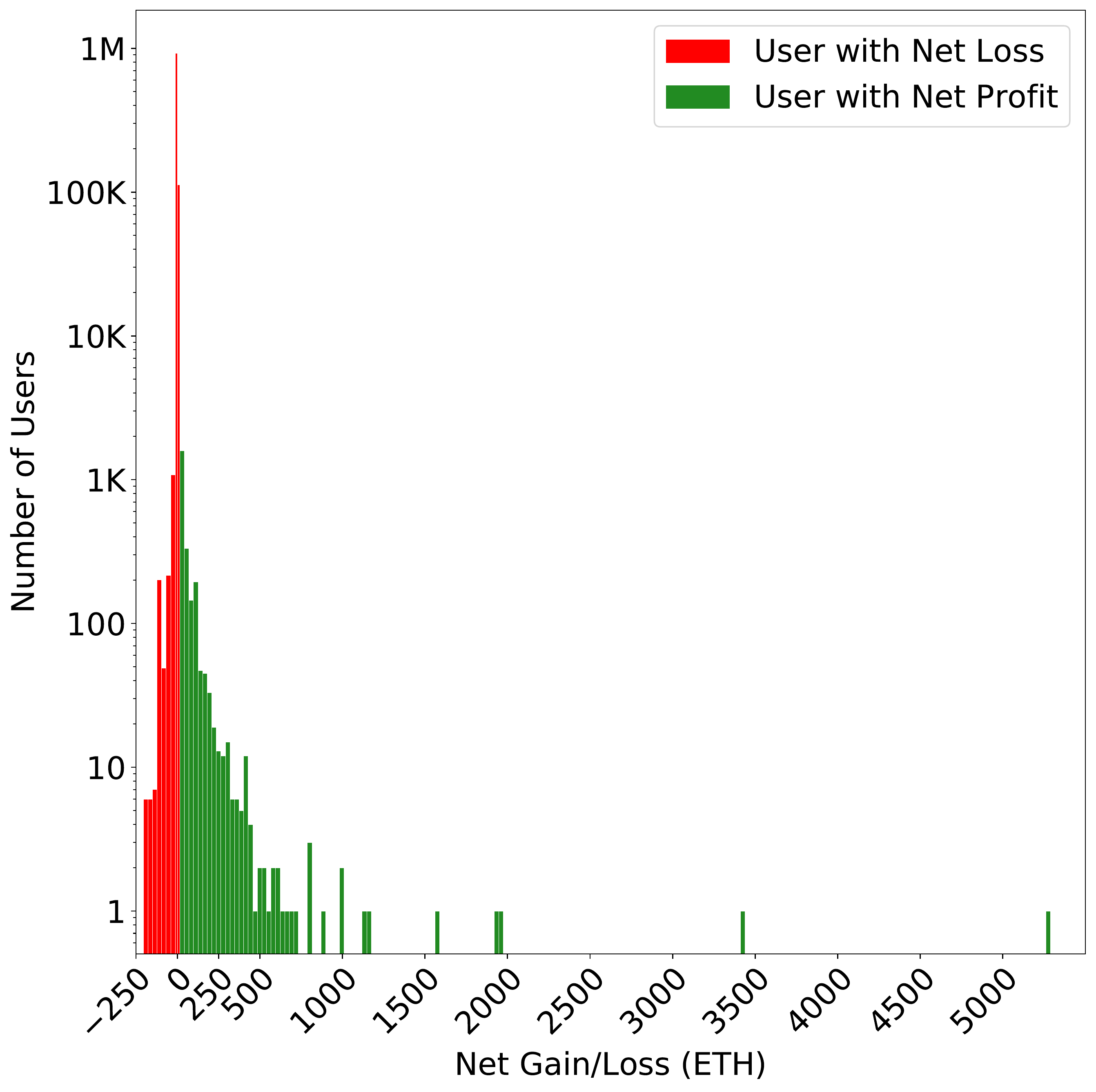}
    \caption[Profit/loss histogram of Ethereum accounts that interacted with the Forsage smart contract]{Profit/loss histogram of Ethereum accounts that interacted with the Forsage smart contract, on a log scale. This graph shows the number of accounts that made a profit or loss for each range of ETH. The majority of accounts incurred a small net loss, less than 1 ETH.}
    \label{fig:profit_histogram_all}
\end{figure}

\begin{figure}
    \centering 
    \includegraphics[width=0.8\linewidth]{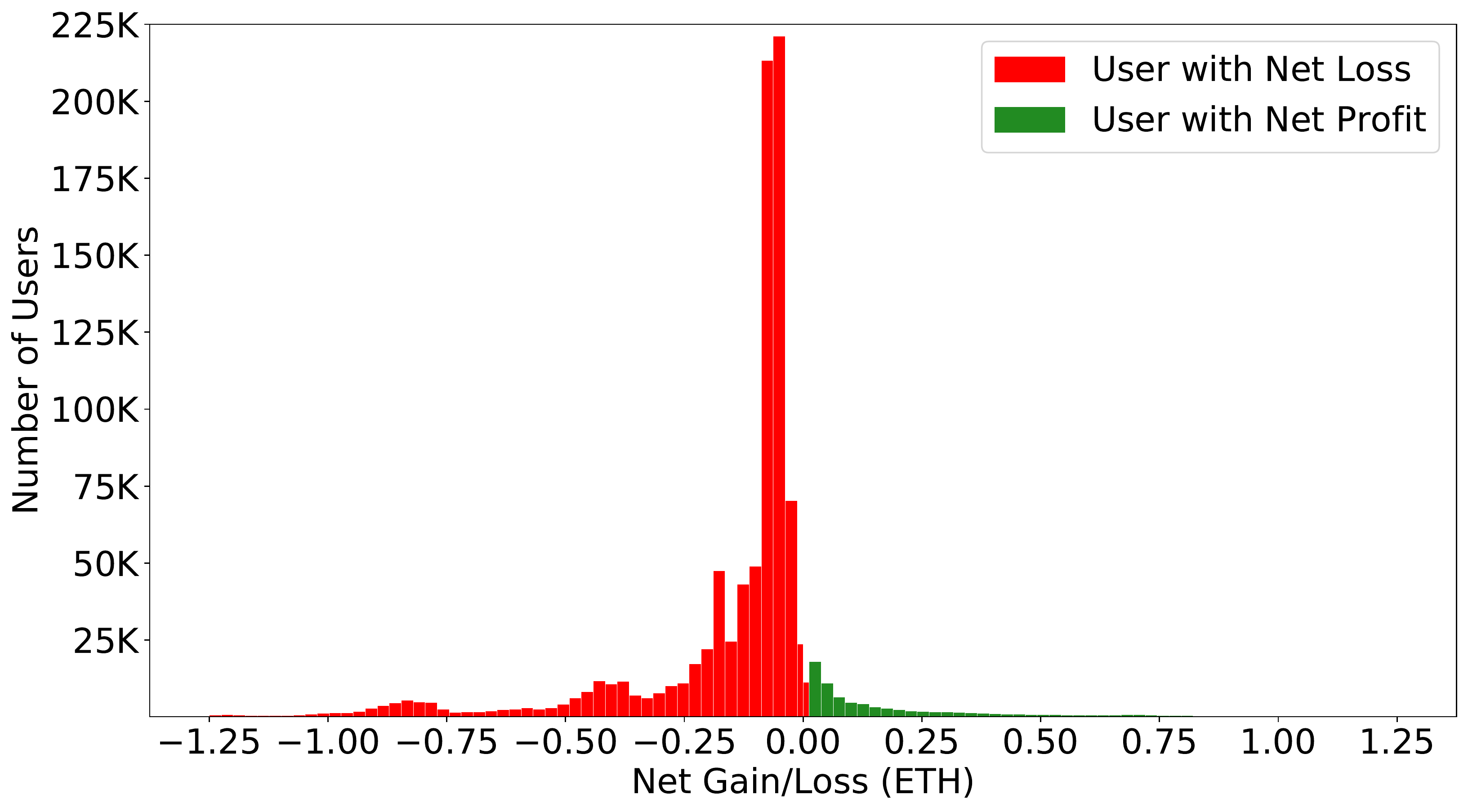}
    \caption[Profit/loss histogram of Ethereum accounts that interacted with the Forsage smart contract across a linear scale]{Profit/loss histogram of Ethereum accounts that interacted with the Forsage smart contract, centered around 0 and on a linear scale.  The vast majority of user accounts that interacted with Forsage lost between 0 and 0.25~ETH, with the peak occurring between 0.038 and 0.063~ETH.}
    \label{fig:profit_histogram_zoomed}
\end{figure}

Figures~\ref{fig:profit_histogram_all} and~\ref{fig:profit_histogram_zoomed} show a histogram of all of the accounts that interacted with the ETH Matrix contract organized by the amount of money either gained or lost by each account (including the amount spent on transaction fees) as of January 14, 2021.  In total, of the 1.04 million Ethereum addresses that took part in the ETH Matrix scheme, only 11.8\% (123,979) earned a profit. 
These profitable accounts made 265,618.52~ETH collectively,  %
and the loss-making accounts (919,194 in total) lost 305,785.44~ETH collectively (0.33~ETH on average).  We revisit these profit-making accounts below. Users incur additional losses from the high gas fees paid for transacting with the contract, as explained in Section~\ref{subsec:gas}.

\begin{table}
\centering
\begin{tabular}{ccc}
\toprule
 Address & Profit (in ETH) & Notes/First Seen \\
\midrule
 \ethaddr{0x81ca1e4de24136ebcf34ca518af87f18fd39d45e} & 5409.6 & Owner of the contract \\
 \ethaddr{0x44fc2e52243cf20ecc91f61ffa33e59fc7e1c148} & 3445.0 & March 22, 2020 \\
 \ethaddr{0xdedba197cb186e6d129110e71138ef6c6ca153d8} & 1954.9 & March 22, 2020 \\
 \ethaddr{0x4aaa7083535965d1cdd44d1407dcb11eec3f576d} & 1943.2 & January 31, 2020 \\
 \ethaddr{0x59b312f6cfe5b1864654d1942c8c979ad830777e} & 1573.0 & June 4, 2020 \\
\bottomrule
\end{tabular}
\caption[Five most profitable accounts that interacted with Forsage]{The five most profitable accounts that interacted with Forsage.}
\label{table:most_profitable_accounts}
\end{table}

\paragraph{Profit-making accounts:} 
\label{subsec:most_profitable}

The five addresses with the highest profits in Forsage can be found in Table~\ref{table:most_profitable_accounts}.  Perhaps unsurprisingly given our discussion in Section~\ref{sec:evaluation}, the most profitable Forsage user is the owner of the contract, who earned 5409.6~ETH, or 2.04\% of the total profits.  Collectively, the five most profitable users made 14,325.7~ETH, or 5.4\% of profits, despite representing only 0.0004\% of users.  The top 1000 users made 50\% of the total profits.

Examination of the five most profitable addresses shows that the most profitable address is another Ethereum contract created by the owner of the ETH Matrix contract.  Of the money received by this contract, 99\% came from ETH Matrix.  
The fourth highest earner sent 9\% of received ETH directly back to Forsage.  In fact, if we follow all the addresses to which this user sent money, we see over 1321~ETH sent back to Forsage eventually. 
Similarly, the fifth highest earner sent 204~ETH directly back to Forsage.

Some of the top addresses interact directly with other known scams, such as Beurax.com and TorqueBot.net, meaning they sent or received coins directly from addresses associated with these scams. The top five profit-making accounts received 6.987 ETH from these scams.

Interestingly, the first transaction sent to the address that deployed Forsage was from \ethaddr{0xb19dA4fd9f9A73A5A564C66D229B1E7219e8bdbe}, which is the Ethereum address that deployed Million.money.  This suggests interaction between smart contract-based scam operators.

Finally, we consider the extent to which users who profited by interacting with the Forsage ETH Matrix contract also interacted with other Forsage contracts.  The ETH xGold contract has 17,560 users, of which 17,129 (97.5\%) also interacted with ETH Matrix.  
Furthermore, the highest earner in xGold was the third highest earner in Matrix,  the fourth highest xGold earner was the seventh highest earner in Matrix, and the eighth highest earner in xGold was the second highest earner in Matrix. 
These three earners (all of which are within the ten wealthiest Matrix users) hold 21.85\% of net profits in xGold. 
This suggests that at least some prominent users of Matrix did indeed migrate over to xGold.

\section{Study of Forsage Community}
\label{sec:community}

\paragraph{Methodology:}
We studied the Forsage community by examining the presence of Forsage on social media. The Forsage website promotes official social media presences on Facebook, Instagram, Telegram, Twitter, and YouTube. All of these services have official APIs to collect data, but some of the research we conducted required manual interaction with the various social websites via a web browser, or more sophisticated data collection techniques like web scraping.

We manually watched YouTube videos to understand the claims that Forsage promotional videos make, as discussed in Section~\ref{sec:youtube}, and made requests to the public YouTube API for view count and other popularity-related data.\footnote{\url{https://developers.google.com/youtube/v3/docs/search/list}} To get a sense of Forsage's Facebook and Instagram presence, we manually browsed various Facebook groups and official Instagram accounts and leveraged the Facebook and Instagram Graph APIs.\footnote{\url{https://developers.facebook.com/docs/graph-api/}} 
Facebook group data is not available on the Graph API so we wrote a custom Python script leveraging the Selenium WebDriver browser automation tool to collect more in-depth data about Forsage Facebook groups and their users.\footnote{\url{https://www.selenium.dev/}} This yielded a dataset of just over 5000 of the most recent members from the largest Facebook group dedicated to Forsage.\footnote{\url{https://www.facebook.com/groups/forsageinformationgroup}}
Using the Twitter API for academic researchers,\footnote{\url{https://developer.twitter.com/en/docs/twitter-api/tweets/search/introduction}} we were able to scrape all tweets with the word ``Forsage" from January 1, 2020 until February 13, 2021. 
We used the official Telegram API \footnote{\url{https://core.telegram.org/}} to collect information about telegram groups related to Forsage.

\paragraph{Community size}
Forsage has a substantial presence on the social network sites that they target. 
This includes:
\begin{itemize}
    \item {\em Facebook:} 131 active Facebook groups with titles or descriptions including {``Forsage,''} containing 403,029 distinct Facebook members.
    \item {\em Instagram:} 24 Instagram accounts with Forsage in the username, disseminating information about Forsage to 24,747 followers of these accounts, with an additional 78,220 posts on the Instagram \#forsage hashtag.
    \item {\em Telegram:} 285,788 people spread across 49 different channels on Telegram dedicated to Forsage.
    \item {\em Twitter:} Our collected Twitter dataset included 85,085 tweets from 21,746 unique accounts, including 513 accounts on Twitter that feature Forsage in the account name. 
    \item {\em YouTube:} 57,551 video results from 325 different YouTube channels. %
\end{itemize}

The Forsage website also features a ``community'' subdomain\footnote{\url{https://community.forsage.io/}} that hosts a tips and tricks section, blog-post style news, a frequently asked questions section, ``academy courses'' that include video lectures on how to be an effective multi-level-marketer, and a Stack-Overflow-like site where users can ask questions and ``Forsage Community Authors'' answer.

A substantial amount of the Forsage online social media ecosystem may be driven by bots. We ran the University of Indiana's Observatory on Social Media (OSoMe) Botometer tool~\cite{botometertool} on our collected dataset of tweets and found that the tool identified roughly 47\% of the Forsage-related tweets we collected as coming from likely bot accounts. For comparison, in March of 2017, Varol et al.~\cite{botometerstudy} used an earlier version of the Botometer tool to perform a measurement study across all of Twitter and found that ``between 9 and 15\% of active Twitter accounts are bots."

\goodbreak
\begin{table*}[h!]
\small
\begin{tabular}{m{15mm}m{75mm}m{15mm}m{20mm}}
\toprule
Type & Claim & \ Appears  & Cumulative Views \\ 
\midrule
Wealth             & Forsage users make money forever.                                               & 3/10    & 425,356 \\
                   & Forsage users make unlimited income.                                            & 3/10    & 449,429 \\
                   & Forsage users make passive income.                                              & 3/10    & 247,344 \\
                   & Forsage users can earn hundreds of ETH in the first few weeks or months.        & 4/10    & 558,617 \\
\midrule
Risk               & Forsage is risk-free for users.                                                 & 3/10    & 393,927 \\
                   & No one can stop Forsage.                                                        & 4/10    & 558,617 \\
                   & Forsage is safe because the contract does not store funds.                      & 4/10    & 530,165 \\
                   & Forsage is scam-proof.                                                          & 3/10    & 393,927 \\
\midrule
Ethereum           & The video explains what Ethereum is for new users.                              & 5/10    & 637,881 \\
Education          & The video explains what a smart contract is for new users.                      & 5/10    & 637,881 \\
\midrule
How to   & Successful Forsage users open at least 3 slots per program to start (0.2 ETH).  & 6/10    & 745,960 \\
use             & Users should buy more slots (send Forsage more money) as soon as they earn.     & 5/10    & 654,727 \\
Forsage                   & The more slots you open (money you send Forsage), the more you will earn.       & 4/10    & 511,858 \\
                   & If you do not keep opening slots (sending money to Forsage), you will not earn. & 5/10    & 444,539 \\ 
 \bottomrule
\end{tabular}
 \caption[Coded claims extracted from the top 10 most viewed Forsage videos]{We coded repeated claims that appear across the top 10 most viewed, English language videos on YouTube, which mention "Forsage" in their title to measure user expectations when joining Forsage. }
\label{table:youtubeclaims}
\end{table*}

\subsection{Analysis of Forsage YouTube Promotion}\label{sec:youtube}

Forsage promotional materials offer a window into users' expectations for the contract. They also provides insight into how mention of the technical properties of blockchain technology is harnessed to manipulate novice users. We find that the information gap between those who understand blockchain technology and the broader community provides opportunities for scammers.

\begin{table}
\centering
\begin{tabular}{lccc}
\toprule
Country & Facebook & Twitter & YouTube \\
\midrule
Nigeria & 84 & 4878 & 3 \\
Philippines & 272 & 668 & 14  \\
India & 97 & 488 & 88 \\
United States & 45 & 1019 & 26 \\
Indonesia & 17 & 203 & 8 \\
TOTAL & 771 & 10200 & 216 \\
\bottomrule 
\end{tabular}
\caption[Top five countries with the highest absolute level of Forsage user engagement]{Top five countries with the highest absolute level of Forsage user engagement. User engagement here is measured as a country's total number of Facebook observed users in the most popular Forsage Facebook group, plus its analogous number of Twitter observed users that tweeted about Forsage in 2020, and YouTube data for the number of YouTube channels with geo-tagged locations that produced videos with Forsage in the title of the video.}%
\label{tab:socialmedia-absolute}
\end{table}

YouTube is a primary promotional channel for Forsage. Each participant joining Forsage is referred to an official YouTube video explaining the program~\cite{youtube1}. We searched YouTube for English language videos with ``Forsage'' in the title and tracked the claims that repeat across videos to measure user expectations for Forsage. The search for most viewed videos about Forsage also returned promotional videos in Tagalog, Russian, Hindi, Tamil, Bangala, Telugu, Indonesian, and Spanish. Quasi-official (they share the same branding) Telegram chat groups for Forsage news exist in English, Spanish, French, Italian, Russian, Arabic, Portuguese, Hindi, Tamil, German, Azerbaijani, and Turkish. 

Recommendation algorithms, like the one used by YouTube for search results, work in terms of popularity measured in views. The most viewed videos on YouTube are the most likely to be seen by users. We selected the top ten videos by views to qualitatively measure what users who search for informational videos about Forsage would see and hear about the program and gain a sense of participant expectations. We did so by coding the claims asserted about Forsage in these videos. We focused on just the top ten videos because coding claims is a labor-intensive, manual process. A researcher watched each video and noted if each video contained any instance of certain claims (see Table~\ref{table:youtubeclaims} and Table~\ref{app:youtube}). Each video was watched and coded twice to ensure accuracy.

The top ten YouTube videos we coded had between 267,008 views (1st) and 61,996 views (10th). Beyond the videos we coded, the 11th most viewed video had just over 50,000 views\footnote{\url{https://youtu.be/aGi5G5mTCUM}} 
and the 20th had 33,000 views.\footnote{\url{https://youtu.be/9vlOYRSLaHI}}

\begin{table}
\centering
\begin{tabular}{lll}
\toprule
Rank & Title                                                                 & {Views}                  \\ \midrule
1    & \begin{tabular}[c]{@{}l@{}}Forsage Overview: Earn Ethereum Daily! \\ \end{tabular}                    & 267008 \\
2    & \begin{tabular}[c]{@{}l@{}}Forsage Presentation - How does Forsage work \\ \end{tabular}  
& 120425    \\
3    & \begin{tabular}[c]{@{}l@{}}Forsage Smart Contract - \$735 Made Without Referring \\  Anyone \end{tabular}            
    & 113677        \\
4    & \begin{tabular}[c]{@{}l@{}}FORSAGE: HOW TO EARN WITHOUT RECRUITING \\ ANYONE IN FORSAGE \end{tabular} 
    & 117931     \\
5    & \begin{tabular}[c]{@{}l@{}}Forsage Smart Contract Review - Is It A SCAM Or Legit \\  Ethereum MLM?  \end{tabular}    & 106261  \\
6    & FORSAGE.io - BIG SPECIAL EVENT     & 91973   \\
7    & 
\begin{tabular}[c]{@{}l@{}}Forsage Smart Contract \$1,778 Made Without Referring \\ a Single Person  \end{tabular} 
  & 91188     \\
8    & Forsage Review - Is Forsage a Scam or Legit?                          & 79264      \\
9    & \begin{tabular}[c]{@{}l@{}}Smartway Forsage REVIEW - First Ever SCAM PROOF \\ Program \end{tabular} 
                & 64923     \\
10   &  \begin{tabular}[c]{@{}l@{}}how to make money on forsage without referring anyone\\   \end{tabular}                    & 61996    \\ \bottomrule
\end{tabular}
 \caption[Top 10 Forsage videos from the official channel ordered by views]{Top 10 Forsage videos from the official channel ordered by views. }\label{app:youtube}
\end{table}

The top 10 ``Forsage'' YouTube videos by views as of December 14, 2020 (see Table~\ref{app:youtube}) fit into three categories: official promotion, user-led recruitment, and user reviews. Two of the videos were official promotion posted to Forsage's YouTube channel~\cite{youtube1, youtube6}. Table~\ref{table:youtubeclaims} shows the repeated claims across the top ten videos. 

In recruiting new users, Forsage promoters pointed to users who earned tens of thousands of dollars per day and hundreds of thousands of dollars per month, showing images of successful users' Forsage dashboards displaying six-figure returns. %
Forsage official promotion videos highlight the immutable nature of the smart contract and the transparency of Ethereum as proof that Forsage cannot be a scam.  %
They also make claims about the life-changing wealth and unstoppable, passive income that users could unlock from the Forsage contract. %

Forsage promotional videos also provide basic explanations of blockchains, Ethereum, smart contracts, and how to use a cryptocurrency wallet to pay the contract, implying that they expect users to be cryptocurrency novices. %
Only one of the top ten videos identifies Forsage as a scam and warns users against using it.  %

Many of the incorrect claims made in the Forsage promotional YouTube videos also appear on the Forsage website and in the questions section of the official Forsage Community website. 
 
\goodbreak
\begin{figure*}[t!]
  \centering
    \includegraphics[width=0.99\linewidth]{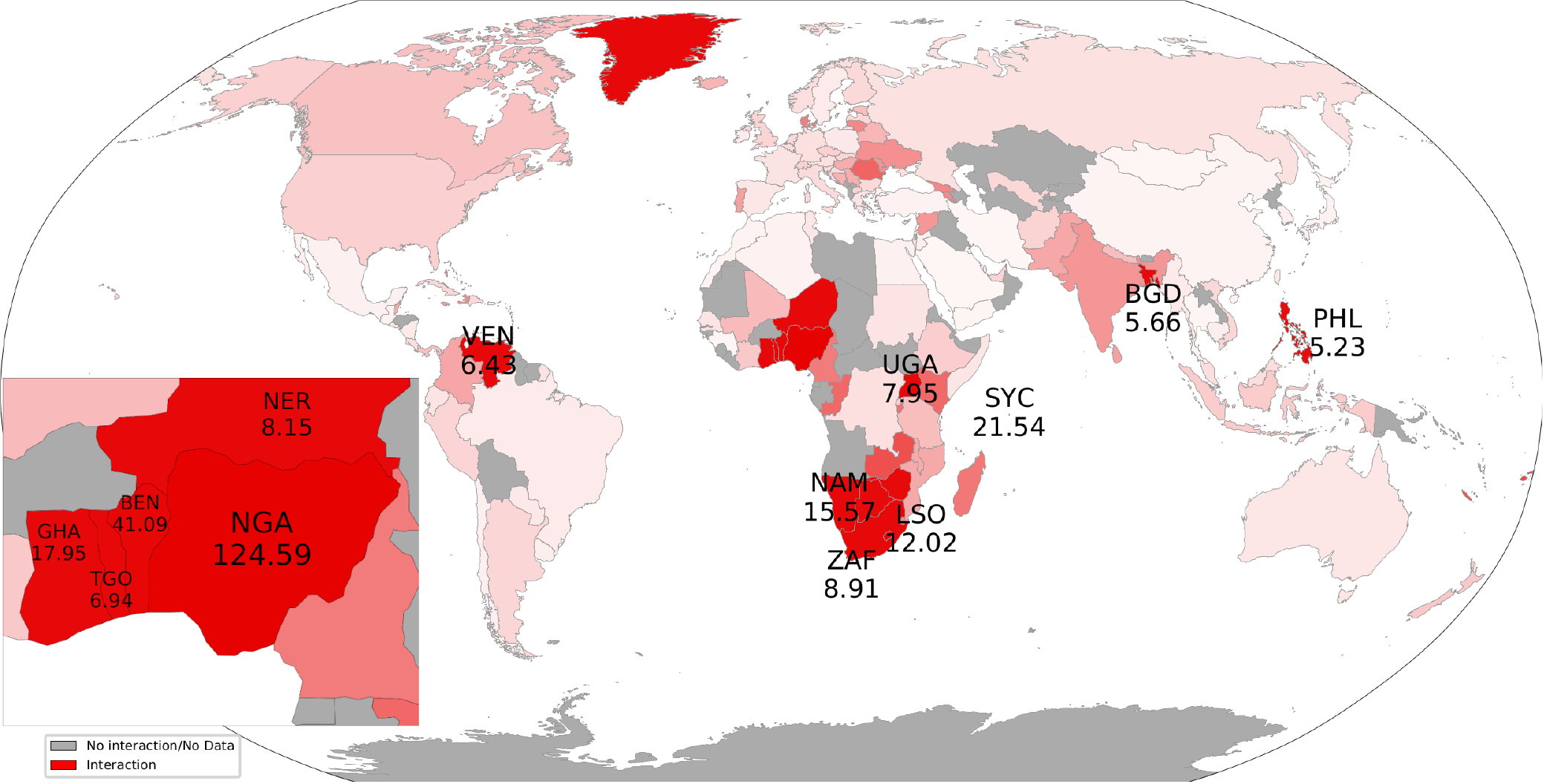}
    \caption[Forsage social media interaction heat map by country.]{Forsage social media interaction heat map by country. Country labels indicate the ISO-alpha-3 name of the country and the number of Forsage users per 100k people in that country.  The data reflects the public location of members in a popular Forsage Facebook group and Twitter users that tweeted about Forsage. Countries depicted in gray had no Forsage interaction. The intensity of color from white to red is scaled linearly from the 0th percentile of data to the 90th percentile, and everything above 90\% of the data is colored the same shade of dark red. This slightly understates the relative depth of penetration in outlier countries like Nigeria.
    }
    \label{fig:heatmapsocialmedia}
\end{figure*}

\subsection{Forsage user geography}
\label{sec:community-geography}

Since transactions on the Ethereum network do not carry any inherent geographic metadata, we turned to social media analysis in order to gain a sense of the geographic placement of people interested in Forsage.  In the data we collected on members of Forsage-related Facebook groups, we found 771 users that publicly listed a country location on their Facebook profile. We also found 10,200 unique Twitter accounts that publicly posted their geographic location.  YouTube does not expose information about geographic location of the consumers of YouTube videos, but YouTube channels that produce videos can choose to include country location in their channel profile.  We summarize this data for the five countries with the highest number of active users in Table~\ref{tab:socialmedia-absolute}.  Despite having a substantial population and being the  nationality of the founders of Forsage, Russia was not a large source of Twitter or Facebook content, although the country did produce a large number of YouTube videos and content about Forsage.

The high number of Forsage users in the Philippines may explain why the Philippines SEC took action to raise awareness about the malicious intent behind Forsage~\cite{psec_warning_2020,psec_forsage_2020}, unlike other countries. Likewise, Nigeria has high penetration rates for both cryptocurrency and Forsage, and has recently banned cryptocurrency payments from its banking sector~\cite{nigeriabanscrypto}.
While each of these five countries had high Forsage activity in absolute terms, they also have large populations.  
We thus normalized our Facebook and Twitter data relative to the specific populations on each service for each country (i.e., the number of people per country divided by a public estimate of the number of Facebook and Twitter users in that country) to get a sense of the number of Facebook and Twitter users, per 100,000 users, that interacted on each platform with the Forsage topic. Statistics for the number of Facebook and Twitter users per country came from Miniwatts Marketing Group, WeAreSocial, and Hootsuite~\cite{miniwatts, hootsuite}. We did not include the YouTube data at this stage as it was too small to be useful.  We gave equal weight to the numbers for Facebook and Twitter to produce the heat map in Figure~\ref{fig:heatmapsocialmedia}.

Our normalized data showed that Forsage is most popular in Nigeria and the African continent, the Philippines, and Venezuela. Greenland, the Seychelles, and some Caribbean islands may appear to have heavy Forsage penetration, but may be outliers due to small population sizes. Google Trends traffic and geographic data agree with our conclusions: Google Trends shows the greatest amount of population-adjusted search traffic in Nigeria and surrounding West African countries, and shows a peak in user search interest in July 2020, which is when we observed a similar peak in transactions involved Forsage in Figure~\ref{fig:transactionsovertime}.

Familiarity with cryptocurrency does not appear to have any positive or negative correlation with interest in Forsage: The 2021 Statista Global Consumer Survey~\cite{cryptopercountry} lists the top countries globally with the reported highest number of cryptocurrency users. 
Vietnam (\#2) and China (\#3) both had relatively high levels of cryptocurrency use, but low levels of interest in Forsage.   Similarly, familiarity with cryptocurrency does not appear to prevent people from falling for the Forsage scam, as in the case of Nigeria and the Philippines (\#1 and \#3 globally for cryptocurrency usage). Nigeria may be a special case, as Statista found that almost a third of Nigerians said they used cryptocurrency, far beyond most countries. It is also an outlier in the data for interest in Forsage. 

\section{Proposed Solutions}
\label{sec:solutions}

\subsection{Targeted education}
From our analysis of Forsage user locations in Section~\ref{sec:community-geography}, the majority of Forsage victims are located in only a few countries. This concentration lends itself well to a targeted education campaign and warnings from local financial leaders about the Forsage scam.  For example, a simple user dashboard showing the number of Forsage users who lose money from the contract---more than 88\% as of January 15th, 2020---could serve as an effective tool to combat disinformation from Forsage promoters about the wealth users can amass. Such statistics may be more effective than general warnings such as that issued by the Philippines SEC (see below).

\subsection{Law enforcement and regulation}
Past cryptocurrency pyramid schemes, including Plustoken, Wetoken, Onecoin, and Bitconnect, have collapsed as a result of government sanctioning, which included the arrest or warrants for the arrest of the founders and leadership~\cite{palmer_2020, haig_2020, bel_2020, madeira_2020}.  Similar attempts have been made around the world in regards to Forsage.  On June 30, 2020, The Philippines Securities and Exchange Commission (PSEC) issued numerous warnings declaring that Forsage was not a registered entity within their jurisdiction and was operating without a license. 
On September 30, 2020 the PSEC released a public announcement, mentioning that Forsage was publicly selling securities as investment contracts without a license~\cite{psec_cadpublic_2020, psec_cadpublic_state_2020, psec_forsage_2020, psec_warning_2020}. The PSEC served a cease-and-desist order. Forsage refused to comply, responding that they
``are outside the Commission’s jurisdiction.'' 
On March 22nd 2021, the Commissioner of Securities and Insurance of the U.S. state of Montana ordered Forsage to cease and desist from operating a pyramid scheme in Montana~\cite{montanaDA, montanaDA2}. 

To date, the authors of this paper are unaware of any public arrests made in relation to the Forsage contract. The contract authors continue to profit and their Ethereum addresses actively submit transactions to the network.

\subsection{Voluntary blocklisting}

Previous research has shown blocklisting can effectively combat scams and illicit activity. Moser et al. found that transaction blocklisting of illicit cryptocurrency funds is an effective additional layer above existing anti-money laundering (AML) and know-your-customer (KYC) requirements for cryptocurrencies~\cite{moser2019effective}. Previous research in illicit online pharmaceutical sales found that the payment processing services are the most fragile part of the scam~\cite{180200}. These services play a similar role in online pharmaceutical sales to fiat-accepting cryptocurrency exchanges in Forsage, suggesting that access to exchanges, which could be revoked with blocklisting, may be the the most fragile part of the scam.  
Crypto Defenders Alliance (CDA)\footnote{\url{https://cryptodefendersalliance.com/}} and CryptoSafe Alliance\footnote{\url{https://www.cryptosafe.org/}} are two examples of groups that operate a blocklist. 

On the other hand, blocklists can be biased and enable forms of censorship, and addresses that are blocked in one region may not be considered suspicious or criminal in other regions.  
To understand how professionals navigate these tensions, we spoke to an anti-money laundering cryptocurrency investigator at a high profile exchange.  This expert expressed a belief that it is the responsibility of law enforcement and regulators to comment 
on whether or not an address should be blocked, and that it would be unfair and unjust to hold a user's funds without an explicit request from law enforcement or a court of competent jurisdiction.  Nevertheless, some exchanges have joined the alliances mentioned above, due to the time and resources required to maintain a dedicated list of blocked addresses themselves.

\section{Future work}

There are a number of areas this work can be extended to. 
Due to limited time and resources we were unable to further delve into the xGold contract, thus we leave the exploration of this contract for future research. 
Since the public release of this research, the Forsage group has 
released two products onto Binance USD blockchain, 
the matrix product and a brand new platform titled \textit{Forsage xXx}.
Future research can look into analysing these two newer products. 
Our research focused on analysing a single blockchain based pyramid scheme, future work can look into analysing other pyramid schemes and related scams in Ethereum overall, in a way to quantify the potential pyramid schemes on the network. 

\section{Conclusions}

We presented an in-depth measurement study of Forsage, a smart-contract pyramid scheme. Forsage is currently active and was at one time the second most actively used contract in Ethereum. 

We found that community claims regarding the open and verifiable nature of Forsage are belied by the contract's considerable complexity.
Our study consequently required a number of different data gathering approaches. It also required the creation of new tools---of potential independent interest and to be open-sourced---to analyze the state of the Forsage contract. Thanks to these tools, our study provides detailed insights into the mechanism design, transaction costs, and other features of Forsage.  

Among our key findings were that the vast majority of Forsage accounts---over 88\%---incurred losses, for a combined total loss of 305,785 ETH. The contract owner, in contrast, earned over 5000 ETH (well over 1M~USD), while a small number of other accounts at the top of the pyramid earned similarly large sums. 

Our analysis of Forsage promotional materials reveals that scammers in the Forsage community have taken advantage of misconceptions and misinformation about blockchain technology, using properties like open-source code and transaction transparency as a source of legitimacy with users who lack the skills necessary to understand the  contract's behavior. Our analysis of Forsage on social media shows geographically distinct communities of scammers and victims, with the scammers based primarily in Russia and victims apparently located mainly in Nigeria, southern Africa, the Philippines, Venezuela, Indonesia, and India.

Public warnings about Forsage by entities such as the Philippines SEC have had little apparent effect. We show that Forsage creators have launched new and currently lucrative Forsage variants, some now on blockchains other than Ethereum. We hope that our findings can help stem this spread. In addition to providing insights that may serve to educate potential victims, our study demonstrates highly concentrated earnings among top-earning accounts, suggesting that targeted blocklisting could be an effective step to slow the growth of Forsage and contracts like it. 

\chapter{Conclusion}
\label{chapterlabelConclusion}

This thesis has presented techniques to empirically analyse privacy and crime in blockchain technologies: analysing the privacy and usage of Zcash; tracking transactions moving across chains; and studying a high profile and ongoing smart-contract pyramid scheme, Forsage. 

In Chapter~\ref{chapterlabelZcash} we presented an analysis of privacy and measurements on Zcash - a privacy-focused cryptocurrency forked from Bitcoin. Our analysis demonstrated that a user's privacy when interacting with the shielded pool is greatly determined by the actions of the surrounding users. 
User addresses can be clustered using known and novel heuristics. The hacker collective, TheShadowBrokers, used Zcash as one resource to sell vulnerabilities and tools. 
This chapter offers one of the first academic insights into the working of this privacy coin, demonstrating that privacy coins do not alleviate the anonymity risks demonstrated in non-privacy cryptocurrencies. 

We then addressed the issue of cross-currency tracking in Chapter~\ref{chapterlabelTracing}. 
We exhibited working examples of using our novel heuristics to follow coin ownership across different cryptocurrencies and examined and clustered entities within the ecosystem. 
This concluded with case studies, revealing that multiple scammers used the system to move their funds, and presenting measurements to showcase how privacy coins are used in the system.

Finally, in Chapter~\ref{chapterlabelForsage} we presented an in-depth measurement study of a 
smart-contract pyramid ecosystem that processed \$267 million USD worth of Ethereum. 
The study was conducted using a multi-faceted approach. We revealed the inner workings of one of the obfuscated smart-contracts, measured the profits and losses, revealed the broadness of the global marketing scheme, identified the failed efforts by law enforcement to shutdown the enterprise, and proposed potential countermeasures. 

\section{Future Directions}

In this section, we list some interesting areas of research that could be explored to continue research efforts in this field. 

\paragraph{Anonymity of Zcash} The research in Chapter~\ref{chapterlabelZcash} was conducted between 2017 and 2018. 
As previously mentioned, the developers have since made multiple changes to Zcash, such as improvements to the underlying cryptography which includes additions of new shielded pools. 
In addition, they have given grants to external developers to create user friendly mobile wallets which allow transactions with the shielded pool. 
Future research might look at potentially repeating our work to identify whether the ecosystem has changed and whether the analytics upon the shielded pool still work.

\paragraph{Researching Scams} In Chapter~\ref{chapterlabelForsage} we focused our analysis on deconstructing the Forsage matrix contract. The platform has since expanded their offerings with additional schemes called \textit{xGold} and into the Binance USD platform with \textit{xXx}. Future research on pyramid schemes could extend our work and analyse the inner workings of these new schemes.

\paragraph{Privacy with Taproot}  Taproot~\cite{wuille_nick_towns_2021} is an upgrade in Bitcoin that is expected to launch in November 2021. As one of the first upgrades to be approved by miners in over four years, its aim is to achieve better transaction privacy and efficiency, and to improve the potential for Bitcoin-based smart contracts. 
A new discussion~\cite{anania_hodler_2020} reveals that this update may not alter the current effectiveness of clustering, however, it will create new opportunities for new clustering heuristics to be applied to Taproot-specific scripts. 
 
\section{Closing Thoughts}

We have concluded that heuristics-based approaches can be applied to analysing privacy and crimes in blockchain-based systems. We demonstrated that privacy-focused cryptocurrencies do not alleviate previous privacy risks and that address tracking and clustering is still possible in these ecosystems. Cross-chain payment systems do not act as a shield against preventing tracking of payments. Those involved in illicit activities in cryptocurrencies can, to some extent, have their transactions analysed and tracked. The methodologies described in this thesis can be used by agencies in order to detect crime, and by developers to test and improve the privacy of payment systems.

\printbibliography
    
\end{document}